\documentclass{aa}  

\usepackage{graphicx}
\usepackage{txfonts}
\usepackage{lipsum}

\usepackage[T1]{fontenc}
\usepackage{ae,aecompl}
\usepackage{enumitem}
\usepackage{float} 
\usepackage[usenames,dvipsnames,table,xcdraw]{xcolor}
\usepackage{xcolor}
\usepackage{xspace}
\usepackage[para]{threeparttable} % tables with footnotes
\usepackage{multirow}
\usepackage{multicol}
\usepackage{pdfpages}
\usepackage[version=4]{mhchem}

\usepackage{subcaption}
\usepackage[colorlinks=true,linkcolor=Blue,citecolor=Blue,urlcolor=black,anchorcolor=black,bookmarks=true,breaklinks=true,hypertexnames=false,bookmarksnumbered=true]{hyperref}
\begin{document} 

\newcommand{\vdag}{(v)^\dagger}
\newcommand\aastex{AAS\TeX}
\newcommand\latex{La\TeX}

\newcommand{\todo}[1]{\textcolor{red}{todo:#1}}
\newcommand{\solar}{$_\odot$}
\newcommand{\zml}{$Z = 10^{-4}$}
\newcommand{\zmu}{$Z = 0.03$}
\newcommand{\solarperyr}{$_\odot \, \textrm{yr}^{-1}$}
\newcommand{\bse}{\texttt{bse}}
\newcommand{\binaryc}{\texttt{binary\_c}}
\newcommand{\nupycee}{\texttt{NuPyCEE}}

\newcommand{\Mchand}{$M_{\rm Ch}$}
\newcommand{\Mwd}{$M_{\rm WD}$}
\newcommand{\Mhsh}{$M_{\rm H \ shell}$}
\newcommand{\Mhesh}{$M_{\rm He \ shell}$}
\newcommand{\Mig}{$M_{\rm ig}$}
\newcommand{\Mdot}{$\dot{M}$}
\newcommand{\ace}{$\alpha_{\rm CE}$}
\newcommand{\lce}{$\lambda_{\rm CE}$}
\newcommand*\diff{\mathop{}\!\mathrm{d}}
\newcommand\rate{\mathcal{R}}
\newcommand{\Z}{$Z$}
\newcommand{\tento}[1]{$10^{#1}$}%tento
\newcommand{\timestento}[2]{$#1 \times 10^{#2}$}
\newcommand{\iso}[2]{\hbox{${}^{#1}{\rm #2}$}}

   \title{Nova contributions to the chemical evolution of the Milky Way}

   \subtitle{}

   \author{
    Alex J. Kemp\inst{1}$^,$\inst{2}\thanks{
    \email{alex.kemp@kuleuven.be}}
    \and
    Amanda I. Karakas\inst{2}$^,$\inst{3}
    \and
    Andrew R. Casey\inst{2}\inst{4}
    \and
    Benoit C\^{o}t\'{e}\inst{5}$^,$\inst{6}
    \and
    Robert G. Izzard\inst{7}
    \and
    Zara~Osborn\inst{2}$^,$\inst{3}
  }

   \institute{Institute for Astronomy (IvS), KU Leuven, Celestijnenlaan 200D, 3001, Leuven, Belgium
        \and
             School of Physics \& Astronomy, Monash University, Clayton 3800, Victoria, Australia
        \and
            Centre of Excellence for Astrophysics in Three Dimensions (ASTRO-3D), Melbourne, Victoria, Australia
        \and
            Center for Computational Astrophysics, Flatiron Institute, New York City, New York, USA
        \and
             Department of Physics and Astronomy, University of Victoria, Victoria, BC V8P 5C2, Canada
        \and
             Konkoly Observatory, Research Centre for Astronomy and Earth Sciences, E\"otv\"os Lor\'and Research Network (ELKH), Konkoly Thege Mikl\'{o}s \'{u}t 15-17, H-1121 Budapest, Hungary
        \and
            Astrophysics Research Group, University of Surrey, Guildford, Surrey GU2 7XH, UK
             }

   \date{}

   \date{Received May 20, 2024; accepted May 21, 2024   }

% \abstract{}{}{}{}{} 
% 5 {} token are mandatory
 
  \abstract
   {The explosive burning that drives nova eruptions results in unique nucleosynthesis that heavily over-produces certain isotopes relative to the solar abundance. However, novae are often ignored when considering the chemical evolution of our Galaxy due to their low ejecta masses. Galactic chemical evolution studies including novae are rare and have previously relied upon simplified treatments for the behaviour of nova populations.}
   {In this work, we use previously computed synthetic nova populations and the galactic chemical evolution code \texttt{OMEGA+} to assess the impact that novae have on the evolution of stable elemental and isotopic abundances.}
   % other than Li, which was covered in a previous paper.}
   {We combine populations of novae computed using the binary population synthesis code \binaryc\ with the galactic chemical evolution code \texttt{OMEGA+} and detailed, white dwarf mass-dependent nova yields to model the nucleosynthetic contributions of novae to the evolution of the Milky Way. We consider three different nova yield profiles, each corresponding to a different set of nova yield calculations.}
   {We examine which nova sites contribute most to which isotopes. Despite novae from low-mass white dwarfs (WDs) dominating nova ejecta contributions, we find that novae occurring on massive WDs are still able to contribute significantly to many isotopes, particularly those with high mass numbers. We find that novae can produce up to 35\% of the Galactic \iso{13}C and \iso{15}N mass by the time the model Galaxy reaches [Fe/H]~=~0, and earlier in the evolution of the Galaxy (between [Fe/H]~=~-2 and -1) novae may have been the dominant source of \iso{15}N. Predictions for [\iso{13}C/Fe], [\iso{15}N/Fe], \iso{12}C/\iso{13}C, and \iso{14}N/\iso{15}N abundances ratios vary by up to 0.2 dex at [Fe/H]~=~0 and by up to 0.7 dex in [\iso{15}N/Fe] and \iso{14}N/\iso{15}N between [Fe/H]~=~-2 and -1 (corresponding approximately to Galactic ages of 170 Myr and 1 Gyr in our model). The Galactic evolution of  other stable isotopes (excluding Li) is not noticeably affected by including novae. For most isotopes, agreement is generally good between the three different yield profiles we consider. Isotopes where agreement is relatively poor include: \iso{3}He (especially at high \Mwd), \iso{7}Li, \iso{18}O, \iso{19}F, and the >1.3 M\solar\ regime of \iso{29}Si, \iso{33}S, \iso{34}S, \iso{35}Cl, and \iso{36}Ar.}
   {}


   \maketitle

\section{Introduction}
The origin of the elements is one of the most fundamental and complex questions in astrophysics \citep{arcones2023}. Our theoretical knowledge of nuclear physics, stellar evolution, galactic evolution, and primordial cosmology all feed into predictions of how much of each isotope in different parts of the Milky Way -- where the vast majority of our observations of stellar abundances take place -- should have as a function of time. Each of these fields requires the support of a vast body of observational work \citep[e.g.,][]{wilson2019short,buder2021} mapping out abundance patterns in diverse environments in order to constrain, validate, and highlight issues with our theoretical models.

The role of novae in galactic chemical evolution (GCE) has traditionally been sidelined due to their low lifetime-integrated ejecta mass when compared to stellar sources such as supernovae and asymptotic giant branch (AGB) stars \citep{gehrz1998,jose1998}. However, the ability of novae to significantly over-produce certain rare isotopes relative to their solar abundance ratios, such as \iso{13}C, \iso{15}N, and \iso{17}O, raises the prospect of novae playing a non-negligible role in the Galactic evolution of certain abundance ratios \citep{gehrz1998,romano2003}. Due primarily to their characteristic overproduction of \iso{13}C and \iso{15}N, novae have also been proposed as the origin of a rare (<1\%) subset of SiC pre-solar grains, the `nova grains' \citep{amari2001,amari2002,lodders2005presolar}. The true origin of at least some of these nova grains remains contentious, primarily due to the difficulty of reproducing abundances of Si and Ti isotopes in nova models \citep{nittler2005,jose2007origin}.

Previous efforts to estimate the importance of novae in the context of galactic chemical evolution \citep[e.g.,][]{romano2003,cescutti2019,grisoni2019} have relied on simplified models for the behaviour of novae and typically focus on a low number of isotopes in detail. In this work, we explore the role that novae play in galactic chemical evolution by combining a metallicity-dependent grid of pre-computed, synthetic nova populations \citep{kemp2022} with the galactic chemical evolution code \texttt{OMEGA+} \citep{cote2018omegaplus}. Using theoretical, \Mwd-dependent nova yields from \cite{jose1998}, \cite{starrfield2009}, \cite{starrfield2020}, and \cite{jose2020}, we examine the contribution of novae to the evolution of 32 different isotopes:  \iso{3}He, \iso{4}He, \iso{12}C, \iso{13}C, \iso{14}N, \iso{15}N, \iso{16}O, \iso{17}O, \iso{18}O, \iso{19}F, \iso{20}Ne, \iso{21}Ne, \iso{22}Ne, \iso{23}Na, \iso{24}Mg, \iso{25}Mg, \iso{26}Al, \iso{26}Mg, \iso{27}Al, \iso{28}Si, \iso{29}Si, \iso{30}Si, \iso{31}P, \iso{32}S, \iso{33}S, \iso{34}S, \iso{35}Cl, \iso{36}Ar, \iso{37}Cl, \iso{38}Ar, \iso{39}K, and \iso{40}Ca.

We compare nova contributions of these isotopes to the Milky Way with asymptotic giant branch (AGB) stars, massive stars (defined as stars that undergo core-collapse supernova explosions), and type Ia supernovae. We also examine which WD masses are the most important to the production of each isotope. Previously this has been discussed in the context of the nucleosynthetic pathways \citep{gehrz1998,jose1998,iliadis2015,starrfield2009,starrfield2020,jose2020,starrfield2024}, but different nova systems produce different amounts of ejecta due to birth and binary stellar evolution considerations. By folding initial mass function (IMF), binary stellar evolution, and nucleosynthesis considerations into our modelling, for the first time we can holistically assess the contributions of different parts of the nova parameter-space. This isotope-specific discussion is prefaced by a discussion of the metallicity-dependent distributions of nova ejecta mass, and how -- and why -- they differ from distributions of nova events. 

In Section \ref{sec:chap5methododolgy} we describe our methodology. In Section \ref{sec:chap5res_ejecta} we present the ejecta-mass distributions of our nova populations, with comparisons to the event-weighted distributions. In Section \ref{sec:chap5res_abund} we present our analysis of nova contributions to the chemical evolution of the Milky Way, with discussion of which nova white dwarf masses contribute most to which isotopes. Finally, in Section \ref{sec:chap5_conc} we present our conclusions.

Machine-readable outputs from the underlying nova populations and machine-readable yield profiles, both suitable for use in galactic chemical evolution calculations, are available online on Zenodo \footnote{\href{https://zenodo.org/records/6898161}{https://zenodo.org/records/6898161}}.

\section{Methodology}
\label{sec:chap5methododolgy}

Relevant details for the underlying nova and binary stellar physics can be found in \cite{kemp2021}, and details about the specific binary populations used in this work can be found in \cite{kemp2022}. A description of our Galactic modelling is included in \cite{kemp2022li}. The methodology used in this paper is nearly identical to that described in \cite{kemp2022li}, which focused only on Li production in novae. We recommend a review of Section 2 of that work, which summarises the binary population models, nova yield sets, and GCE modelling methodology. In this section, we restrict ourselves to a brief description of details not previously discussed in \cite{kemp2022li} and details specific to this work.

\cite{kemp2022li} describe and employ five different theoretical yield profiles. Two of these yield profiles, R2017 and R2017simple, only have \iso{7}Li yields associated with them, as \cite{rukeya2017} did not report yields for other isotopes. For this reason, these yield sets are excluded from our analysis in this work. We instead only consider the remaining three theoretical yield sets: J1998 \citep{jose1998}, S2009/2020 \citep{starrfield2009,starrfield2020}, and J2020 \citep{jose2020}; their composition is unchanged. Since the completion of the original analysis relevant to both \cite{kemp2022li} and this work, an additional relevant yield set has been released in \cite{starrfield2024}, providing updated O/Ne WD nova yields relative to the previously available \cite{starrfield2009} yields. The formal inclusion of a new yield-set integrating these yields is beyond the scope of the current work. However, a comparison between the O/Ne WD yields from \cite{starrfield2009} and \cite{starrfield2024} does not lead us to believe that such an inclusion would affect the conclusions of this work. In particular, we note that while Li production is increased in these new models, it is still far short of observations of nova ejecta \citep{tajitsu2015,molaro2016,tajitsu2016,selvelli2018,molaro2020,arai2021,izzo2022}.

% We briefly discuss the  impact of these new yields both in the context of the solar Li abundance and the other stable isotopes discussed in this work, but do not re-calculate or provide a new yield-set integrating these yields.

When reporting the abundances and masses of the stable isotopes \iso{18}O, \iso{22}Ne, and \iso{26}Mg, we account for the decays of their radioactive parent isotopes \iso{18}F, \iso{22}Na, and \iso{26}Al. We do not account for the evolution of these radioactive isotopes in detail, and instead assume instantaneous decay into their stable daughter nuclei. This assumption is easily justifiable for \iso{18}F ($t_{1/2}\approx 110$ minutes) and \iso{22}Na ($t_{1/2}\approx 2.5$ years), but not ideal for the longer half-life of \iso{26}Al ($t_{1/2}\approx$ \timestento{7}{5} years). Correctly accounting for the half-life of \iso{26}Al would result in a slight reduction of our \iso{26}Mg abundances; however, as we find that novae do not contribute to \iso{26}Mg at anywhere near the level of AGB or massive stars (see Section \ref{sec:chap5res_abund}, we do not consider this source of error to be concerning when considering \iso{26}Mg yields.

\begin{figure*}
\centering
\includegraphics[width=1\textwidth]{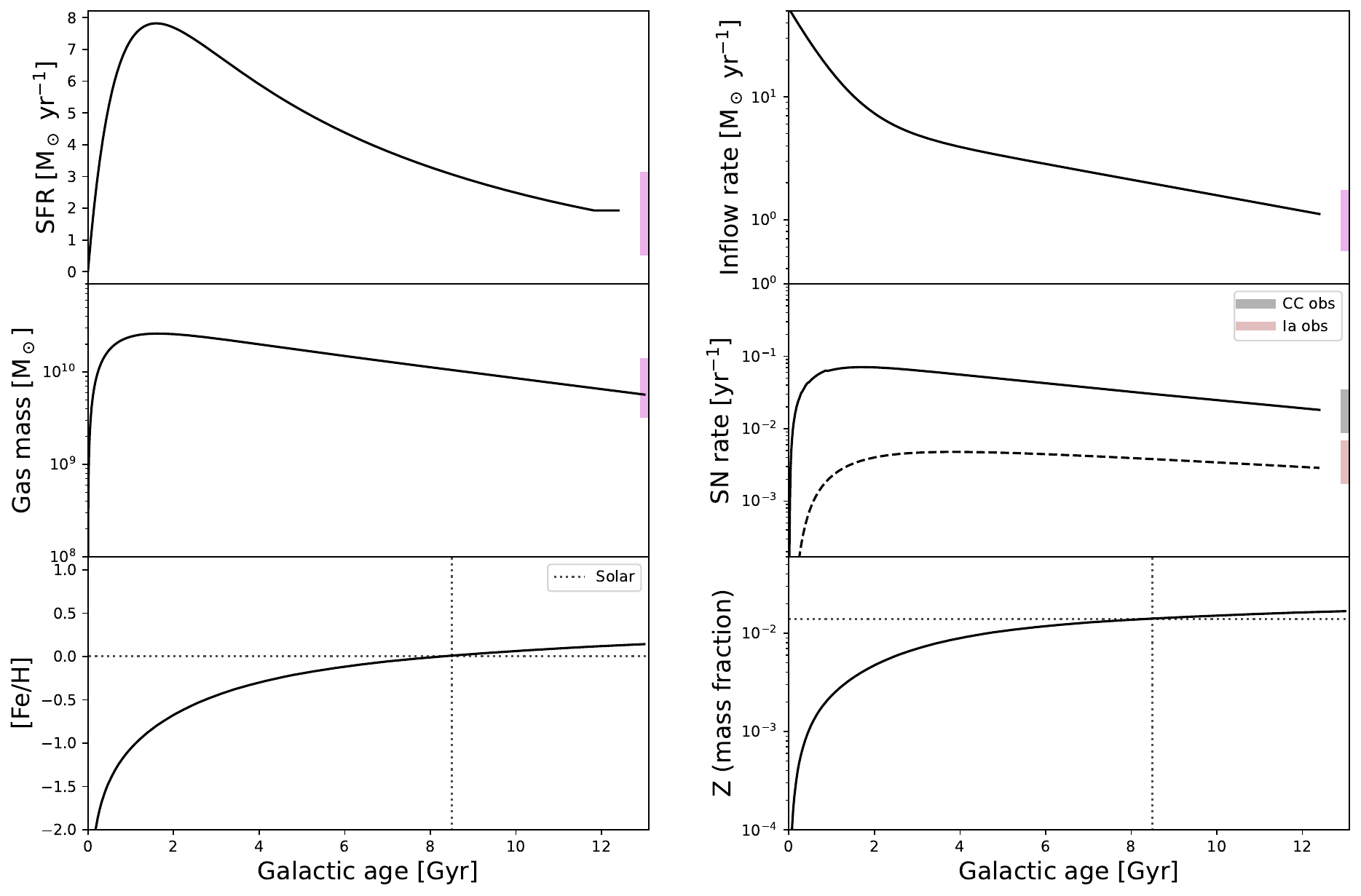} 
\caption{Bulk galactic properties of our baseline model of the Milky Way. The bounds of the shaded vertical bands are drawn from \cite{flynn2006,robitaille2010,chomiuk2011,marasco2012,lehner2011,prantzos2011,kubryk2015}, which the model matches very well.}
\label{fig:mwcalib1}
\end{figure*}

\begin{figure*}
\centering
\includegraphics[width=1\textwidth]{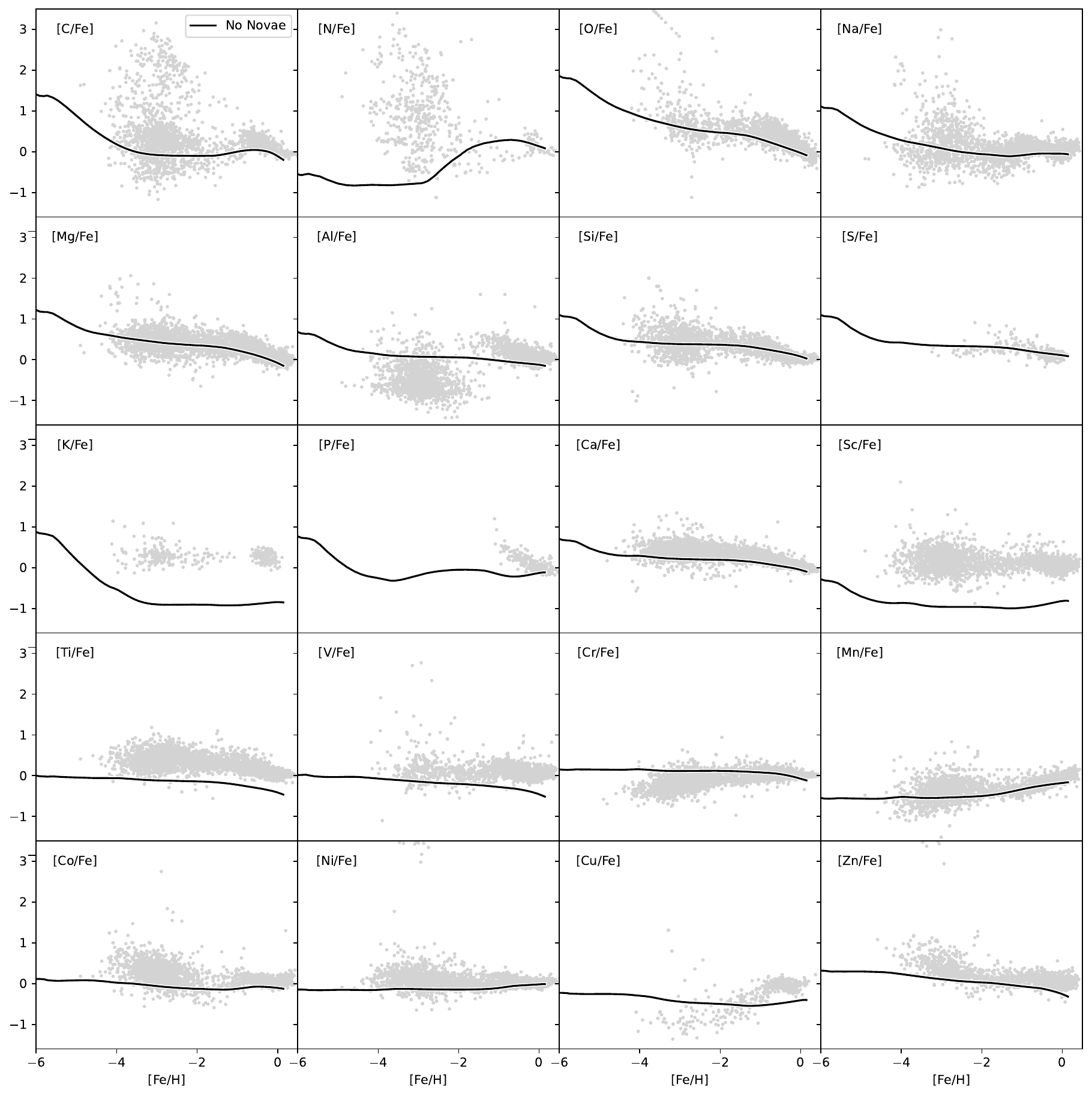} 
\caption[Galactic abundances of key elements used to validate our baseline model of the Milky Way]{Galactic abundances of key elements used to validate our baseline model of the Milky Way, plotted with observational abundance data from NuPyCEE's STELLAB module, where available (all but [P/Fe]; [P/Fe] abundance data are taken from \citealt{roederer2014,maas2019,maas2022,nandakumar2022}). The bulk galactic properties and elemental abundances for most key elements are satisfactorily reproduced. Notable exceptions such as K and Sc have been long established as problematic elements for contemporary stellar evolution and GCE codes \citep[see, for example, ][]{kobayashi2020origin}.}
\label{fig:mwcalib2}
\end{figure*}

Our Galactic chemical evolution models are run using the \texttt{OMEGA+} chemical evolution code \citep{cote2018omegaplus}. Fig. \ref{fig:mwcalib1} shows, as a function of Galactic age, the evolution of the star formation rate (SFR), Galactic inflow rate, gas mass, supernova (SN) rate, [Fe/H] abundance ratios, and metallicity ($Z$) of our baseline model of the Milky Way without novae. The current Galactic age is taken to be 13 Gyr, and the current solar age is taken to be 4.5 Gyr. Fig. \ref{fig:mwcalib2} shows the evolution of the ratios of several important elements relative to iron plotted against [Fe/H] of this baseline model. The observational abundance data (grey points) are plotted alongside  are from NuPyCEE's STELLAB\footnote{\url{https://github.com/NuGrid/NuPyCEE/blob/master/stellab_data/abundance_data_library.txt}, \cite{cote2017omega}} module where available. The bulk galactic properties and elemental abundances for most key elements are satisfactorily reproduced; notable exceptions such as K and Sc have long been established as problematic elements in contemporary stellar evolution and GCE codes \citep[e.g.,][]{kobayashi2020origin}. We therefore assert that our model galaxy is good enough for a meaningful comparison between novae and other stellar sources.

\section{Results: Ejecta-weighted nova distributions}
\label{sec:chap5res_ejecta}

In \cite{kemp2021} and \cite{kemp2022}, nova evolution was assessed almost exclusively with the goal of studying distributions of nova events. These works considered when the nova eruptions occurred and the binary conditions surrounding them at that time. Although eruptions from different binary systems were weighted differently (according to the birth probability of the progenitor system), novae were otherwise treated equally regardless of, for example, how much ejecta the eruption released into the interstellar medium (ISM).

\begin{figure*}
\centering
\includegraphics[width=1\textwidth]{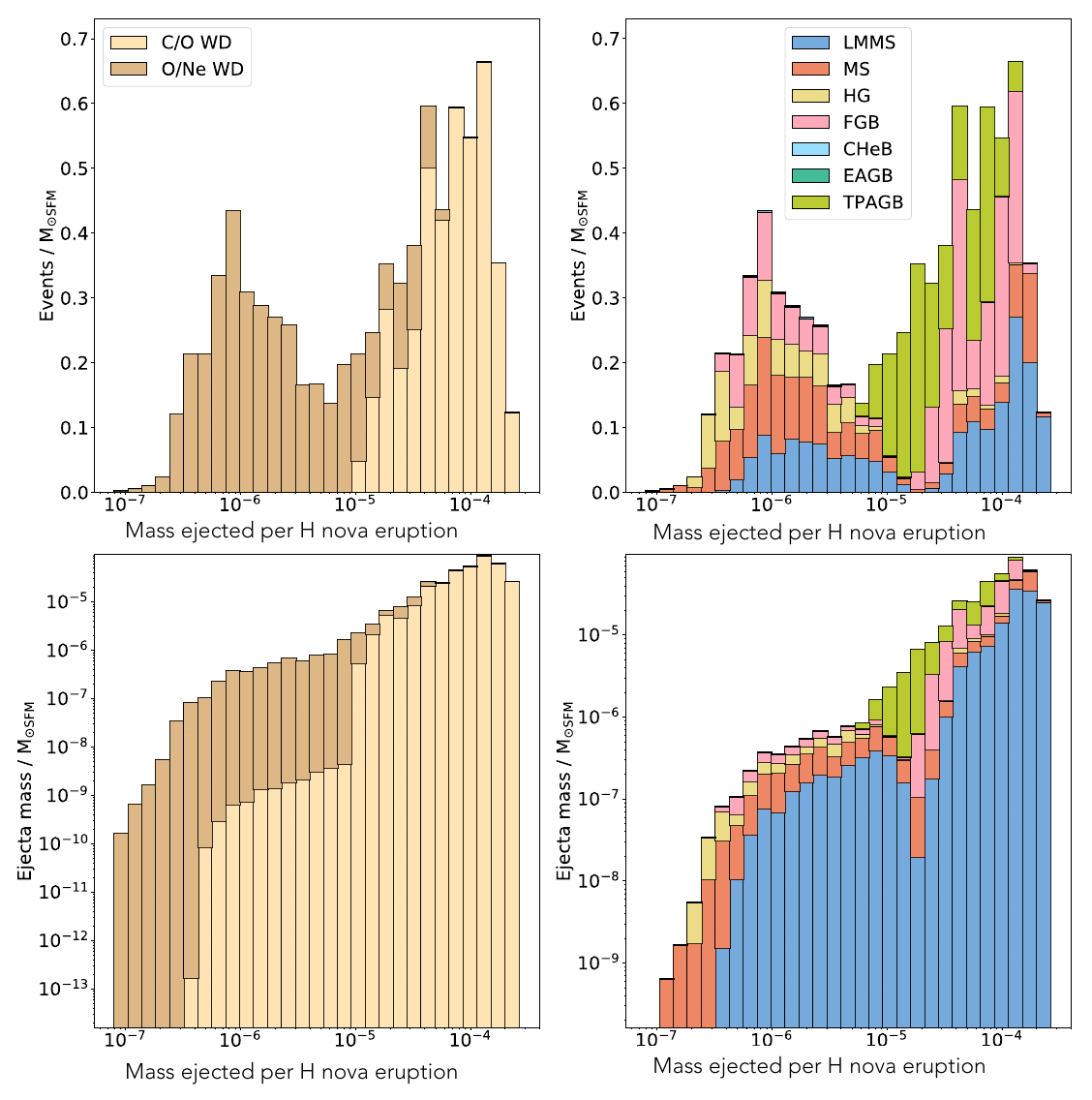}
\caption[Distribution of mass ejected per H nova eruption, $Z$~=~\tento{-3}.]{Distribution of mass ejected per H nova eruption, $Z$~=~\tento{-3}. Top panels count the number of events in each bin while the bottom panels track the ejecta mass in each bin. Coloured according to the WD's composition (left) and the donor stellar type  \citep{hurley2000,hurley2002}. The donor stellar types are, in descending order: low-mass main sequence stars, main sequence stars, Hertzsprung Gap stars, first-ascent giant branch stars, core He burning stars, early asymptotic giant branch stars, and thermally pulsing asymptotic giant branch stars.}
\label{fig:hist_dmproczm3}
\end{figure*}

\begin{figure*}
\centering
\includegraphics[width=1\textwidth]{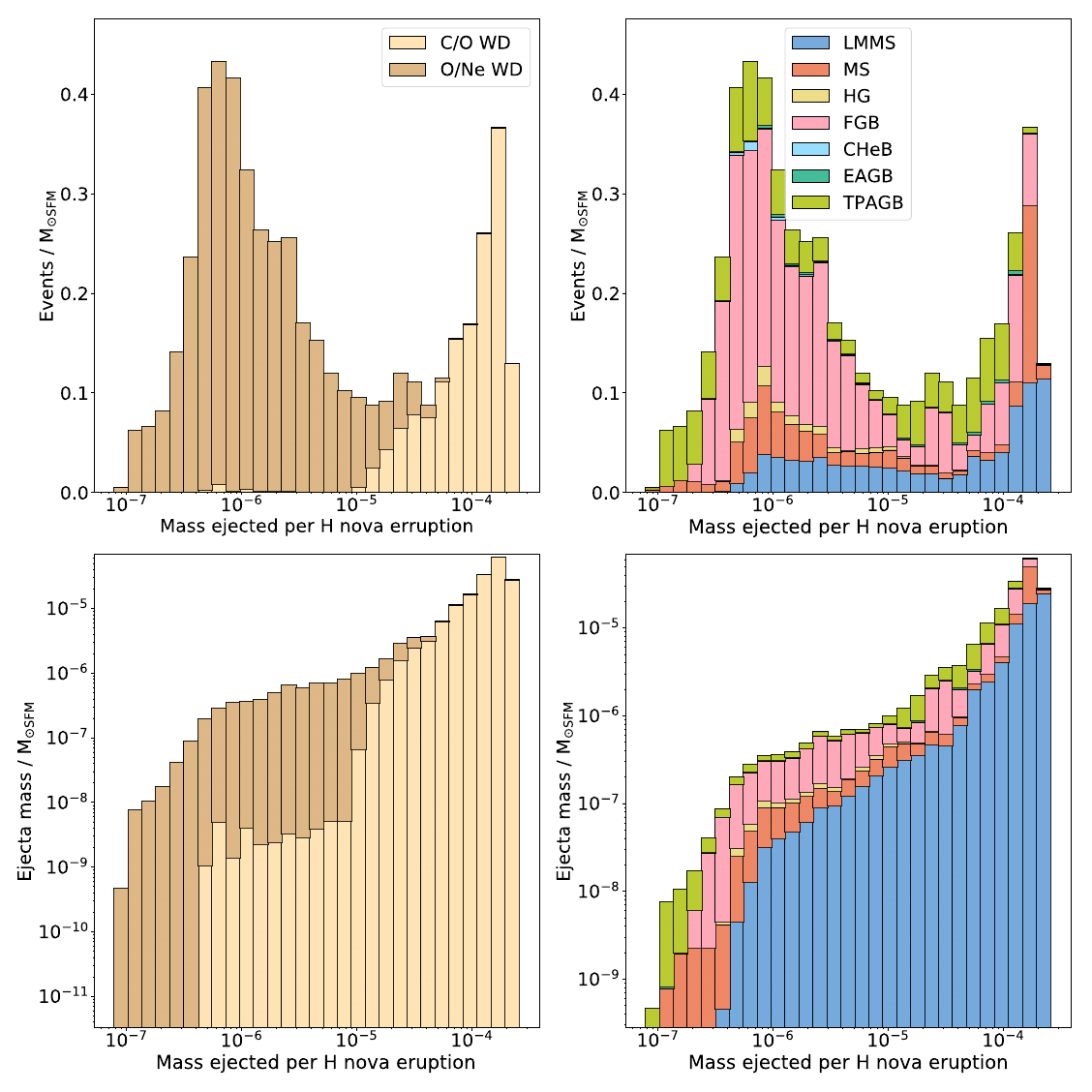}
\caption{Distribution of mass ejected per H nova eruption, $Z$~=~0.02.}
\label{fig:hist_dmprocz0p02}
\end{figure*}

In this work, we consider nova contributions to the origin of the elements. To assume that each nova eruption carries with it equal amounts of ejecta mass is a gross simplification. The mass ejected from different nova eruptions varies significantly depending on the binary properties, as shown in Figs. \ref{fig:hist_dmproczm3} and \ref{fig:hist_dmprocz0p02}. These figures present the distributions of ejecta mass for nova populations computed at $Z$~=~\tento{-3} and $Z$~=~0.02, respectively. The amount of mass ejected in a nova explosion varies by over 3 orders of magnitude. The left panels of Figs. \ref{fig:hist_dmproczm3} and \ref{fig:hist_dmprocz0p02}, coloured by the WD composition (C/O or O/Ne), show that higher ejecta mass eruptions occur on the lower mass C/O WDs. The higher mass O/Ne WDs are largely restricted to far lower ejecta masses per eruption.

The lower panels of Figs. \ref{fig:hist_dmproczm3} and \ref{fig:hist_dmprocz0p02} highlight which eruptions contribute the most ejecta to the ISM by tracking the ejecta mass in each bin, rather than counting the nova eruptions. The vast majority of nova ejecta is produced from C/O WDs, which typically eject more material per eruption. The underlying physics behind this behaviour -- chiefly the far higher critical ignition masses required for novae in low-mass WD systems -- is described in Appendix A of \cite{kemp2022li}.

%%%%%DMPROC%%%%%zm3

%%%%%DMPROC%%%%%z0p02

\begin{figure*}
\centering
\includegraphics[width=1\textwidth]{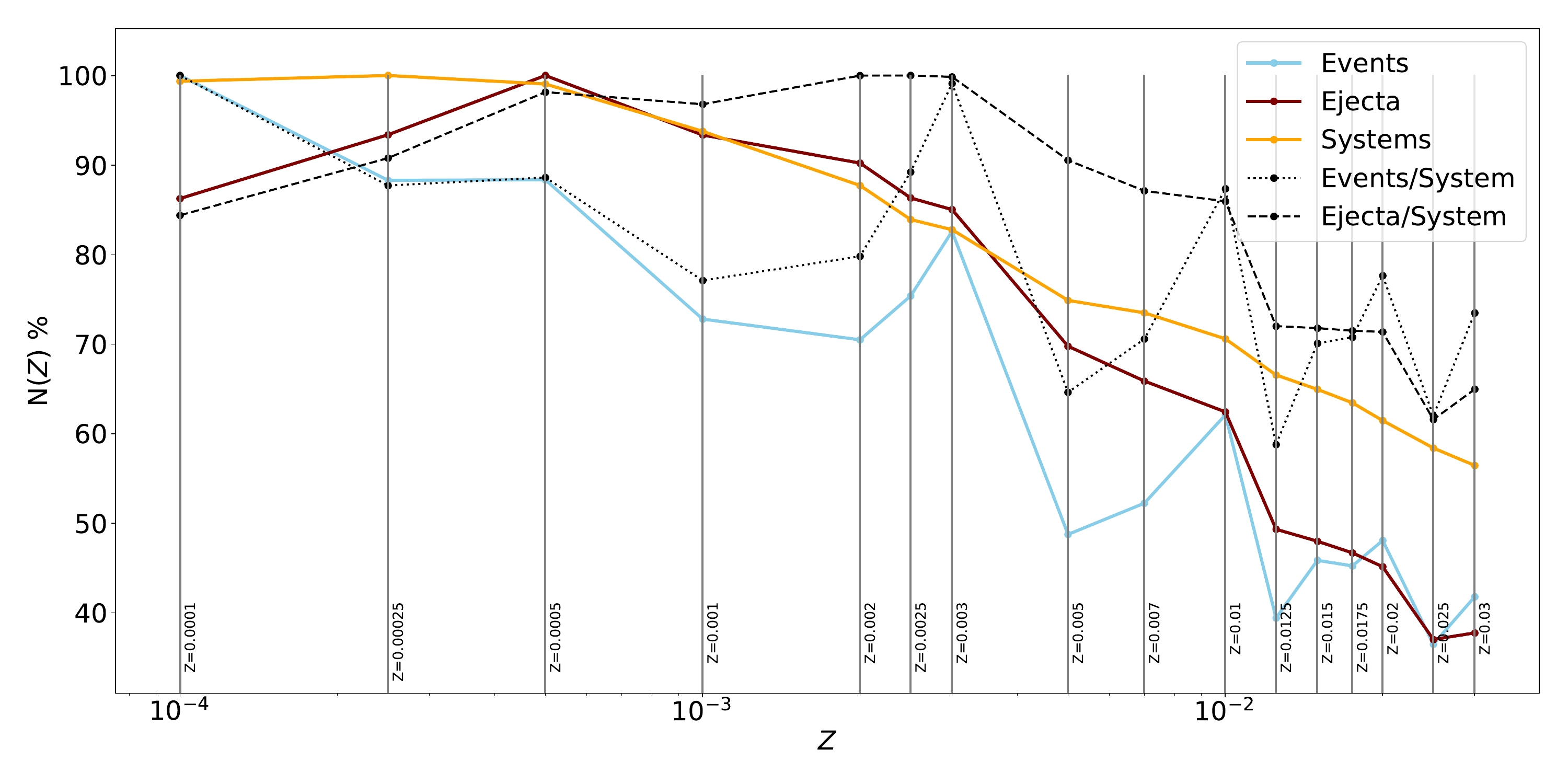} 
\caption{Fractional changes N($Z$)~=~$ X(Z)/\mathrm{max}(X(Z))$ of the aggregate number (summed over 15 Gyr) of nova events, nova systems, and the number of nova events per system as a function of the metallicity. There is a significant anti-correlation between all three of these quantities, with the total nova productivity from the highest metallicity systems roughly half that of the lowest. The ejecta mass tracks the event productivity well, particularly for metallicities above $Z=0.01$. This figure is an updated version of Figure 5 in \citep{kemp2022}, including the mass of ejected material.}
\label{fig:zsumtabpercentupdated}
\end{figure*}

The overall influence of metallicity on nova productivity is presented in Fig. \ref{fig:zsumtabpercentupdated}, analogous to Fig. 5 from \cite{kemp2022} but including the ejecta mass and mean ejecta mass per system. Table \ref{tab:zsumtabupdated} is a similarly updated version of Table 3 from \cite{kemp2022}.

The ejecta mass tracks the event productivity fairly well, implying that the influences governing the metallicity dependence of nova productivity are not dominated by influences specific to ejecta mass considerations. The main discrepancy between the scaling of ejecta mass and nova event counts with metallicity is that the nova event counts exhibit far more scatter. As described in \cite{kemp2022}, the origin of much of the scatter in the number of events is due to a single, well-defined channel that is sensitive to the initial primary mass, and can be considered numerical. This source of scatter is largely absent in the ejecta mass plot because high-mass WDs contribute far less ejecta, making this channel relatively unimportant.

Comparing the two metallicity cases presented in Figs. \ref{fig:hist_dmproczm3} and \ref{fig:hist_dmprocz0p02}, increasing the metallicity reduces the ejecta contributions of the C/O WDs, following the reduced relative importance of low-mass WDs at higher metallicities when considering event counts (see \citealt{kemp2022}). Figures  \ref{fig:hist_dmproczm3} and \ref{fig:hist_dmprocz0p02} highlight the increased importance of giant donor stars at high metallicity, particularly for low ejecta-mass novae occurring on the massive O/Ne WDs.

\begin{figure*}
\centering
\begin{subfigure}{0.45\textwidth}
\centering
\includegraphics[width=1\columnwidth]{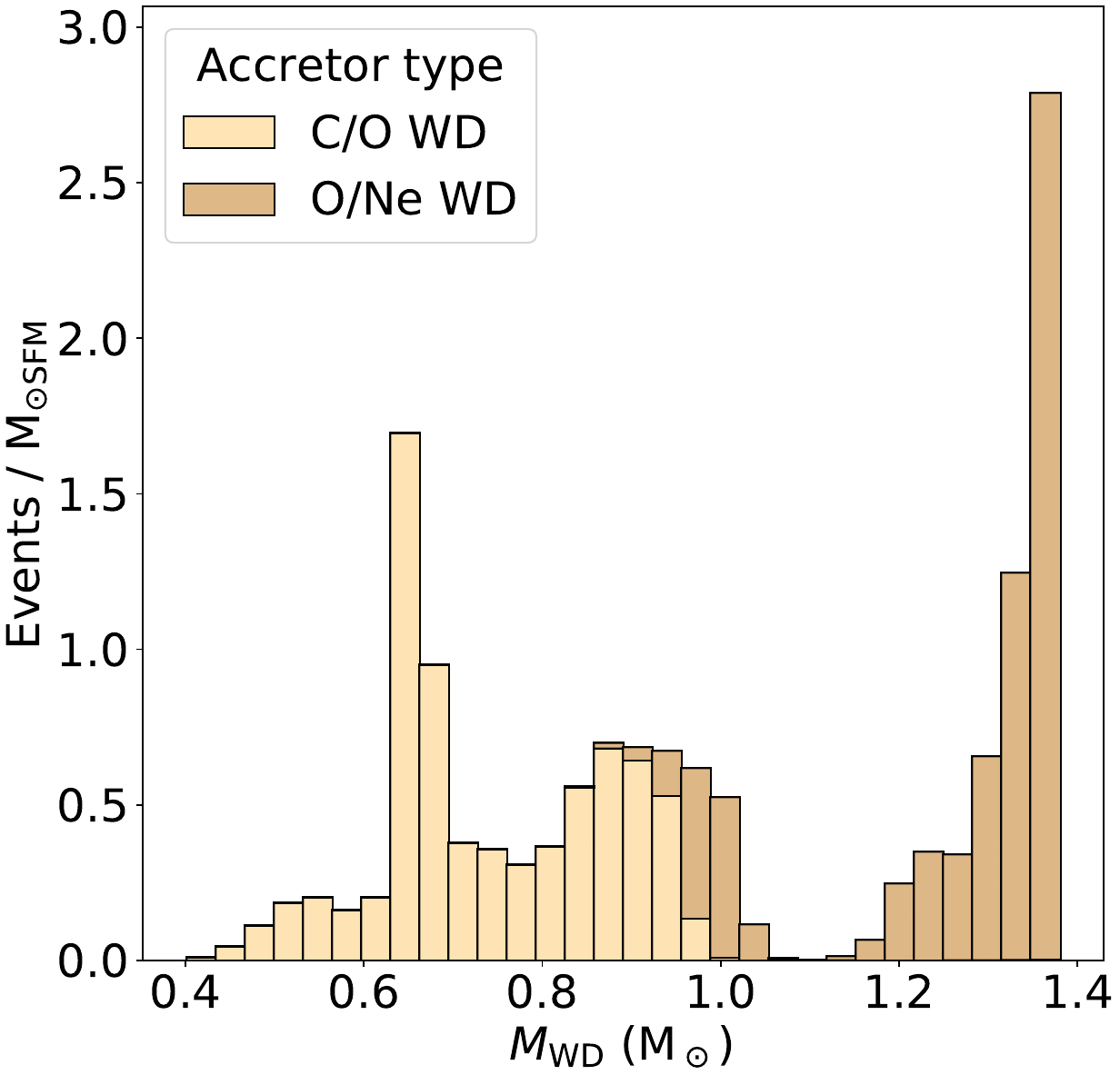}
\caption{}
\end{subfigure}%
\begin{subfigure}{0.45\textwidth}
\centering
\includegraphics[width=1\columnwidth]{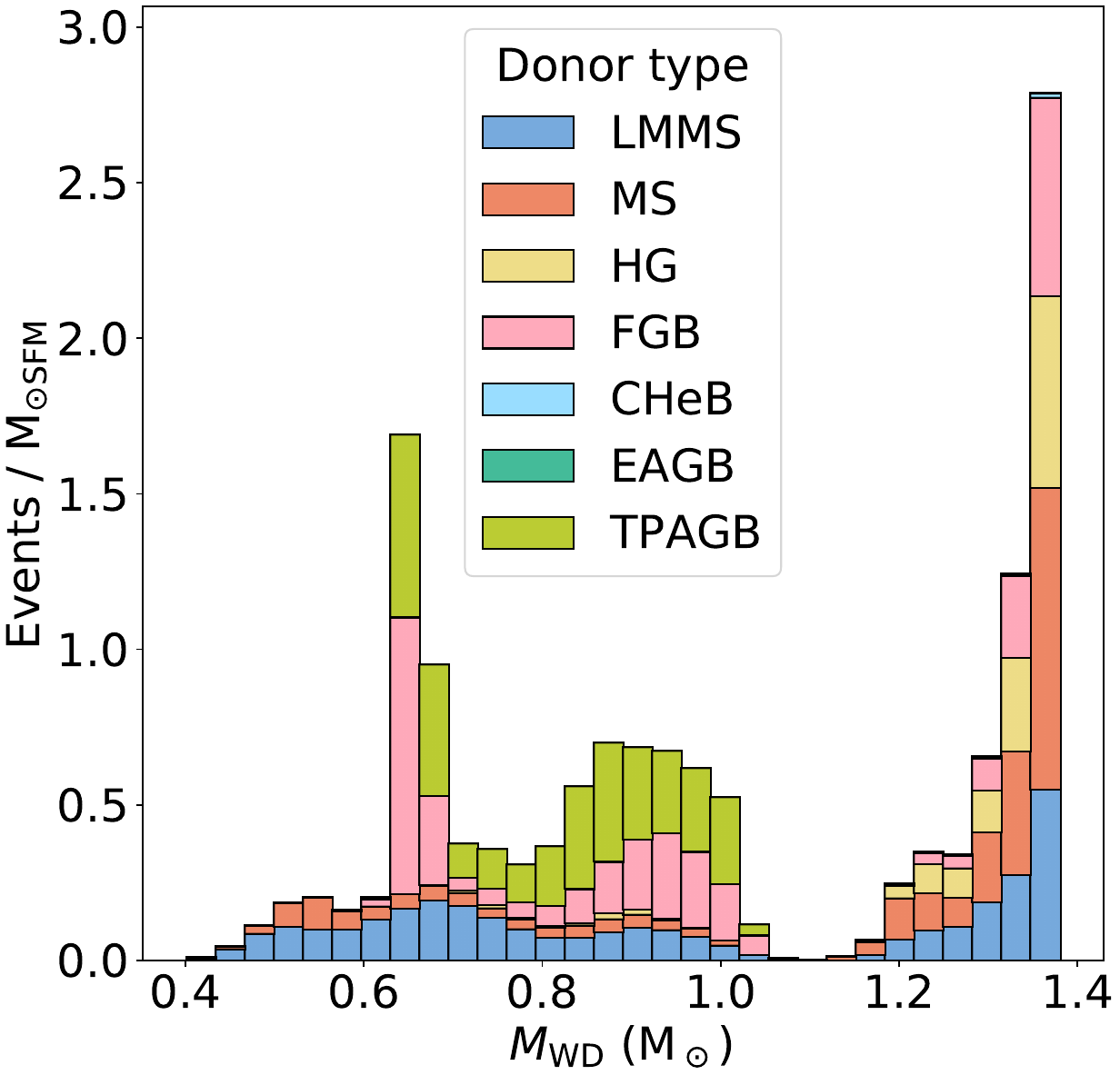}
\caption{}
\end{subfigure}

\begin{subfigure}{0.45\textwidth}
\centering
\includegraphics[width=1\columnwidth]{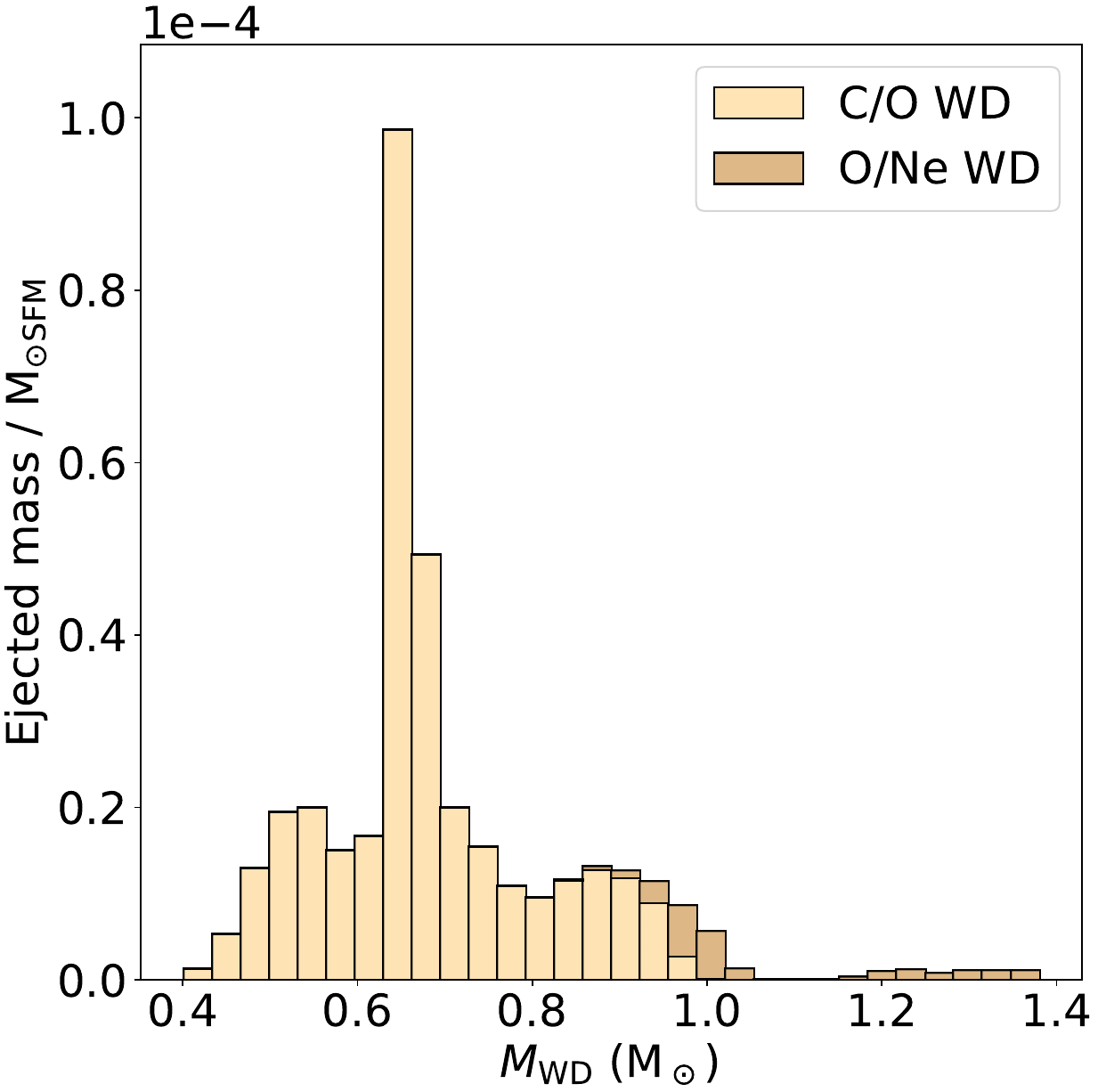}
\caption{}
\end{subfigure}%
\begin{subfigure}{0.45\textwidth}
\centering
\includegraphics[width=1\columnwidth]{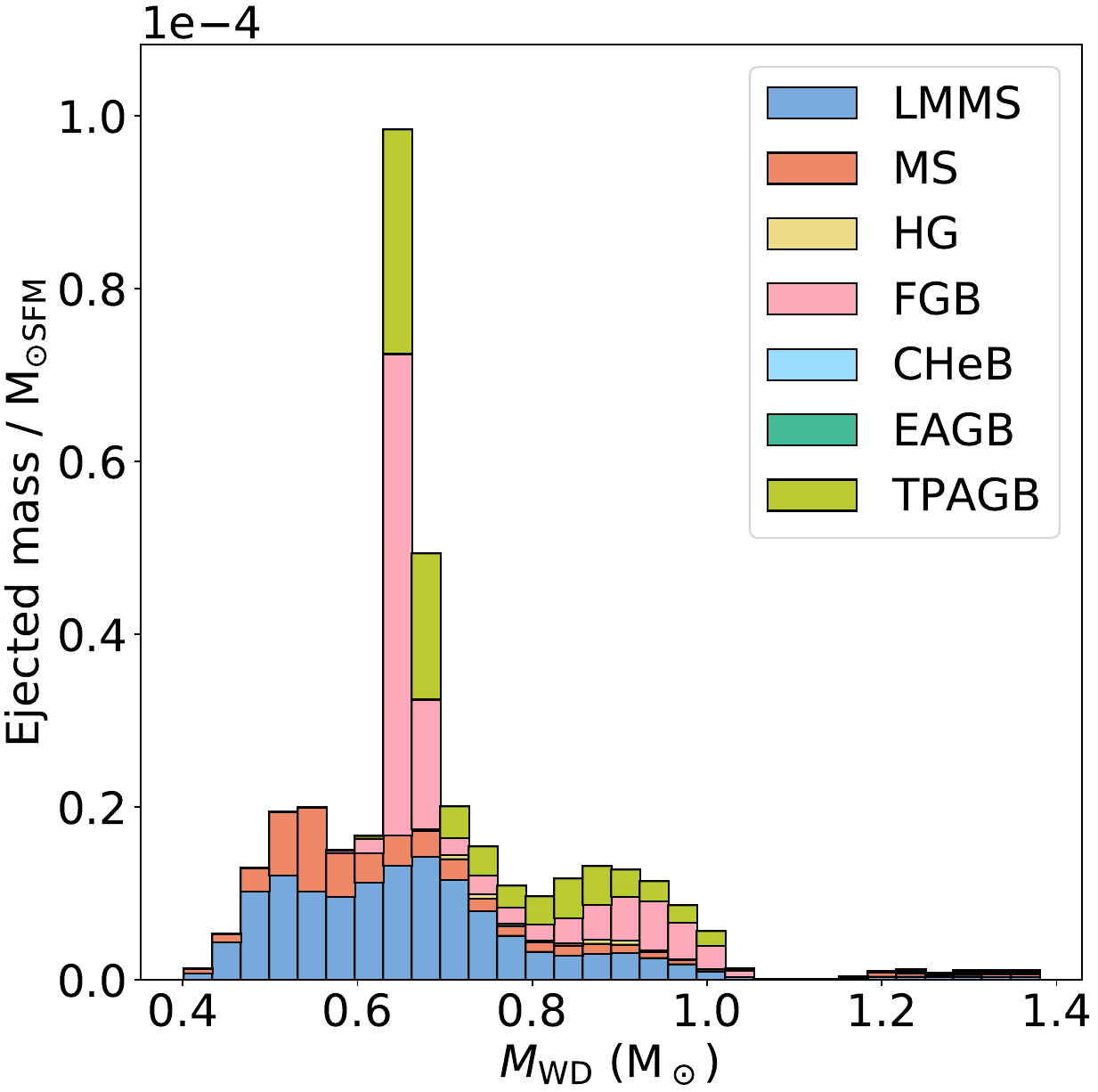}
\caption{}
\end{subfigure}

\caption{Nova WD mass distributions for $Z$~=~\tento{-3}, weighting by number of novae (top) and ejecta mass (bottom) and colouring by WD composition (left) and donor stellar type (right).}
\label{fig:hist_mwdzm3}
\end{figure*}

\begin{figure*}
\centering
\begin{subfigure}{0.5\textwidth}
\centering
\includegraphics[width=1\columnwidth]{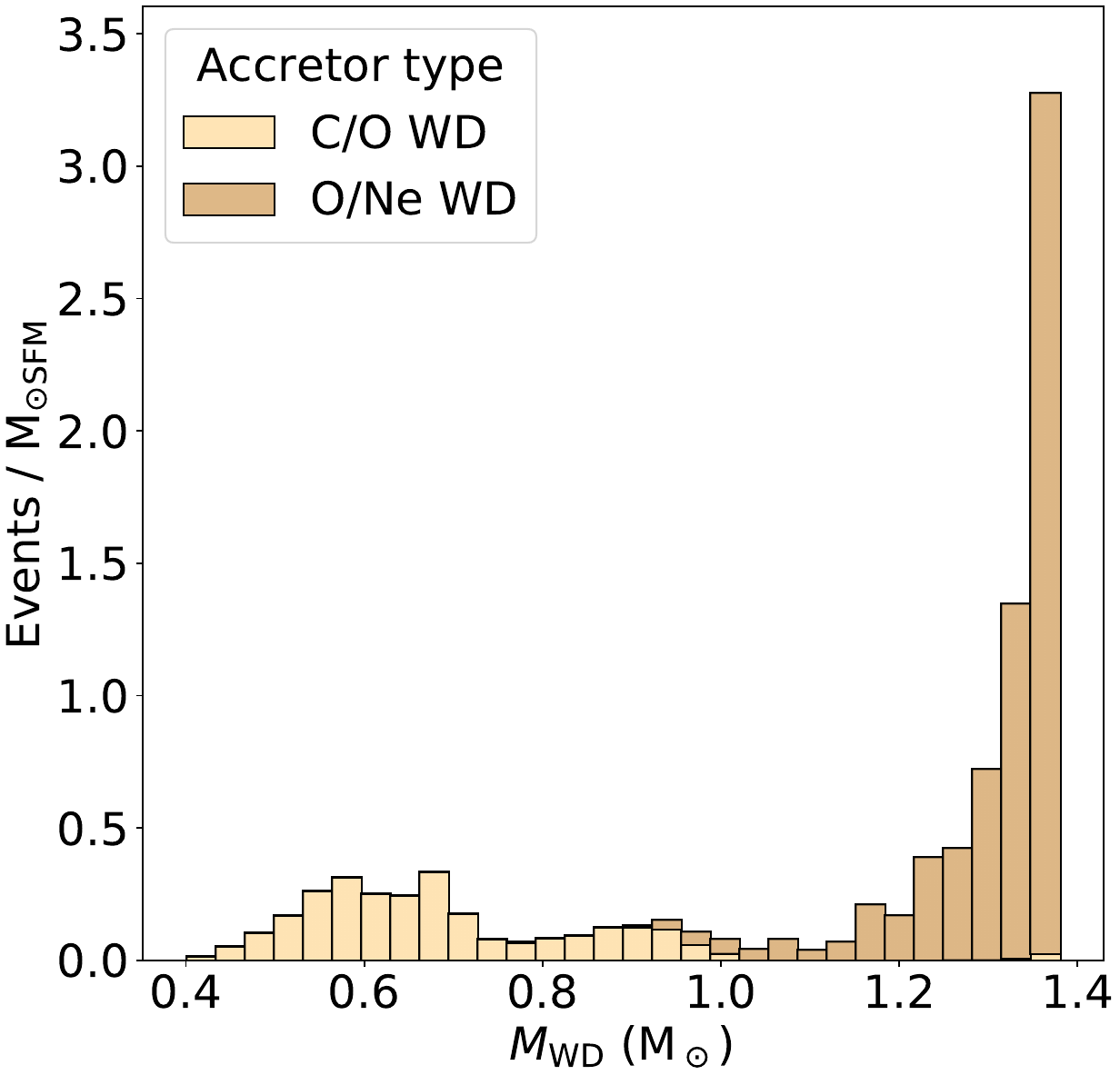}
\caption{}
\end{subfigure}%
\begin{subfigure}{0.5\textwidth}
\centering
\includegraphics[width=1\columnwidth]{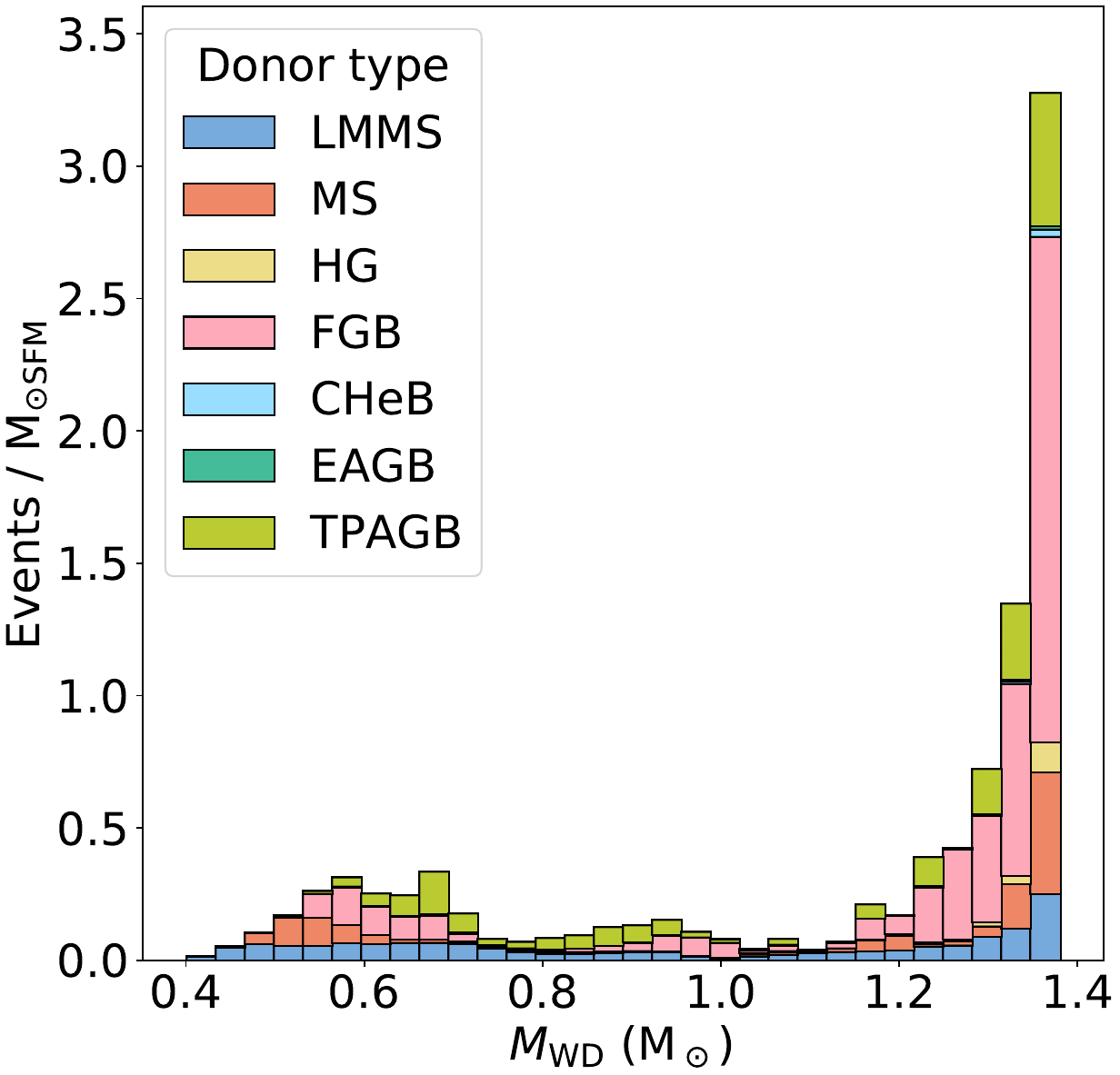}
\caption{}
\end{subfigure}

\begin{subfigure}{0.5\textwidth}
\centering
\includegraphics[width=1\columnwidth]{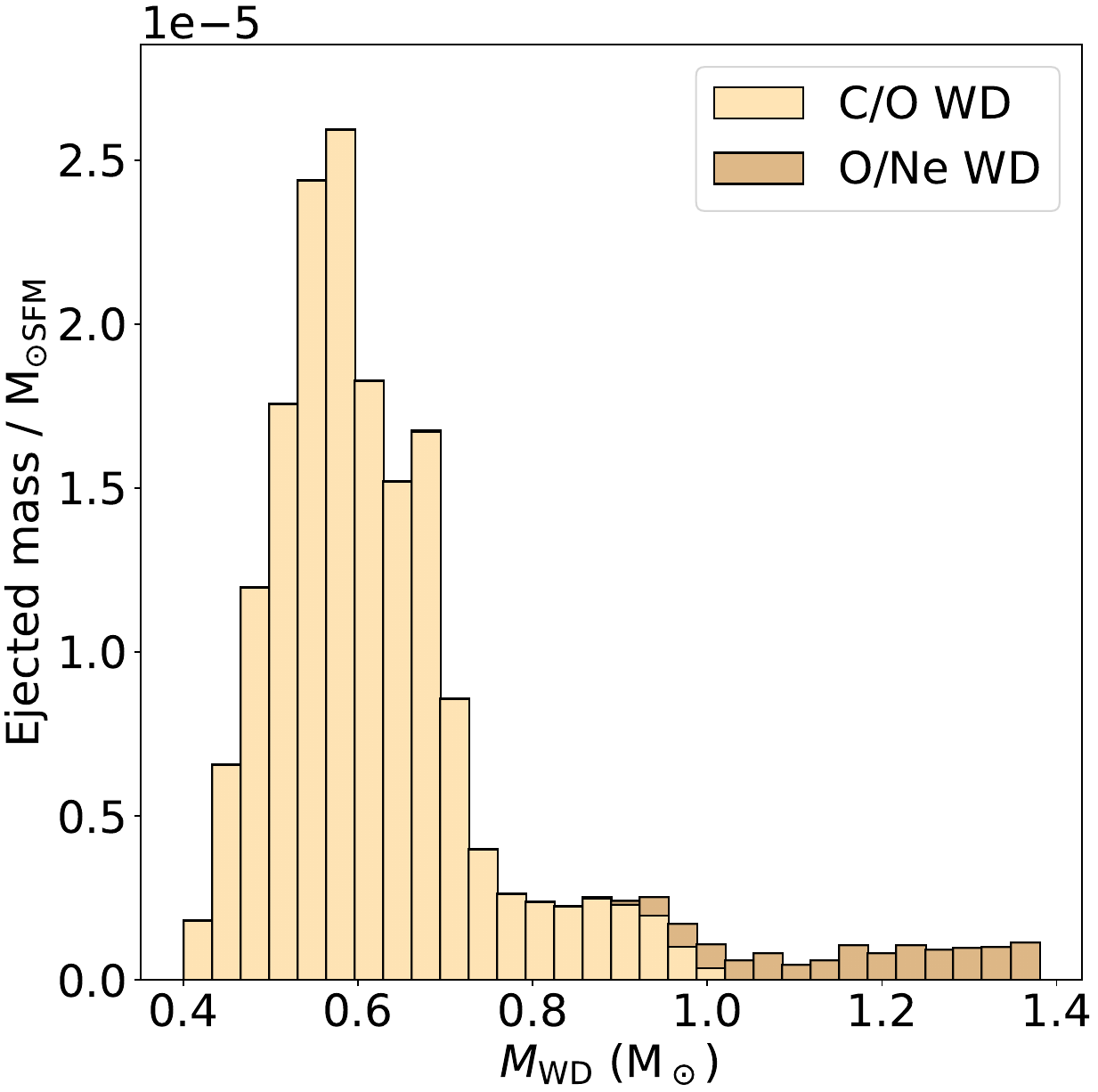}
\caption{}
\end{subfigure}%
\begin{subfigure}{0.5\textwidth}
\centering
\includegraphics[width=1\columnwidth]{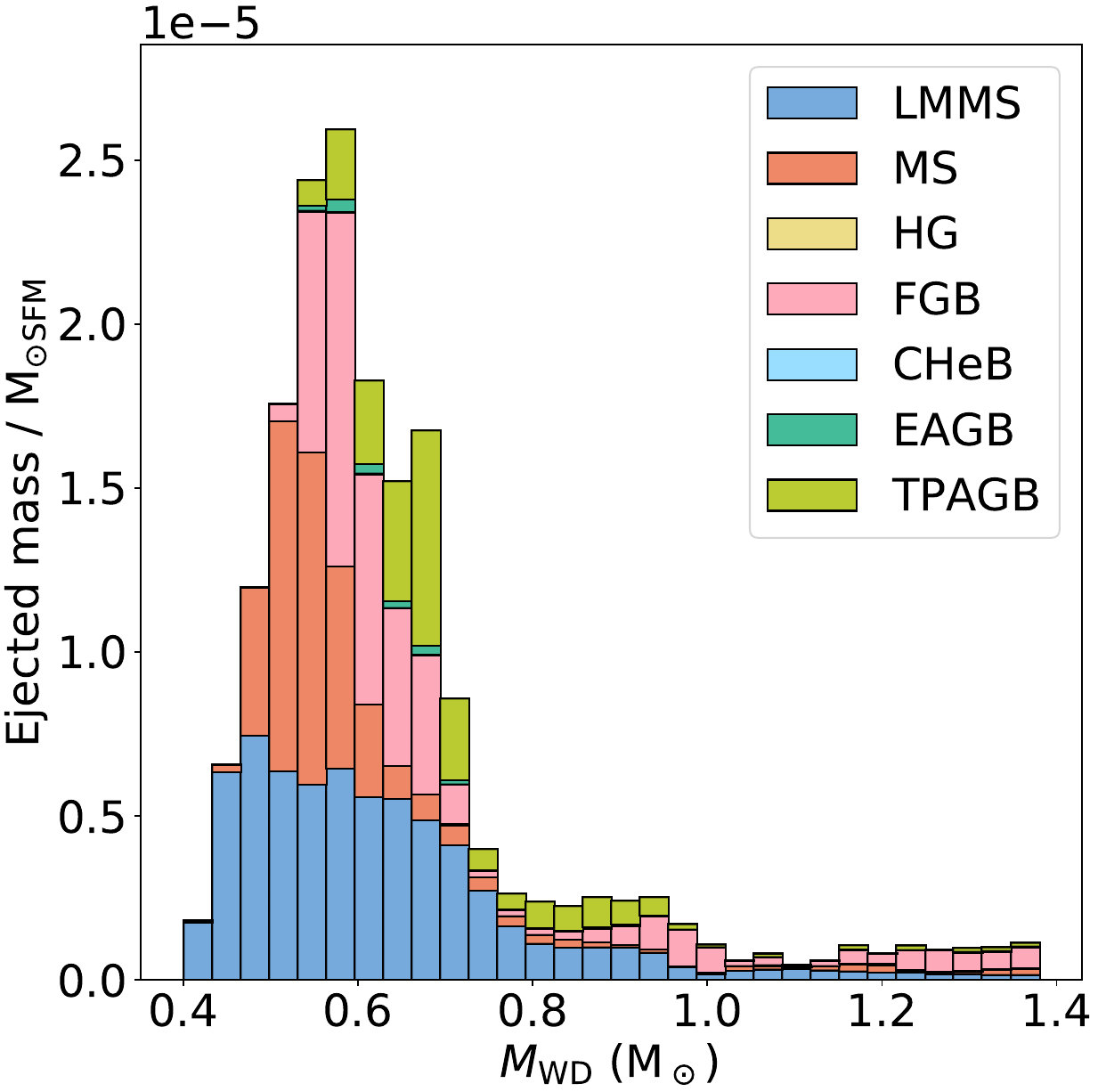}
\caption{}
\end{subfigure}

\caption{Nova WD mass distributions for $Z$~=~0.02.}
\label{fig:hist_mwdz0p02}
\end{figure*}

It was noted in Appendix A of \cite{kemp2022li} that the WD mass distribution showed the greatest variation when comparing nova events to ejecta masses. Figs. \ref{fig:hist_mwdzm3} and \ref{fig:hist_mwdz0p02} demonstrate this variation graphically. Although high-mass WDs contribute significantly to overall nova counts, they contribute very little to overall nova ejecta mass contributions.
This is mostly due to the lower critical ignition masses associated with high-mass WDs (responsible for their high nova counts in the first place) now acting against them when considering ejecta-mass weighted distributions. The result is that the initial mass function's \citep{kroupa2001} preference for low-mass WD systems causes low-mass WDs to dominate the ejecta mass distribution.

\cite{kemp2022} discuss the emergence of a bimodal feature in the WD mass distribution of nova events at low metallicities. In short, as metallicity reduced, contributions from low-mass WDs increased in importance, particularly in a broad peak from 0.8-1.0 M\solar\ and in a narrow peak from 0.6-0.7 M\solar. Examining the lower panels of Figs. \ref{fig:hist_mwdzm3} and \ref{fig:hist_mwdz0p02}, we see similar behaviour in the ejecta-mass distributions, although low WD-mass features become more pronounced. Even in the $Z$~=~0.02 population, where contributions to total nova counts from lower mass WDs are low compared to lower metallicity populations, most of the nova ejecta is produced from WDs between 0.5 and 0.8 M\solar. Coupled with the introduction of the sharp peak in nova event contributions around 0.6-0.7 M\solar at low metallicities, this produces a prominent spike in the $Z$~=~\tento{-3} distribution of ejecta mass. The influence of the emergence of the second, broader peak in nova counts present at low metallicity (around 0.8-1.0 M\solar) is less pronounced, but still discernable in Fig. \ref{fig:hist_mwdzm3}. The right-hand panels of Figs. \ref{fig:hist_mwdzm3} and \ref{fig:hist_mwdz0p02} confirm the common evolutionary origin of these peaks: increased nova productivity driven by giant (FGB and TPAGB) donor stars (see also discussion in \citealt{kemp2022}).

\begin{figure*}
\centering
\includegraphics[width=0.9\textwidth]{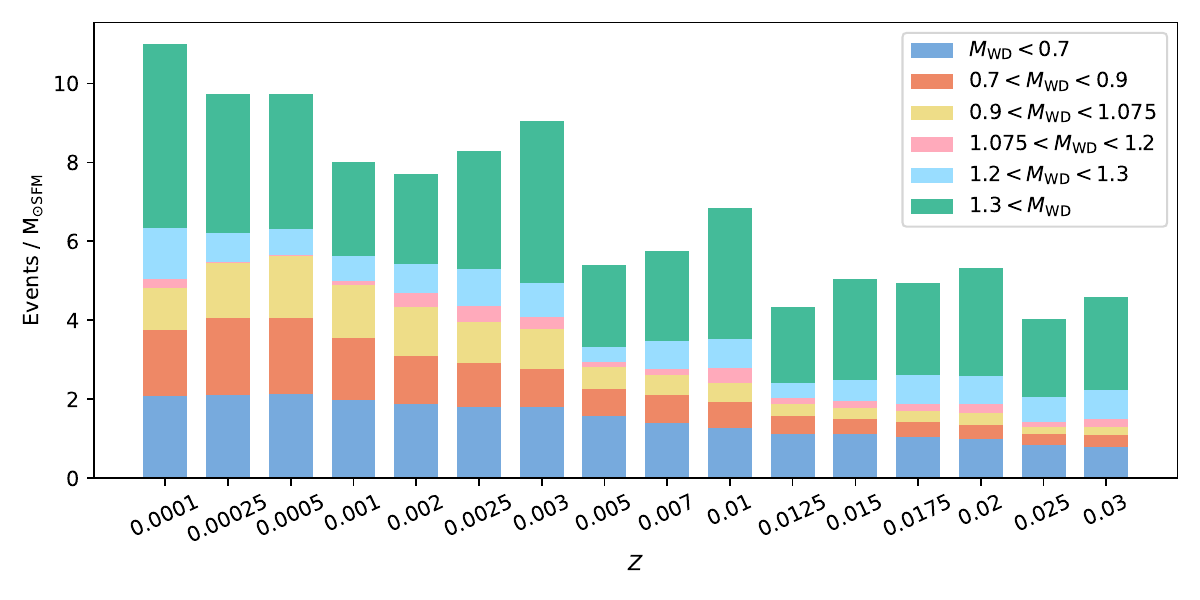} 
\caption{Aggregate number (summed over 15 Gyr) of nova events per unit star forming material as a function of metallicity, broken down by WD mass bracket. The data is tabulated in Table \ref{tab:zsumtab_events_mwd}. Contributions from high-mass WDs are ubiquitous across metallicity, while intermediate mass WD contributions are higher at low metallicity.}
\label{fig:zsumtab_mwd_event}
\end{figure*}

\begin{figure*}
\centering
\includegraphics[width=0.9\textwidth]{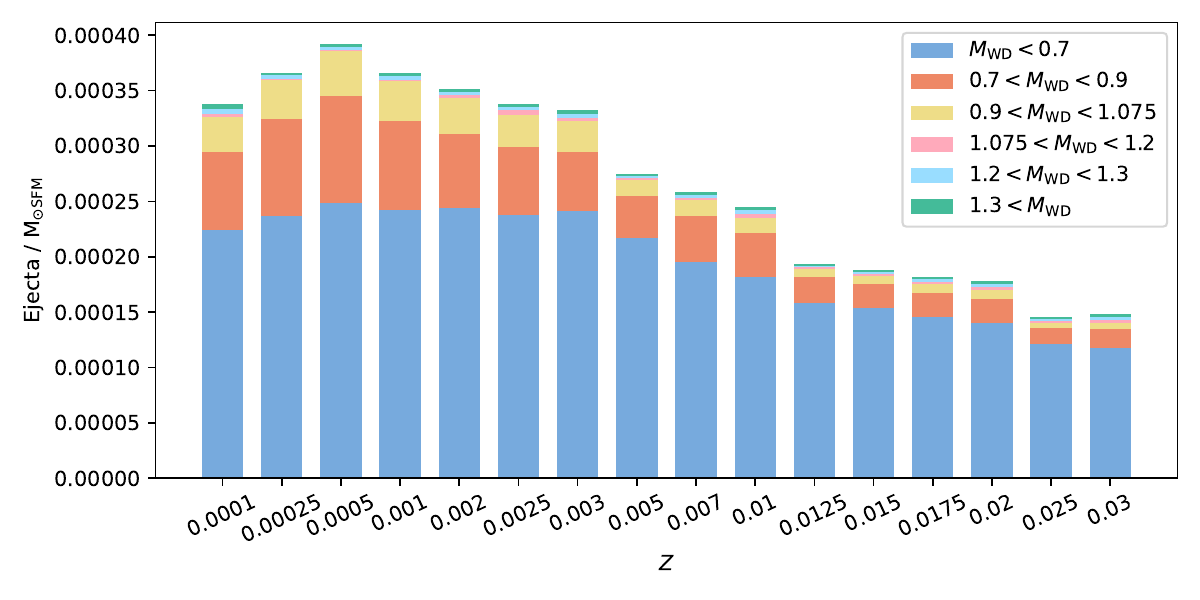} 
\caption{Aggregate number (summed over 15 Gyr) of nova ejecta per unit star forming material as a function of metallicity, broken down by WD mass bracket. The data is tabulated in Table \ref{tab:zsumtab_ejecta_mwd}. In contrast with Fig. \ref{fig:zsumtab_mwd_event}, which shows the event distribution, low-mass WD contributions dominate the ejecta mass profiles at all metallicities, although an increase in intermediate-mass WD contributions also be seen at low metallicities.}
\label{fig:zsumtab_mwd_ejecta}
\end{figure*}

Figs. \ref{fig:zsumtab_mwd_event} and \ref{fig:zsumtab_mwd_ejecta} and Tables \ref{tab:zsumtab_events_mwd} and \ref{tab:zsumtab_ejecta_mwd} provide a more comprehensive view of how different mass WDs contribute to ejecta and event counts at different metallicities. The WD mass brackets are tailored to the WD masses with detailed simulations of nova abundances \citep{jose1998,starrfield2009,jose2020,starrfield2020,starrfield2024}. Large contributions from high-mass WDs are ubiquitous across all metallicities when considering event counts. In contrast, relative contributions from intermediate mass WDs increase at lower metallicities, reflecting the emergence of the low-intermediate mass bimodal feature. When considering ejecta mass contributions shown in Figure \ref{fig:hist_mwdz0p02}, low-mass WD contributions dominate across all metallicities, while high-mass WD contributions remain very small. The emergence of the bimodal feature is visible in the rise in contributions from intermediate mass WDs at lower metallicities.

\begin{table*}
\centering
\caption{Updated version of Table 3 from \citep{kemp2022}, presenting the aggregate number (summed over 15 Gyr) of nova events, nova systems, and the number of nova events per system for each metallicity.}
\resizebox{\textwidth}{!}{\begin{tabular}{lrrrrr}
\textbf{$Z$} & \textbf{Events / M$_{\odot \rm SFM}$} & \textbf{Ejecta / M$_{\odot \rm SFM}$} & \textbf{Systems / M$_{\odot \rm SFM}$} & \textbf{Events / System} & \textbf{Ejecta / System} \\ \hline
0.0001 & 11.00393 & 0.000338 & 0.013471 & 816.8 & 0.025 \\
0.00025 & 9.715102 & 0.000366 & 0.013556 & 716.6 & 0.027 \\
0.0005 & 9.721356 & 0.000391 & 0.013428 & 723.9 & 0.029 \\
0.001 & 8.002775 & 0.000365 & 0.0127 & 630.1 & 0.028 \\
0.002 & 7.710307 & 0.000351 & 0.011822 & 652.1 & 0.029 \\
0.0025 & 8.324096 & 0.000339 & 0.011417 & 729 & 0.029 \\
0.003 & 9.060606 & 0.000332 & 0.011191 & 809.6 & 0.029 \\
0.005 & 5.372794 & 0.000273 & 0.010169 & 528.3 & 0.026 \\
0.007 & 5.742047 & 0.000257 & 0.009954 & 576.8 & 0.025 \\
0.01 & 6.843398 & 0.000245 & 0.009589 & 713.6 & 0.025 \\
0.0125 & 4.346741 & 0.000193 & 0.009043 & 480.6 & 0.021 \\
0.015 & 5.053725 & 0.000188 & 0.00882 & 572.9 & 0.021 \\
0.0175 & 4.972212 & 0.000182 & 0.008595 & 578.4 & 0.021 \\
0.02 & 5.300315 & 0.000177 & 0.008352 & 634.5 & 0.021 \\
0.025 & 4.016605 & 0.000145 & 0.007923 & 506.9 & 0.018 \\
0.03 & 4.606218 & 0.000148 & 0.007669 & 600.5 & 0.019
\end{tabular}}
\tablefoot{
The events, ejecta, and systems columns are normalised per unit mass of star forming material (M$_{\odot \rm SFM}$).}
\label{tab:zsumtabupdated}
\end{table*}

%%%%%%MWD%%%%%zm3

%%%%%%MWD%%%%%z0p02

\begin{table*}[]
\begin{tabular}{llllllll}
\textbf{$Z$} & \textbf{\Mwd\textless{}0.7}& \textbf{0.7\textless{}\Mwd\textless{}0.9 }& \textbf{0.9\textless{}\Mwd\textless{}1.075} & \textbf{1.075\textless{}\Mwd\textless{}1.2}& \textbf{1.2\textless{}\Mwd\textless{}1.3} & \textbf{1.3\textless{}\Mwd} &\textbf{All} \\ \hline
0.0001 & 2.072137 & 1.680010 & 1.068057 & 0.220397 & 1.301370 & 4.650666 & 10.99264 \\
0.00025 & 2.111505 & 1.937149 & 1.401279 & 0.017795 & 0.744554 & 3.506661 & 9.718943 \\
0.0005 & 2.136337 & 1.931354 & 1.548759 & 0.044335 & 0.652409 & 3.406756 & 9.719950 \\
0.001 & 1.981952 & 1.559799 & 1.345211 & 0.100697 & 0.636987 & 2.379447 & 8.004093 \\
0.002 & 1.883467 & 1.219091 & 1.232788 & 0.346584 & 0.731531 & 2.293006 & 7.706468 \\
0.0025 & 1.800458 & 1.116510 & 1.048383 & 0.407171 & 0.936668 & 2.984717 & 8.293906 \\
0.003 & 1.797227 & 0.974580 & 0.997121 & 0.306053 & 0.874978 & 4.109652 & 9.059610 \\
0.005 & 1.579250 & 0.672812 & 0.572286 & 0.115527 & 0.376039 & 2.072693 & 5.388607 \\
0.007 & 1.398327 & 0.721754 & 0.500943 & 0.141881 & 0.722456 & 2.259692 & 5.745053 \\
0.01 & 1.270525 & 0.669482 & 0.469553 & 0.373532 & 0.753295 & 3.301617 & 6.838003 \\
0.0125 & 1.131020 & 0.448098 & 0.295811 & 0.157540 & 0.375305 & 1.935096 & 4.342870 \\
0.015 & 1.108896 & 0.390628 & 0.290160 & 0.171892 & 0.534657 & 2.556340 & 5.052573 \\
0.0175 & 1.041641 & 0.378872 & 0.276474 & 0.193417 & 0.716184 & 2.342184 & 4.948772 \\
0.02 & 0.982697 & 0.367568 & 0.295257 & 0.227497 & 0.705073 & 2.737965 & 5.316057 \\
0.025 & 0.839674 & 0.267833 & 0.190316 & 0.122393 & 0.636342 & 1.964759 & 4.021317 \\
0.03 & 0.795438 & 0.288432 & 0.221746 & 0.205154 & 0.731924 & 2.358992 & 4.601686
\end{tabular}
\caption{Total number of events per unit star forming material produced by each WD mass bracket at each metallicity. Fig. \ref{fig:zsumtab_mwd_event} presents this information graphically.}
\label{tab:zsumtab_events_mwd}
\end{table*}

\begin{table*}[]
\caption{Total mass ejected per unit star forming material produced by each WD mass bracket at each metallicity.}
\begin{tabular}{llllllll}
\textbf{$Z$} & \textbf{\Mwd\textless{}0.7}& \textbf{0.7\textless{}\Mwd\textless{}0.9 }& \textbf{0.9\textless{}\Mwd\textless{}1.075} & \textbf{1.075\textless{}\Mwd\textless{}1.2}& \textbf{1.2\textless{}\Mwd\textless{}1.3} & \textbf{1.3\textless{}\Mwd} &\textbf{All} \\ \hline
0.0001 & 2.241E-04 & 7.060E-05 & 3.148E-05 & 2.497E-06 & 5.163E-06 & 4.062E-06 & 3.379E-04 \\
0.00025 & 2.371E-04 & 8.680E-05 & 3.640E-05 & 1.951E-07 & 3.277E-06 & 2.367E-06 & 3.662E-04 \\
0.0005 & 2.489E-04 & 9.629E-05 & 4.053E-05 & 5.127E-07 & 3.096E-06 & 2.589E-06 & 3.919E-04 \\
0.001 & 2.418E-04 & 8.120E-05 & 3.596E-05 & 8.463E-07 & 3.148E-06 & 2.732E-06 & 3.656E-04 \\
0.002 & 2.444E-04 & 6.674E-05 & 3.187E-05 & 2.768E-06 & 3.101E-06 & 2.543E-06 & 3.514E-04 \\
0.0025 & 2.378E-04 & 6.169E-05 & 2.882E-05 & 3.769E-06 & 3.472E-06 & 2.762E-06 & 3.383E-04 \\
0.003 & 2.410E-04 & 5.349E-05 & 2.766E-05 & 3.206E-06 & 3.646E-06 & 3.303E-06 & 3.323E-04 \\
0.005 & 2.166E-04 & 3.785E-05 & 1.520E-05 & 1.181E-06 & 1.736E-06 & 2.142E-06 & 2.747E-04 \\
0.007 & 1.953E-04 & 4.184E-05 & 1.398E-05 & 1.819E-06 & 3.053E-06 & 2.098E-06 & 2.581E-04 \\
0.01 & 1.817E-04 & 3.954E-05 & 1.343E-05 & 4.038E-06 & 3.184E-06 & 3.154E-06 & 2.450E-04 \\
0.0125 & 1.581E-04 & 2.391E-05 & 7.222E-06 & 1.354E-06 & 1.371E-06 & 1.661E-06 & 1.936E-04 \\
0.015 & 1.539E-04 & 2.135E-05 & 7.357E-06 & 1.767E-06 & 2.062E-06 & 1.990E-06 & 1.884E-04 \\
0.0175 & 1.458E-04 & 2.180E-05 & 7.321E-06 & 2.293E-06 & 2.675E-06 & 2.100E-06 & 1.820E-04 \\
0.02 & 1.401E-04 & 2.151E-05 & 8.104E-06 & 2.700E-06 & 2.893E-06 & 2.563E-06 & 1.779E-04 \\
0.025 & 1.214E-04 & 1.439E-05 & 4.373E-06 & 1.449E-06 & 2.282E-06 & 1.543E-06 & 1.454E-04 \\
0.03 & 1.177E-04 & 1.665E-05 & 6.026E-06 & 2.607E-06 & 2.959E-06 & 2.192E-06 & 1.482E-04
\end{tabular}
\tablefoot{Fig. \ref{fig:zsumtab_mwd_ejecta} presents this information graphically.}
\label{tab:zsumtab_ejecta_mwd}
\end{table*}

\begin{figure*}
\centering
\begin{subfigure}{0.5\textwidth}
\centering
\includegraphics[width=1\columnwidth]{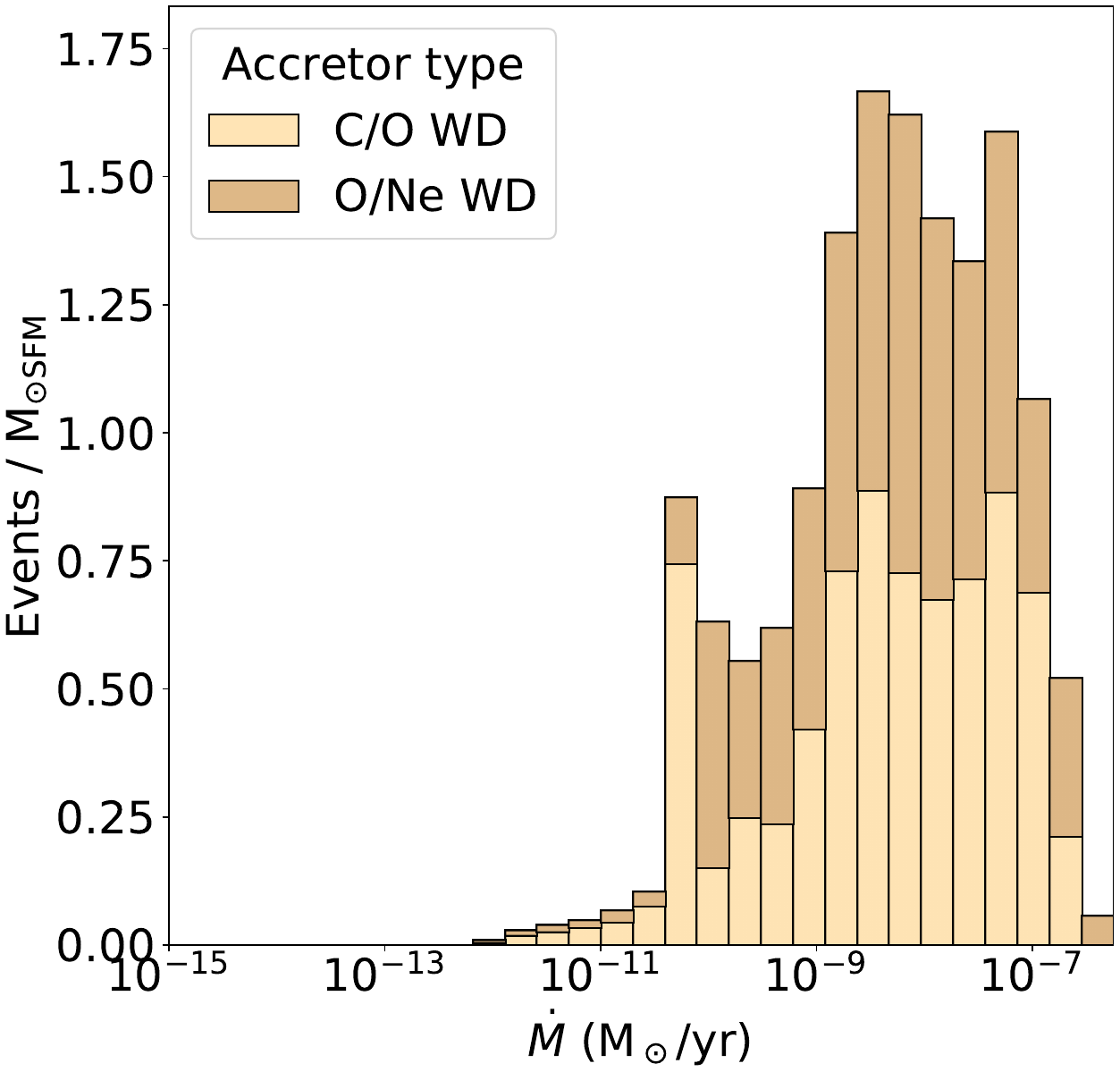}
\caption{}
\end{subfigure}%
\begin{subfigure}{0.5\textwidth}
\centering
\includegraphics[width=1\columnwidth]{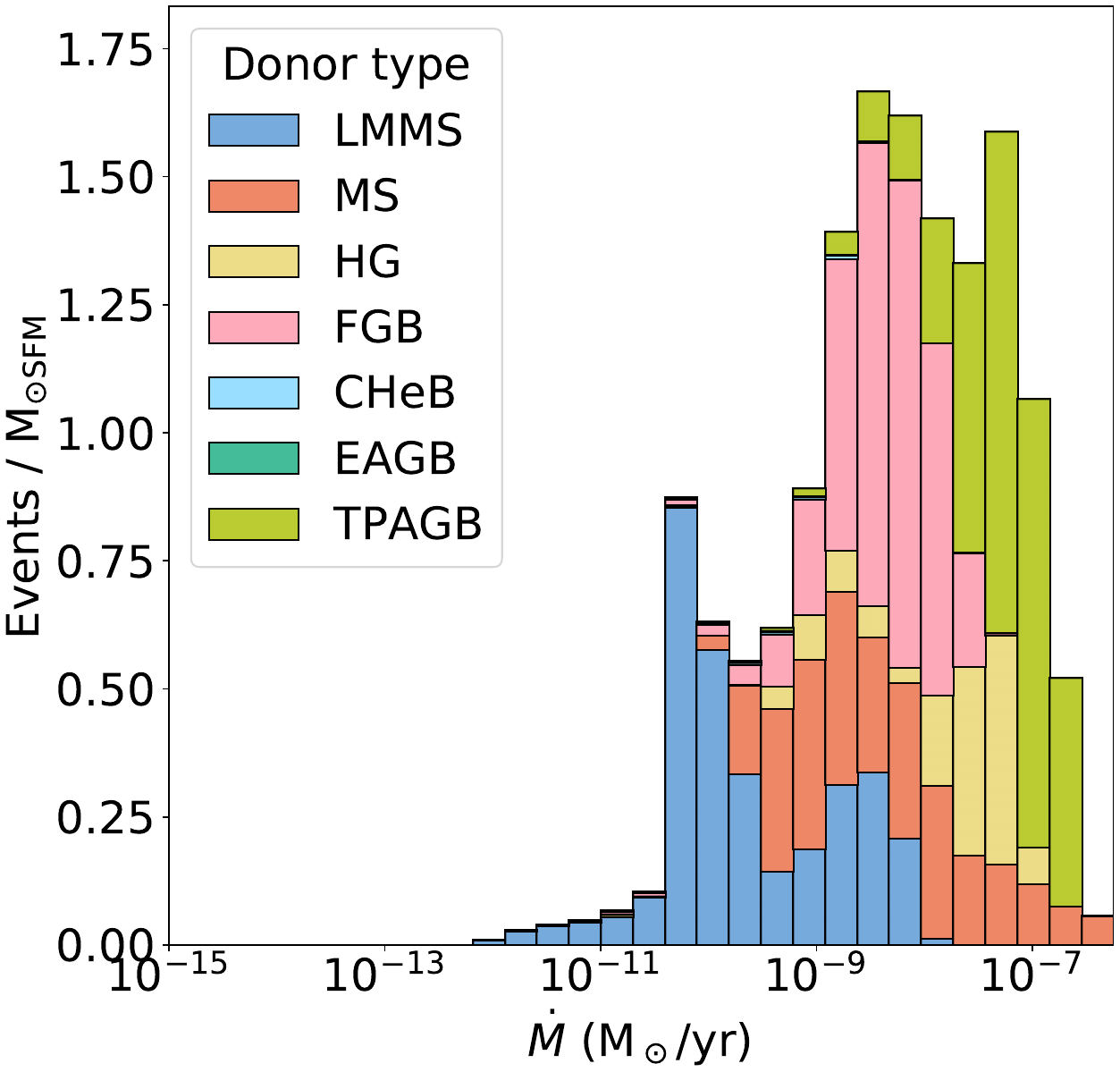}
\caption{}
\end{subfigure}

\begin{subfigure}{0.5\textwidth}
\centering
\includegraphics[width=1\columnwidth]{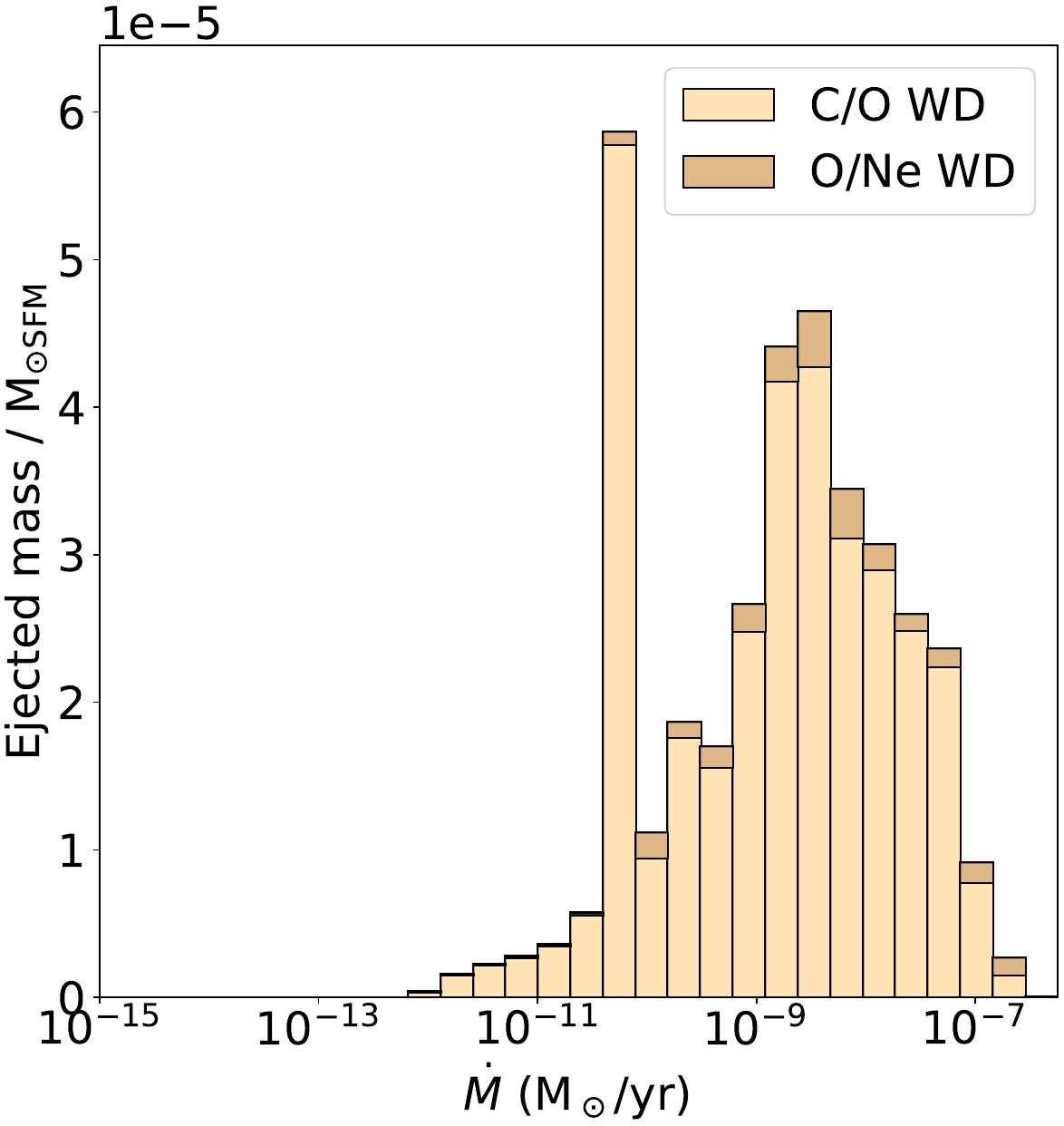}
\caption{}
\end{subfigure}%
\begin{subfigure}{0.5\textwidth}
\centering
\includegraphics[width=1\columnwidth]{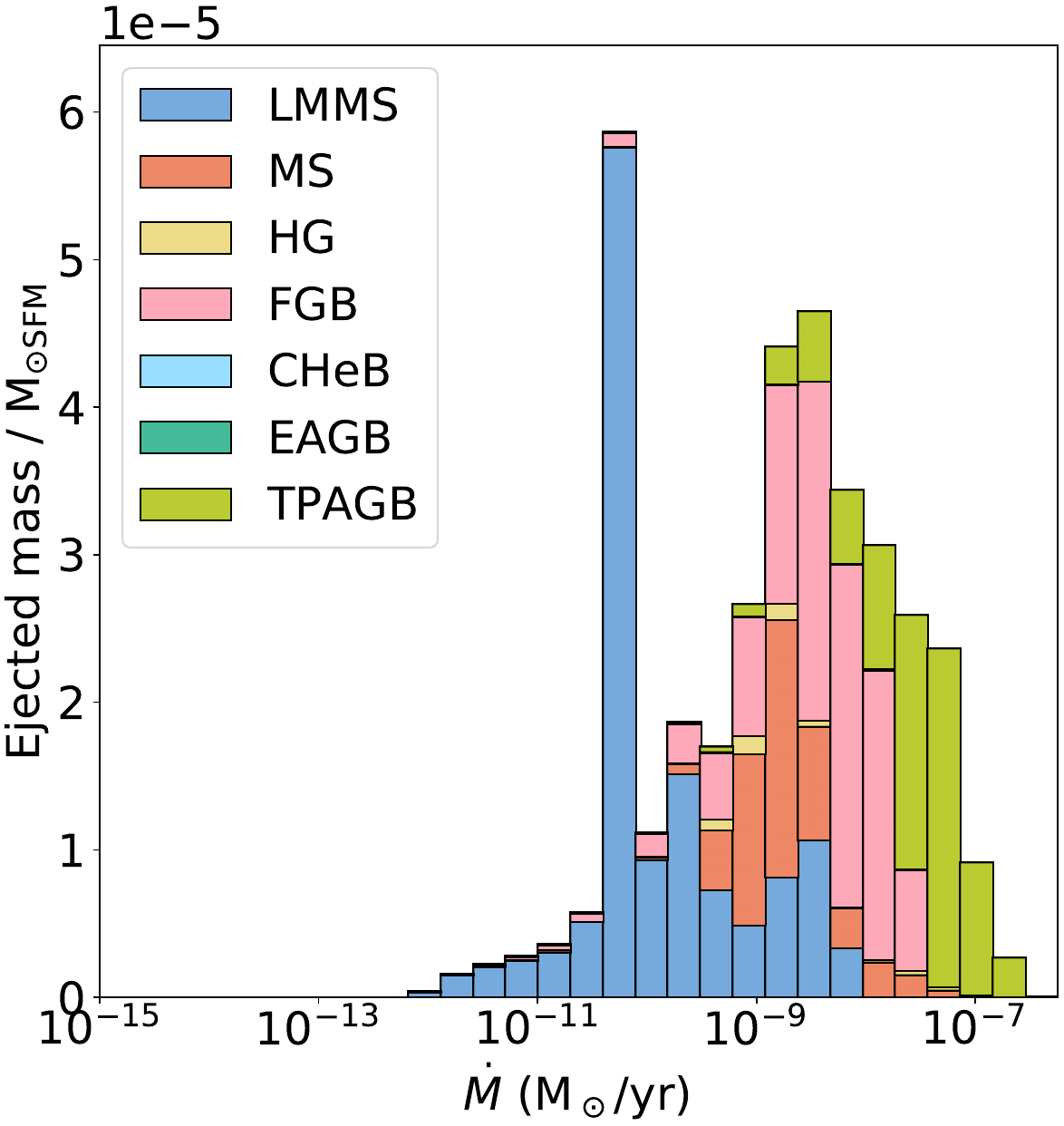}
\caption{}
\end{subfigure}

\caption{Nova accretion rate (\Mdot) distributions for $Z$~=~\tento{-3}, weighting by number of novae (top) and ejecta mass (bottom) and colouring by WD composition (left) and donor stellar type (right).}
\label{fig:hist_mdotzm3}
\end{figure*}

\begin{figure*}
\centering
\begin{subfigure}{0.5\textwidth}
\centering
\includegraphics[width=1\columnwidth]{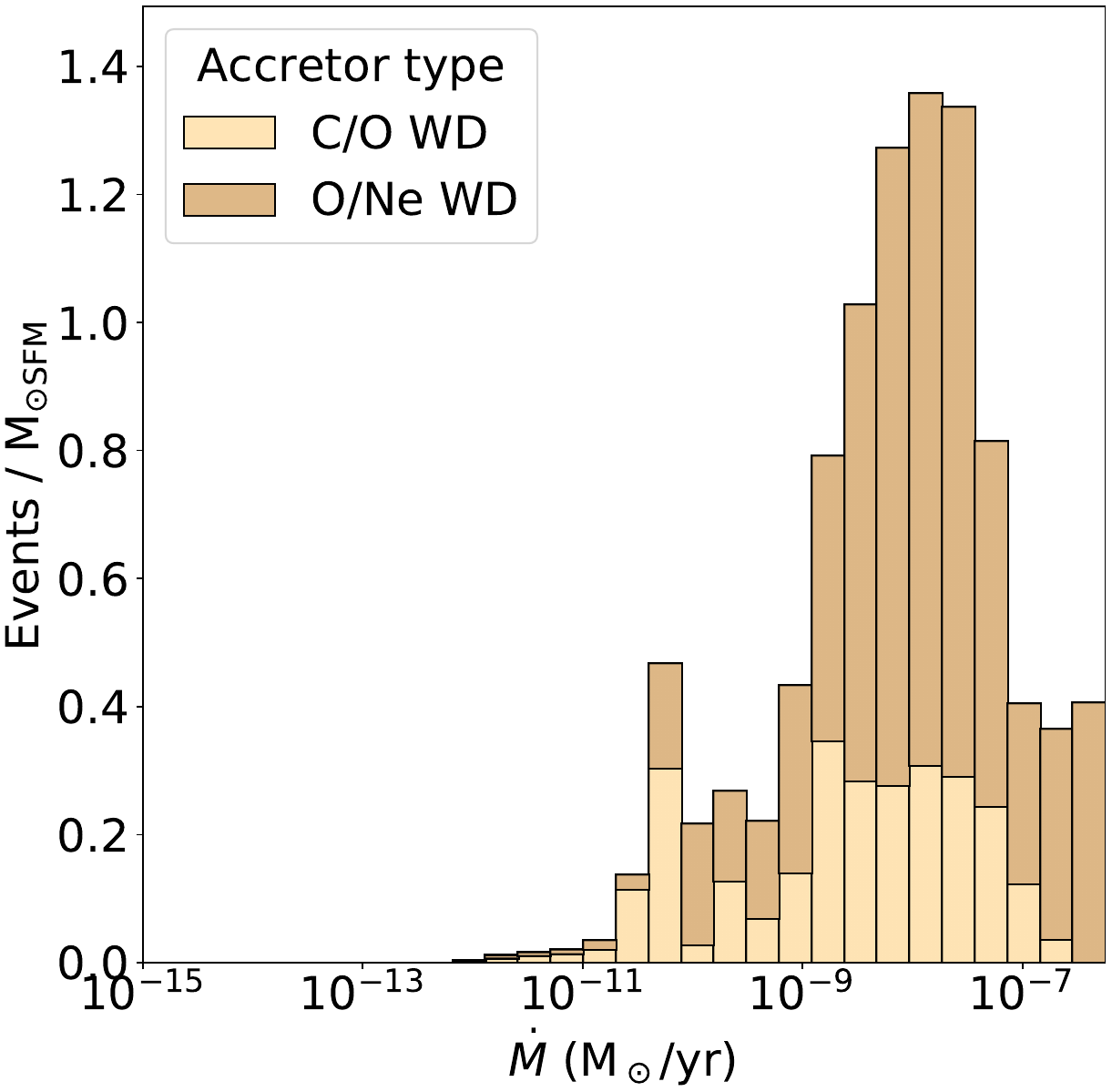}
\caption{}
\end{subfigure}%
\begin{subfigure}{0.5\textwidth}
\centering
\includegraphics[width=1\columnwidth]{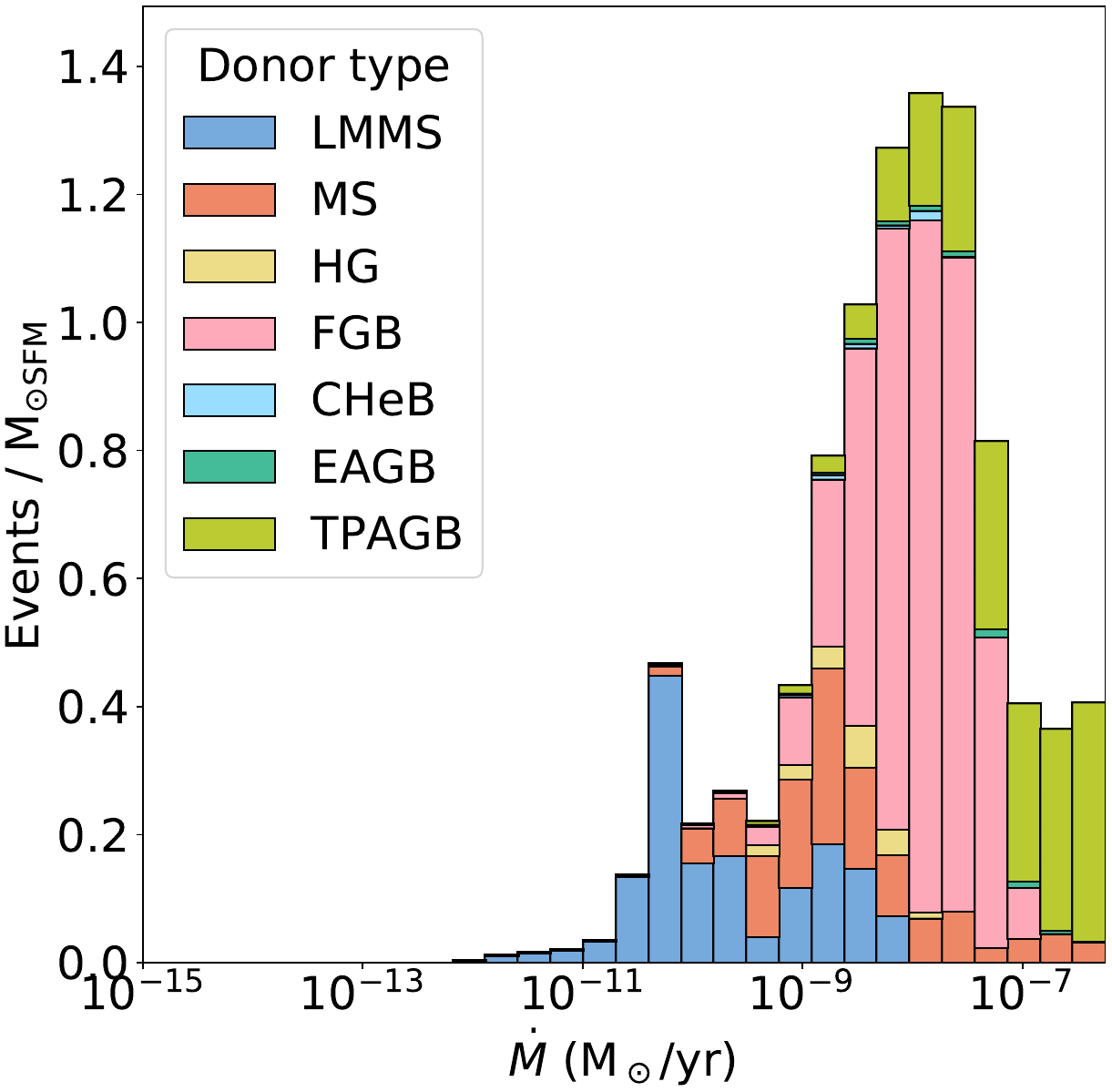}
\caption{}
\end{subfigure}

\begin{subfigure}{0.5\textwidth}
\centering
\includegraphics[width=1\columnwidth]{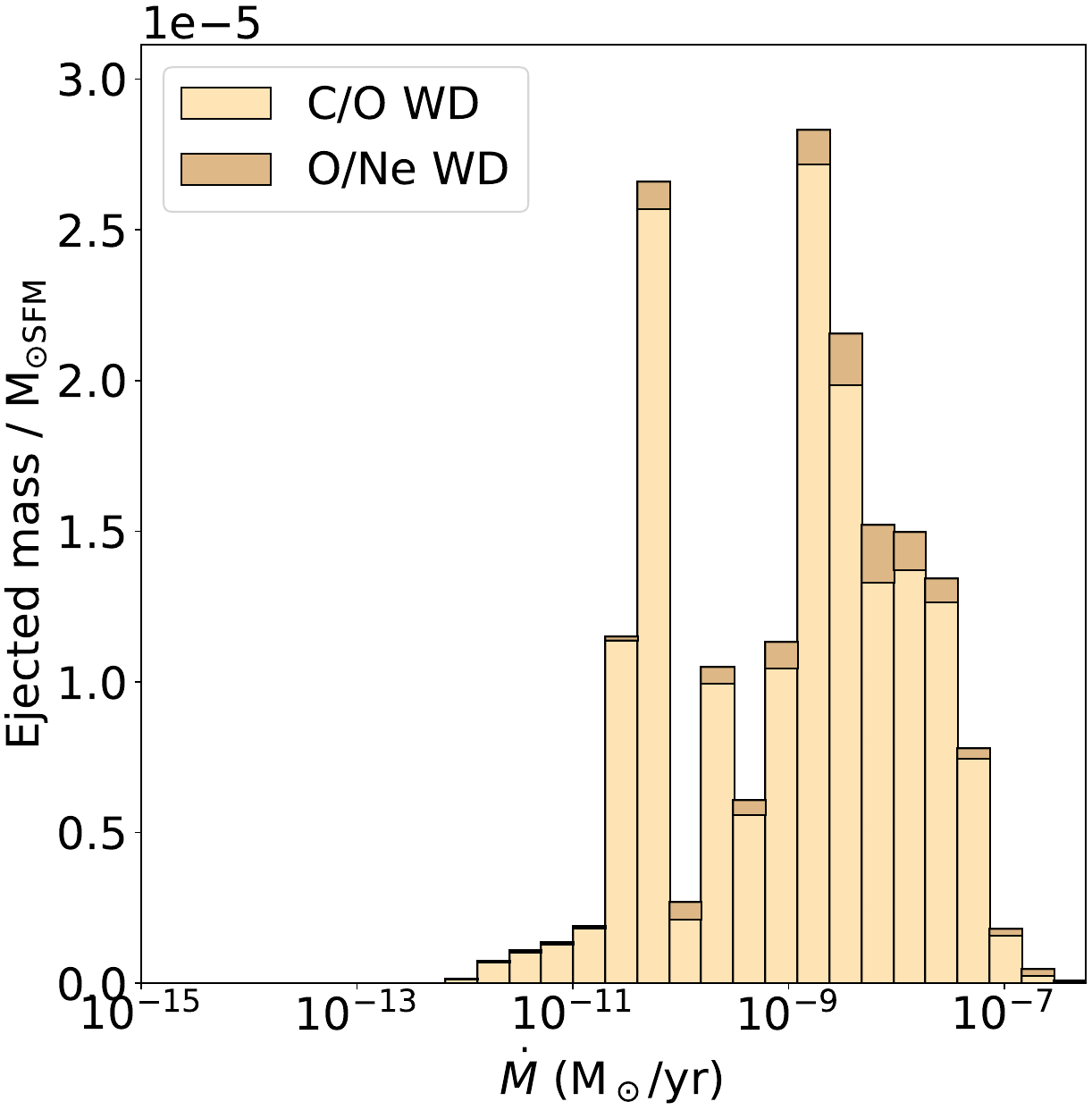}
\caption{}
\end{subfigure}%
\begin{subfigure}{0.5\textwidth}
\centering
\includegraphics[width=1\columnwidth]{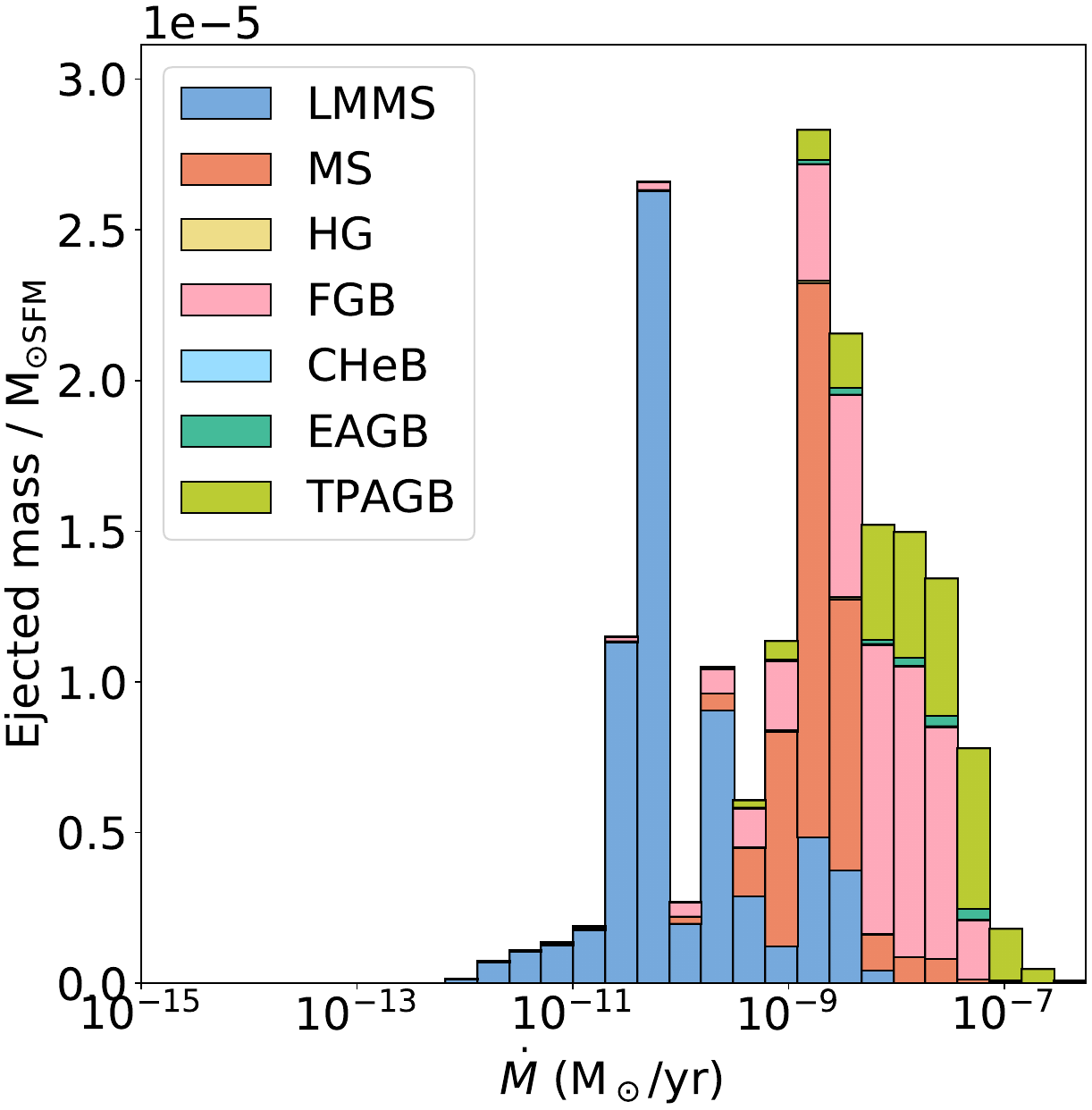}
\caption{}
\end{subfigure}

\caption{Nova accretion rate (\Mdot) distributions for $Z$~=~0.02.}
\label{fig:hist_mdotz0p02}
\end{figure*}

Distributions of the accretion rate at the time of nova eruption are shown in Figs. \ref{fig:hist_mdotzm3} and \ref{fig:hist_mdotz0p02}. Shifts in the \Mdot\ parameter-space are less significant than in \Mwd-space when comparing the event-weighted and ejecta-weighted distributions. The main differences are a reduction in the importance of systems with \Mdot\ $\gtrsim$~\tento{-8}~M\solarperyr, and an enhancement of very slowly accreting systems (\Mdot~$\lesssim$ \tento{-10}~M\solarperyr).

Particularly evident in Figs. \ref{fig:hist_mdotzm3} and \ref{fig:hist_mdotz0p02} -- although also visible in Figs. \ref{fig:hist_dmproczm3}, \ref{fig:hist_dmprocz0p02}, \ref{fig:hist_mwdzm3} and \ref{fig:hist_mwdz0p02} -- is the shifting relevance of main sequence (LMMS and MS) vs giant (FGB and TPAGB) donor stellar types. This is especially clear in at $Z$~=~0.02 (Figs. \ref{fig:hist_dmprocz0p02}, \ref{fig:hist_mwdz0p02} and \ref{fig:hist_mdotz0p02}), where main sequence donors make up a larger portion of the low-mass WD nova population. Disentangling the root cause of this shifting is somewhat problematic. Systems involving main sequence donor stars tend to have lower accretion rates, driving up the critical ignition (and ejecta) mass. Therefore, we expect MS donor stars to be more significant when considering the origin of nova ejecta. However, it is clear from Figs. \ref{fig:hist_mwdzm3} and \ref{fig:hist_mwdz0p02} that the productivity is highly correlated with \Mwd, a property with which the critical ignition mass is far more sensitive. Therefore, most of the shifting between donor types is likely just what the shift in \Mwd\ looks like in donor space. Many of the novae produced by giant donor stars occurred on massive O/Ne WDs (Fig. \ref{fig:hist_mwdz0p02}), and those systems produce very little ejecta mass, reducing the apparent importance of giant donor stars overall. The influence of accretion rates on critical ignition masses undoubtedly also plays a role, but it is likely a secondary one, and difficult to isolate.

%%%%%%MDOT%%%%%zm3

%%%%%%MDOT%%%%%z0p02

% %%%%%%TIME%%%%%%zm3
% \begin{figure*}
% \centering
% \begin{subfigure}{0.5\textwidth}
% \centering
% \includegraphics[width=1\columnwidth]{pics/z0p001/AD_TIME_accH_EW.pdf}
% \caption{}
% \end{subfigure}%
% \begin{subfigure}{0.5\textwidth}
% \centering
% \includegraphics[width=1\columnwidth]{pics/z0p001/AD_TIME_donH_EW.pdf}
% \caption{}
% \end{subfigure}

% \begin{subfigure}{0.5\textwidth}
% \centering
% \includegraphics[width=1\columnwidth]{pics/z0p001/AD_TIME_accH_ISMW.pdf}
% \caption{}
% \end{subfigure}%
% \begin{subfigure}{0.5\textwidth}
% \centering
% \includegraphics[width=1\columnwidth]{pics/z0p001/AD_TIME_donH_ISMW.pdf}
% \caption{}
% \end{subfigure}

% \caption{Nova delay-time (top) and ejecta delay-time (bottom) distributions for $Z$ = \tento{-3}, colouring by WD composition (left) and donor stellar type (right).}
% \label{fig:hist_dtdzm3}
% \end{figure*}

% %%%%%%TIME%%%%%%z0p02
% \begin{figure*}
% \centering
% \begin{subfigure}{0.5\textwidth}
% \centering
% \includegraphics[width=1\columnwidth]{pics/z0p02/AD_TIME_accH_EW.pdf}
% \caption{}
% \end{subfigure}%
% \begin{subfigure}{0.5\textwidth}
% \centering
% \includegraphics[width=1\columnwidth]{pics/z0p02/AD_TIME_donH_EW.pdf}
% \caption{}
% \end{subfigure}

% \begin{subfigure}{0.5\textwidth}
% \centering
% \includegraphics[width=1\columnwidth]{pics/z0p02/AD_TIME_accH_ISMW.pdf}
% \caption{}
% \end{subfigure}%
% \begin{subfigure}{0.5\textwidth}
% \centering
% \includegraphics[width=1\columnwidth]{pics/z0p02/AD_TIME_donH_ISMW.pdf}
% \caption{}
% \end{subfigure}

% \caption{As Fig. \ref{fig:hist_dtdzm3}, for $Z$ = 0.02.}
% \label{fig:hist_dtdz0p02}
% \end{figure*}

\begin{figure*}
\vspace{-0.4cm}
\centering
\begin{subfigure}{0.45\textwidth}
\centering
\includegraphics[width=0.88\columnwidth]{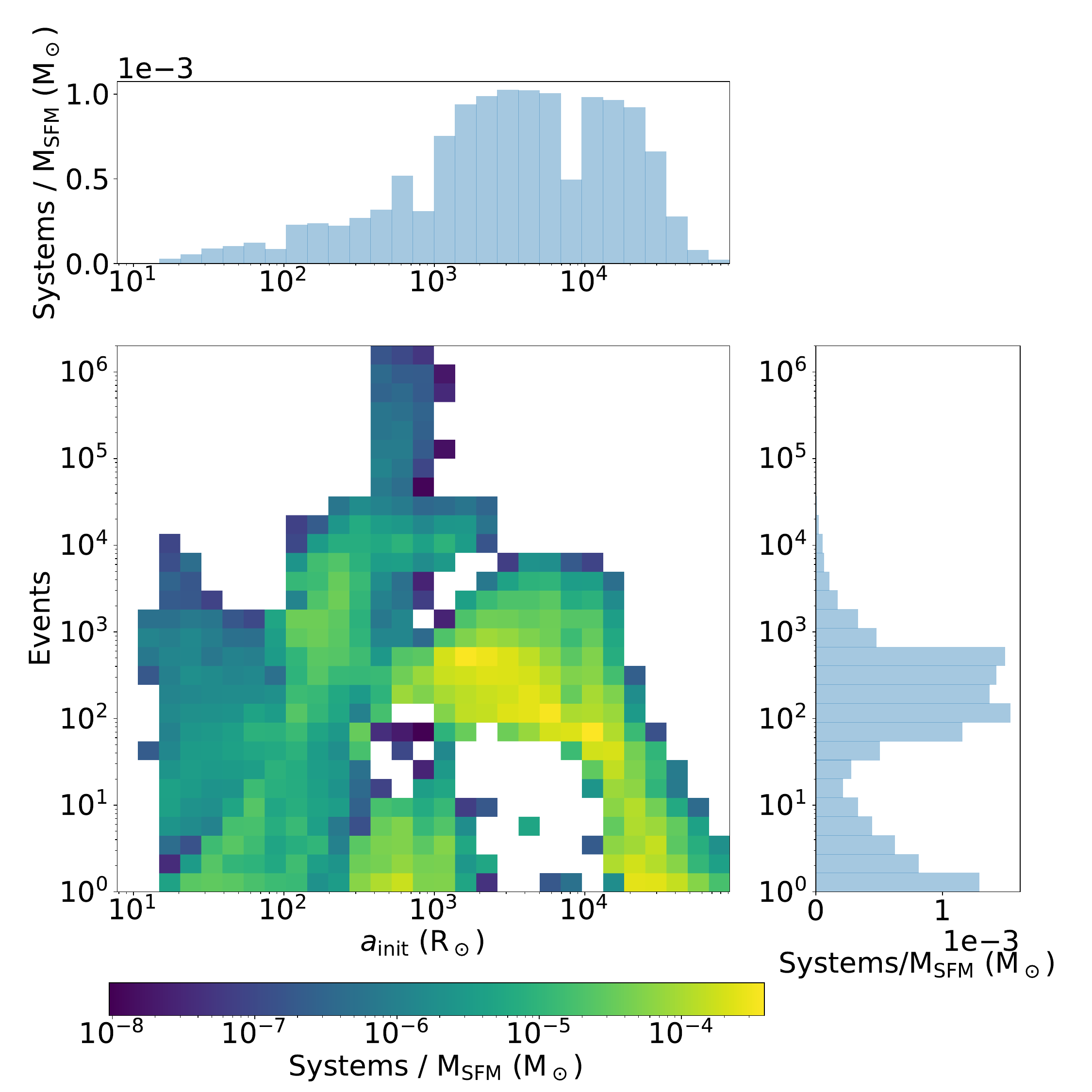}
\caption{}
\end{subfigure}%
\begin{subfigure}{0.45\textwidth}
\centering
\includegraphics[width=0.88\columnwidth]{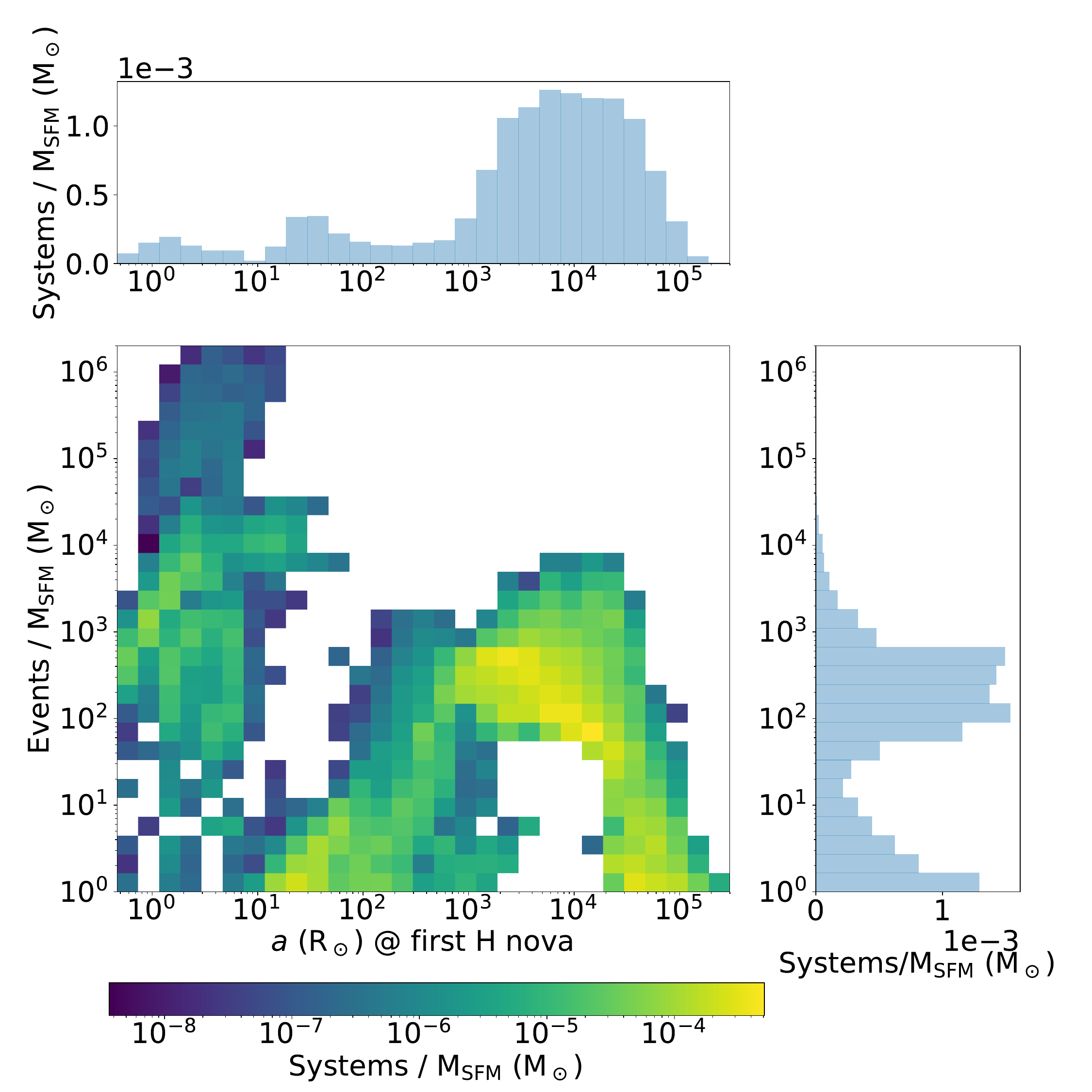}
\caption{}
\end{subfigure}

\begin{subfigure}{0.45\textwidth}
\centering
\includegraphics[width=0.88\columnwidth]{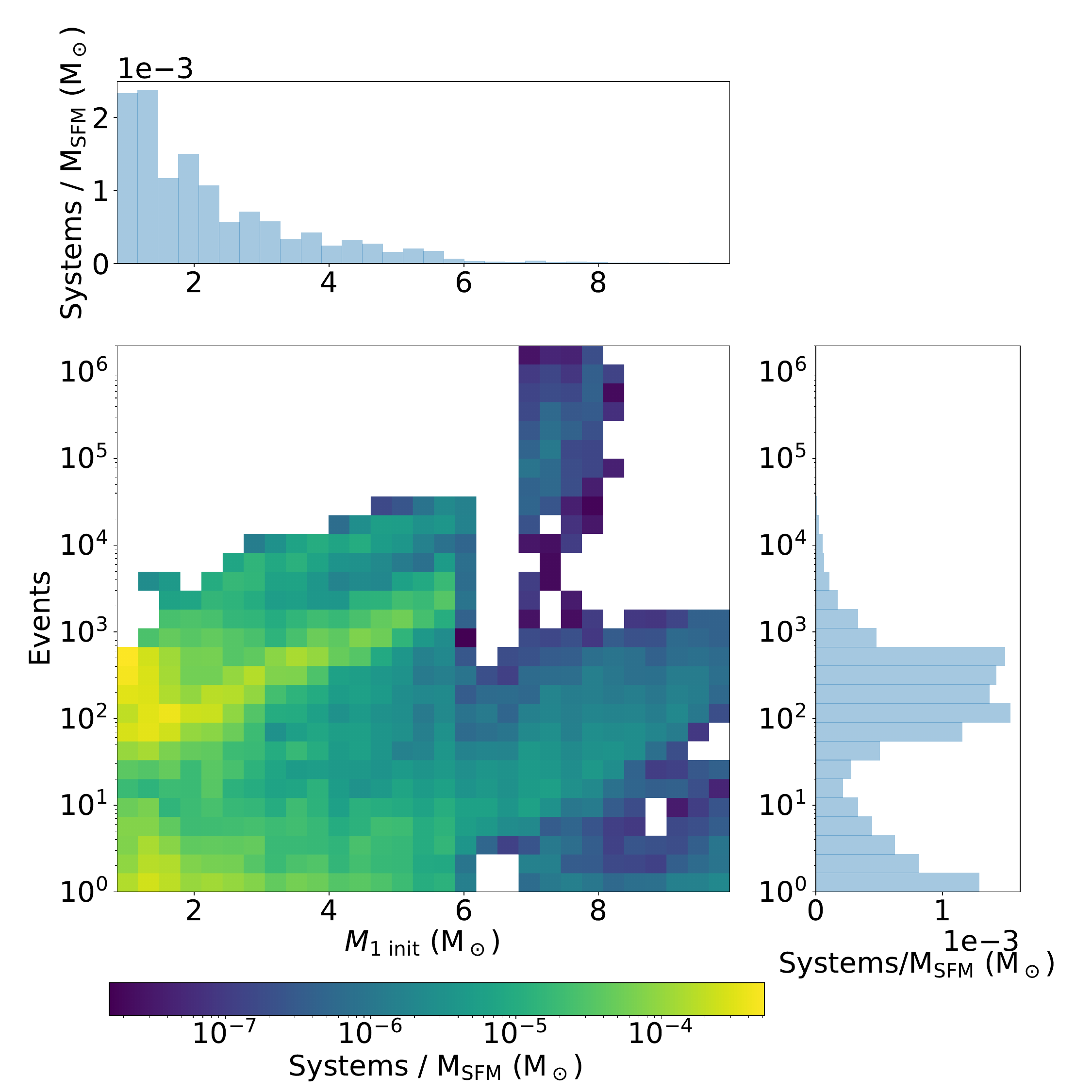}
\caption{}
\end{subfigure}%
\begin{subfigure}{0.45\textwidth}
\centering
\includegraphics[width=0.88\columnwidth]{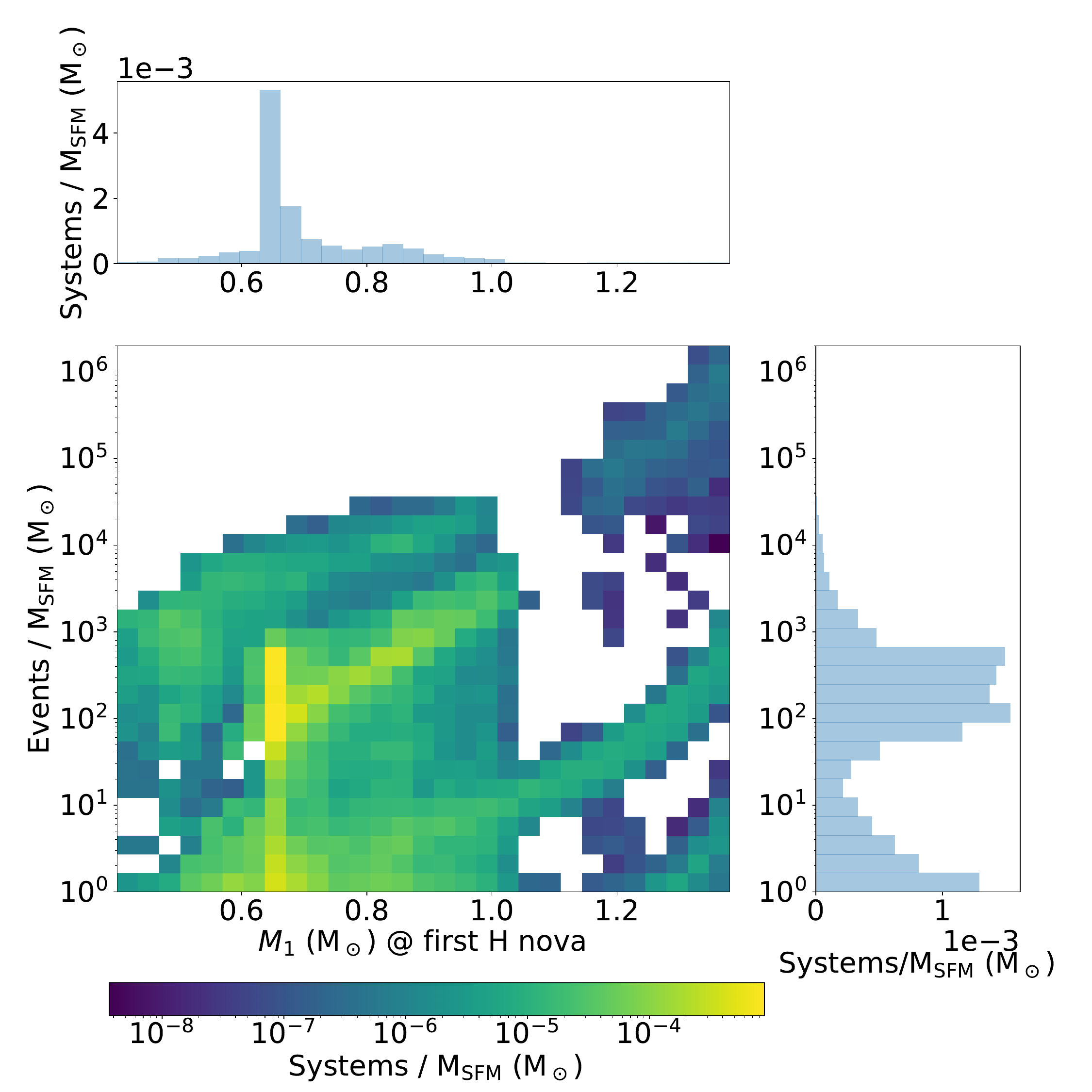}
\caption{}
\end{subfigure}

\begin{subfigure}{0.45\textwidth}
\centering
\includegraphics[width=0.88\columnwidth]{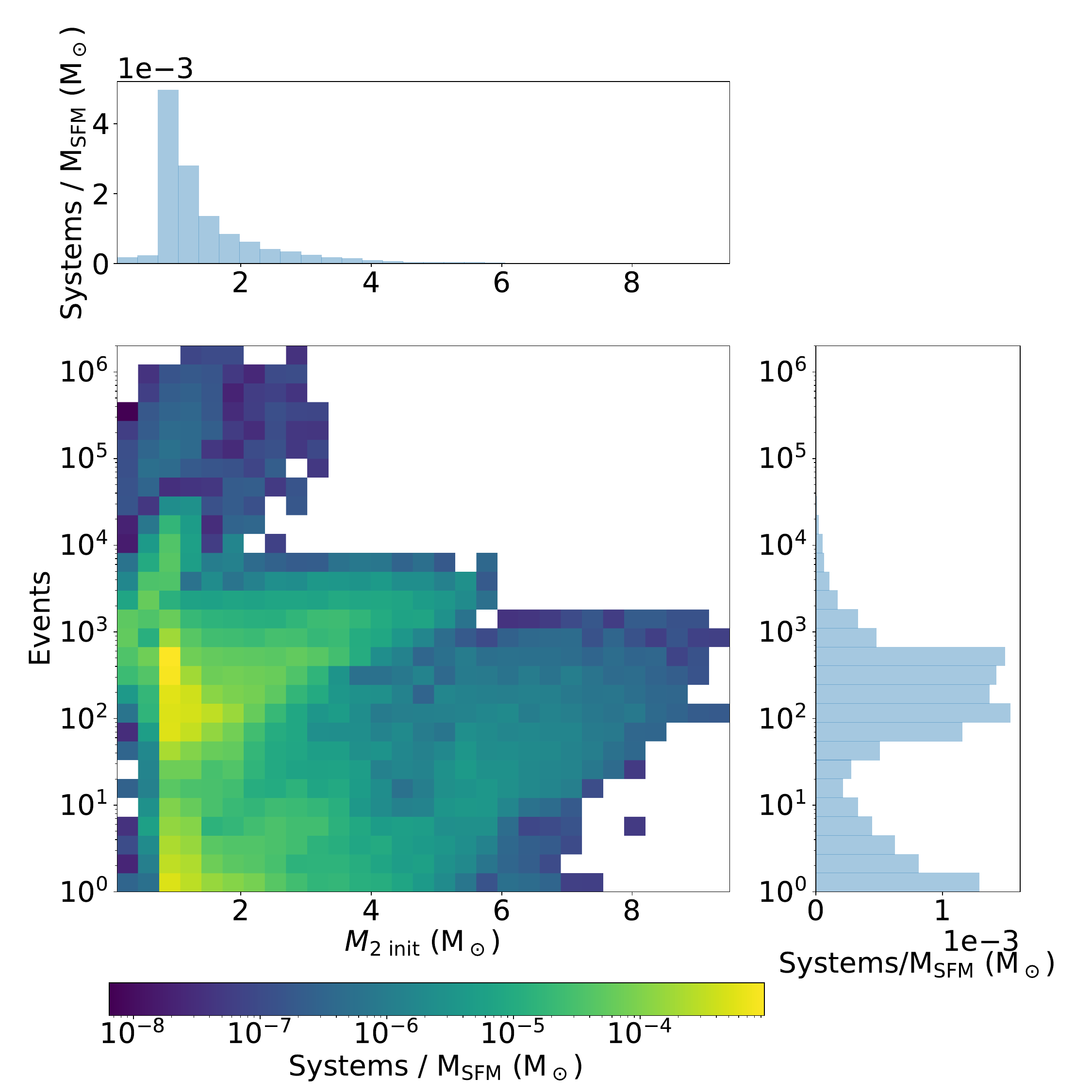}
\caption{}
\end{subfigure}%
\begin{subfigure}{0.45\textwidth}
\centering
\includegraphics[width=0.88\columnwidth]{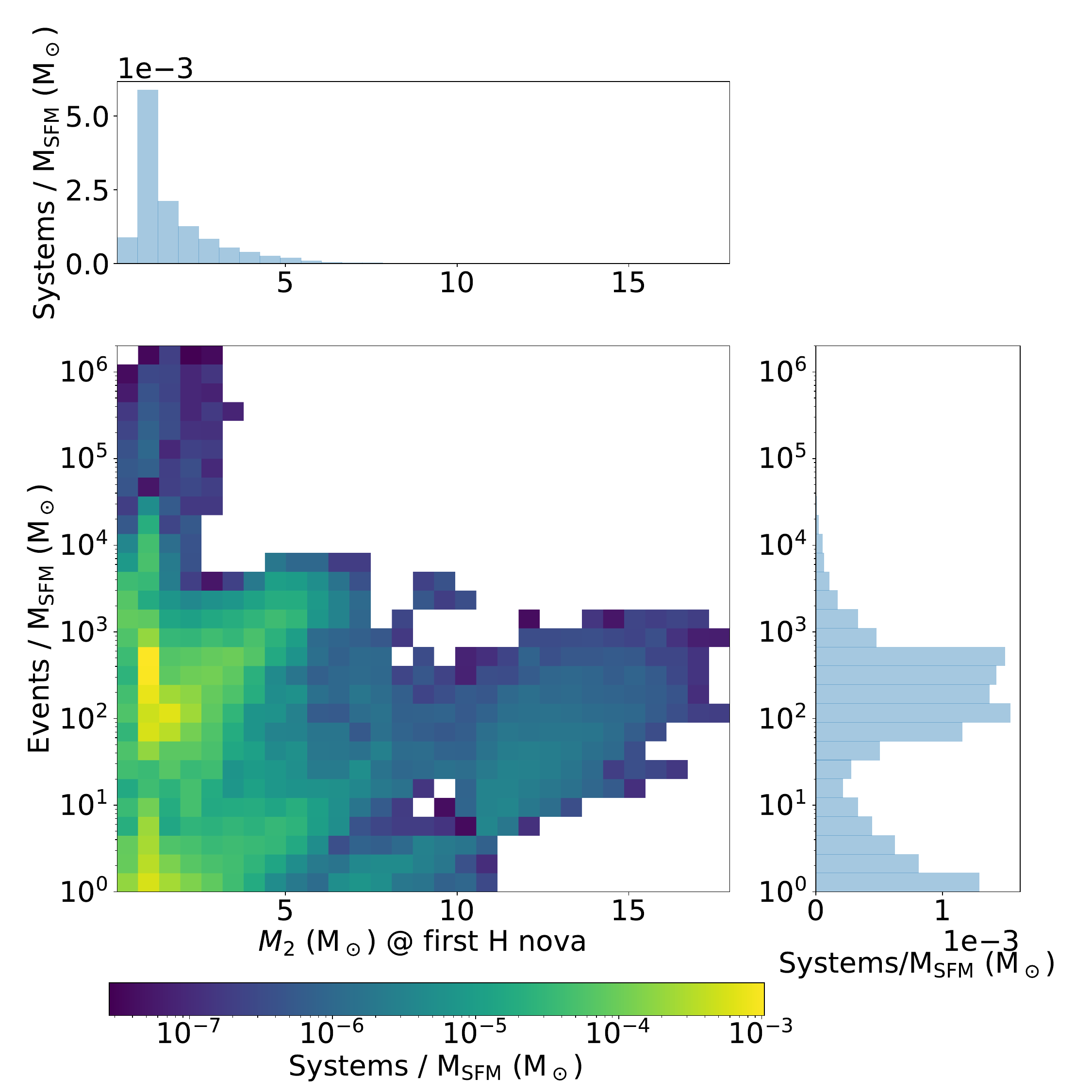}
\caption{}
\end{subfigure}

\caption{Total number of novae produced in each nova system plotted against the initial system properties (left) and the system properties at the time of the first H nova (right). Distributions shown are from the $Z$~=~\tento{-3} population, and coloured by the number of systems per mass of star forming material.}
\label{fig:claweventzm3}
\end{figure*}

\begin{figure*}
\vspace{-0.4cm}
\centering
\begin{subfigure}{0.45\textwidth}
\centering
\includegraphics[width=0.88\columnwidth]{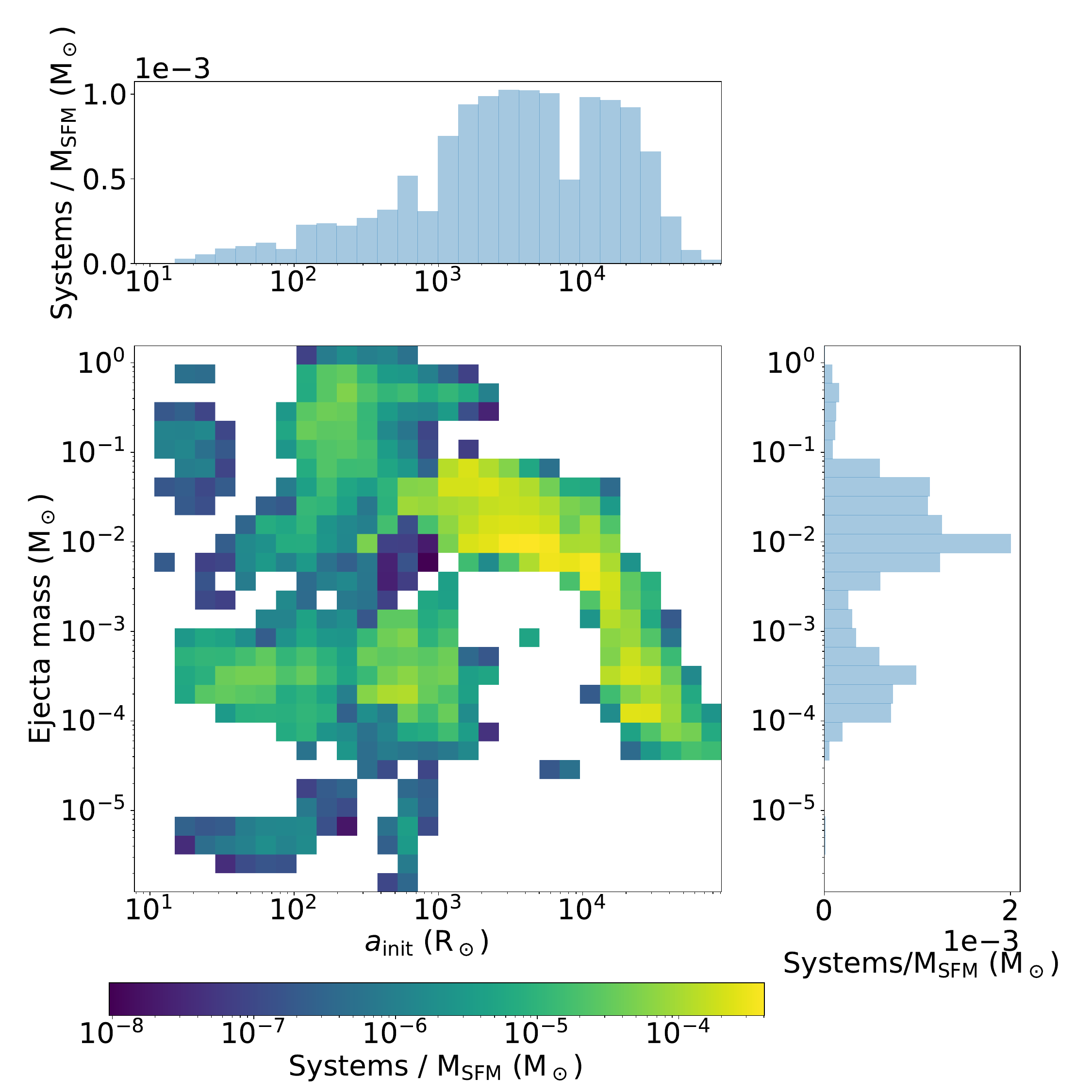}
\caption{}
\end{subfigure}%
\begin{subfigure}{0.45\textwidth}
\centering
\includegraphics[width=0.88\columnwidth]{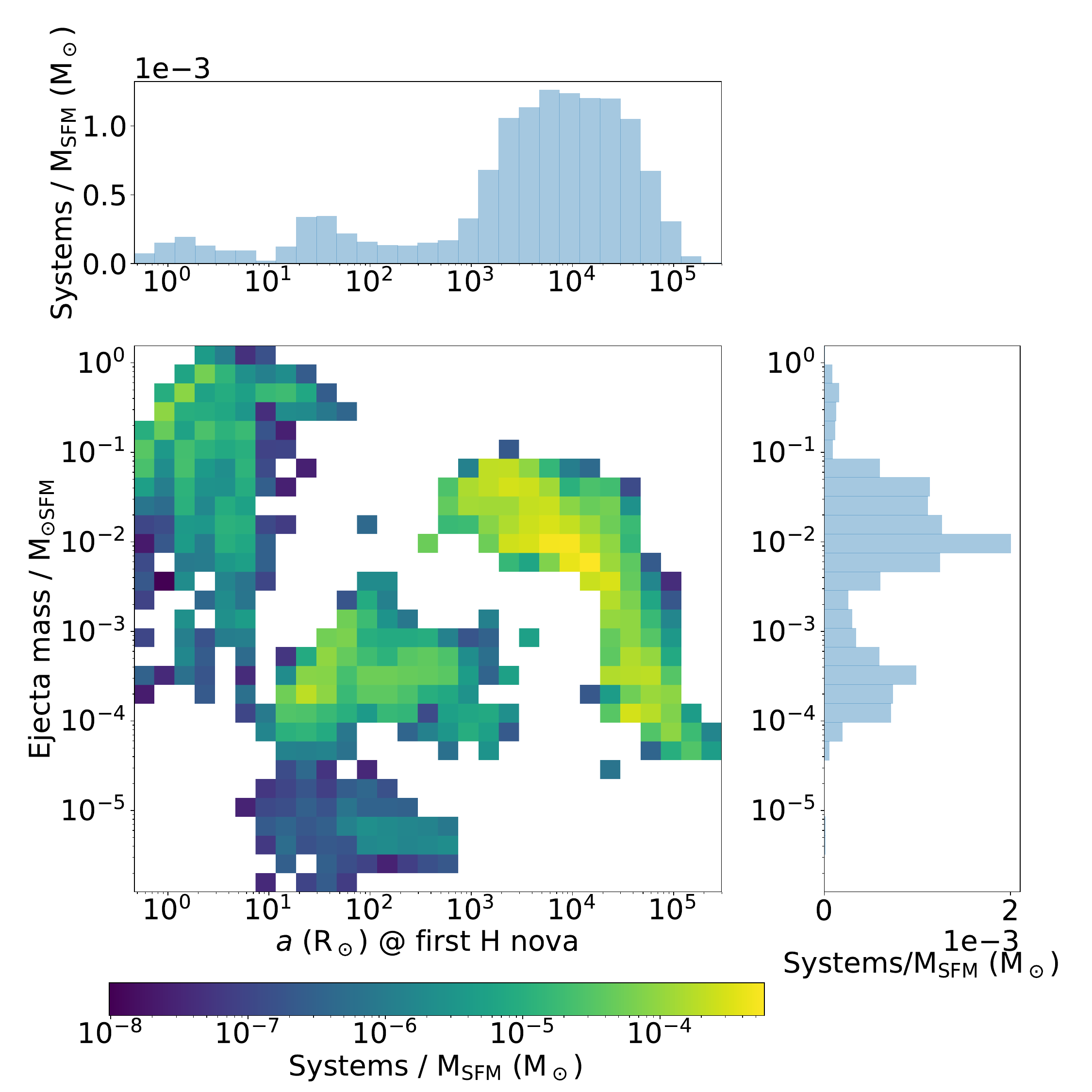}
\caption{}
\end{subfigure}

\begin{subfigure}{0.45\textwidth}
\centering
\includegraphics[width=0.88\columnwidth]{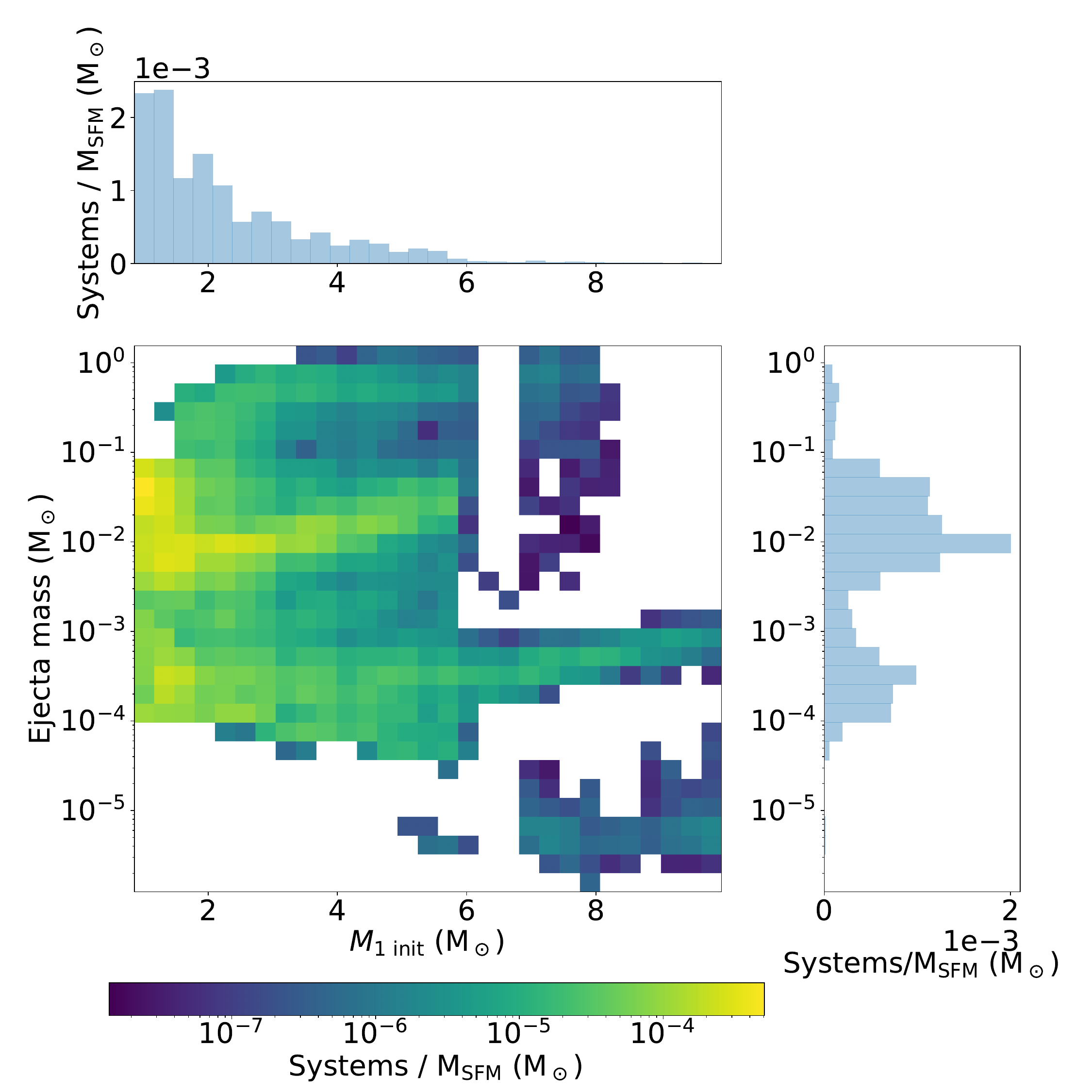}
\caption{}
\end{subfigure}%
\begin{subfigure}{0.45\textwidth}
\centering
\includegraphics[width=0.88\columnwidth]{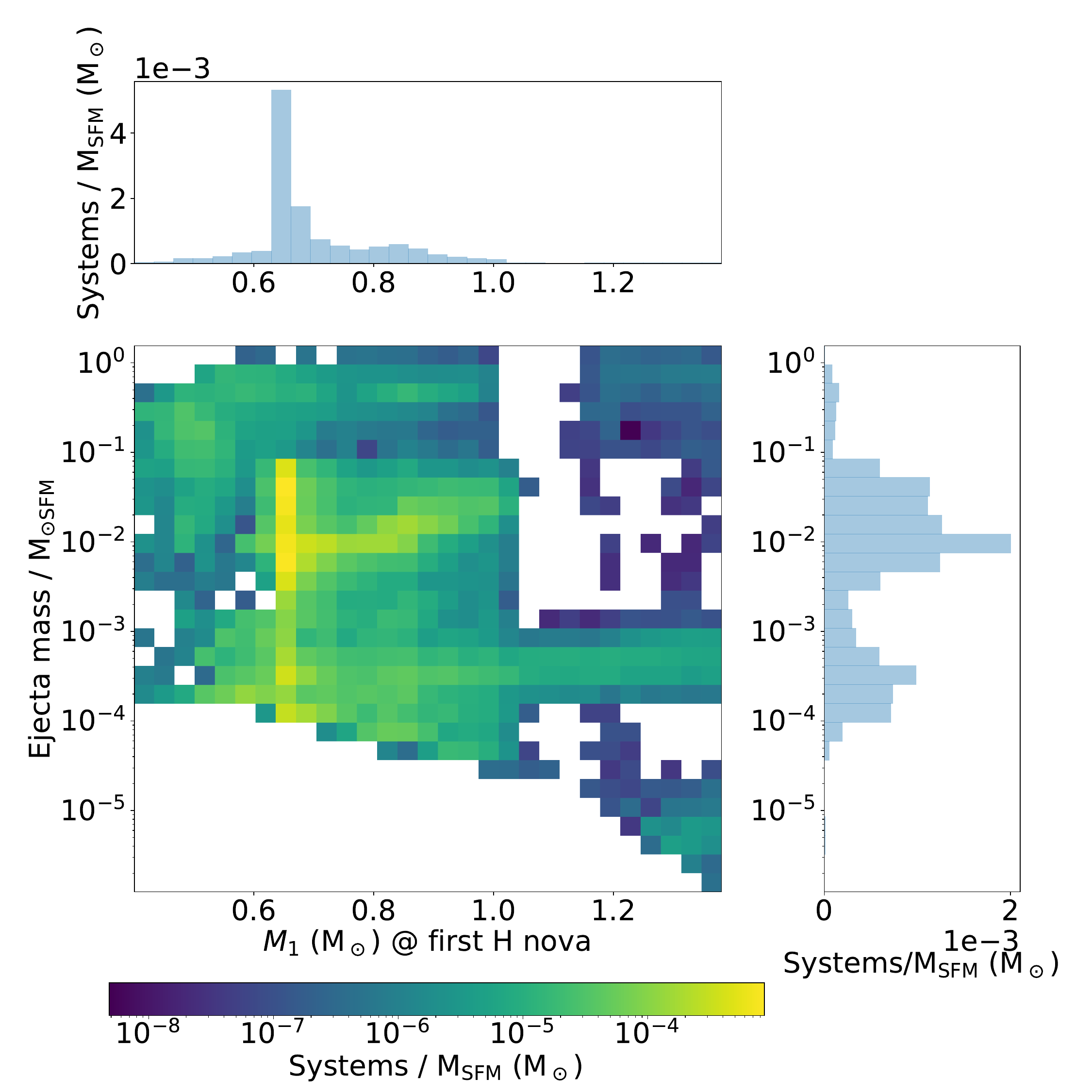}
\caption{}
\end{subfigure}

\begin{subfigure}{0.45\textwidth}
\centering
\includegraphics[width=0.88\columnwidth]{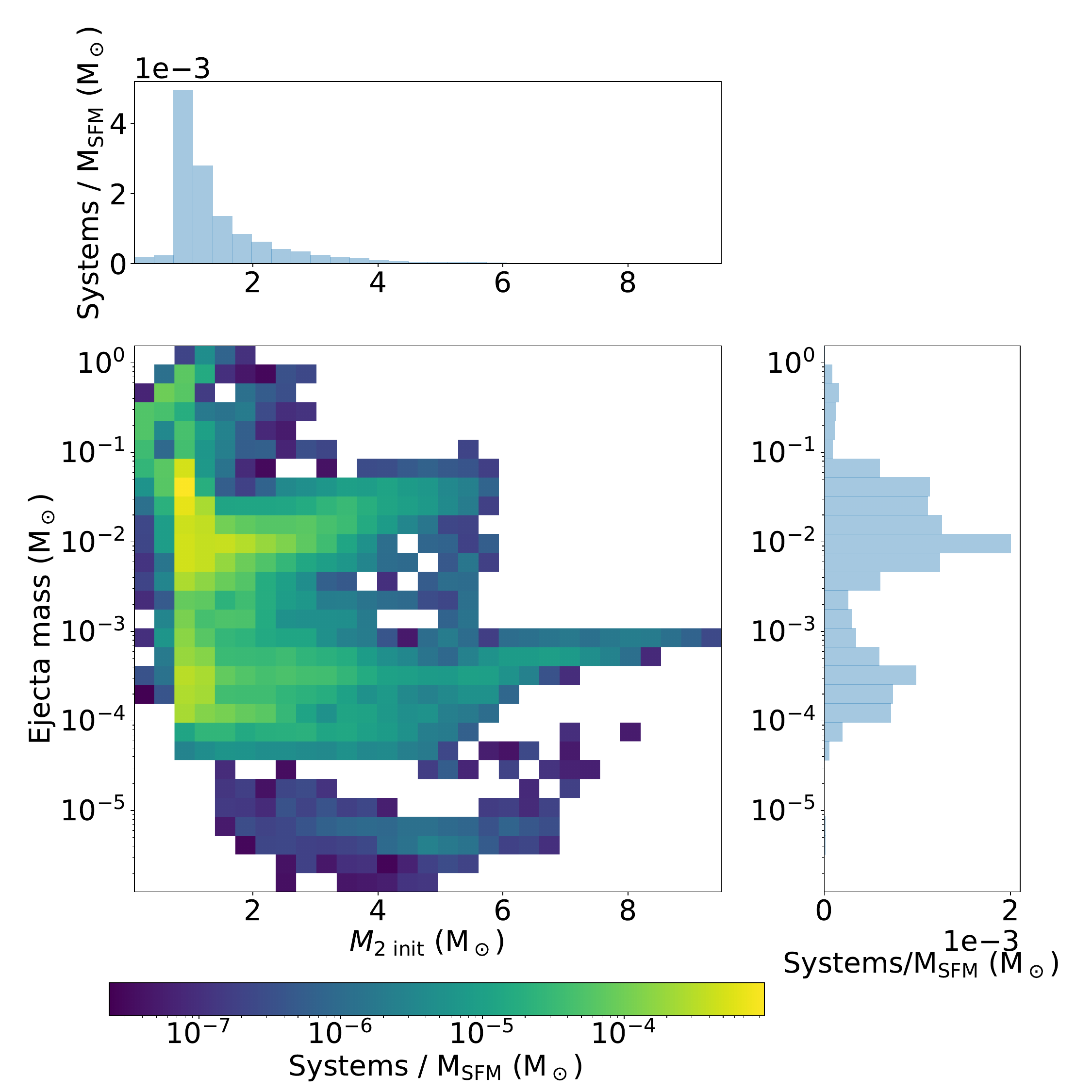}
\caption{}
\end{subfigure}%
\begin{subfigure}{0.45\textwidth}
\centering
\includegraphics[width=0.88\columnwidth]{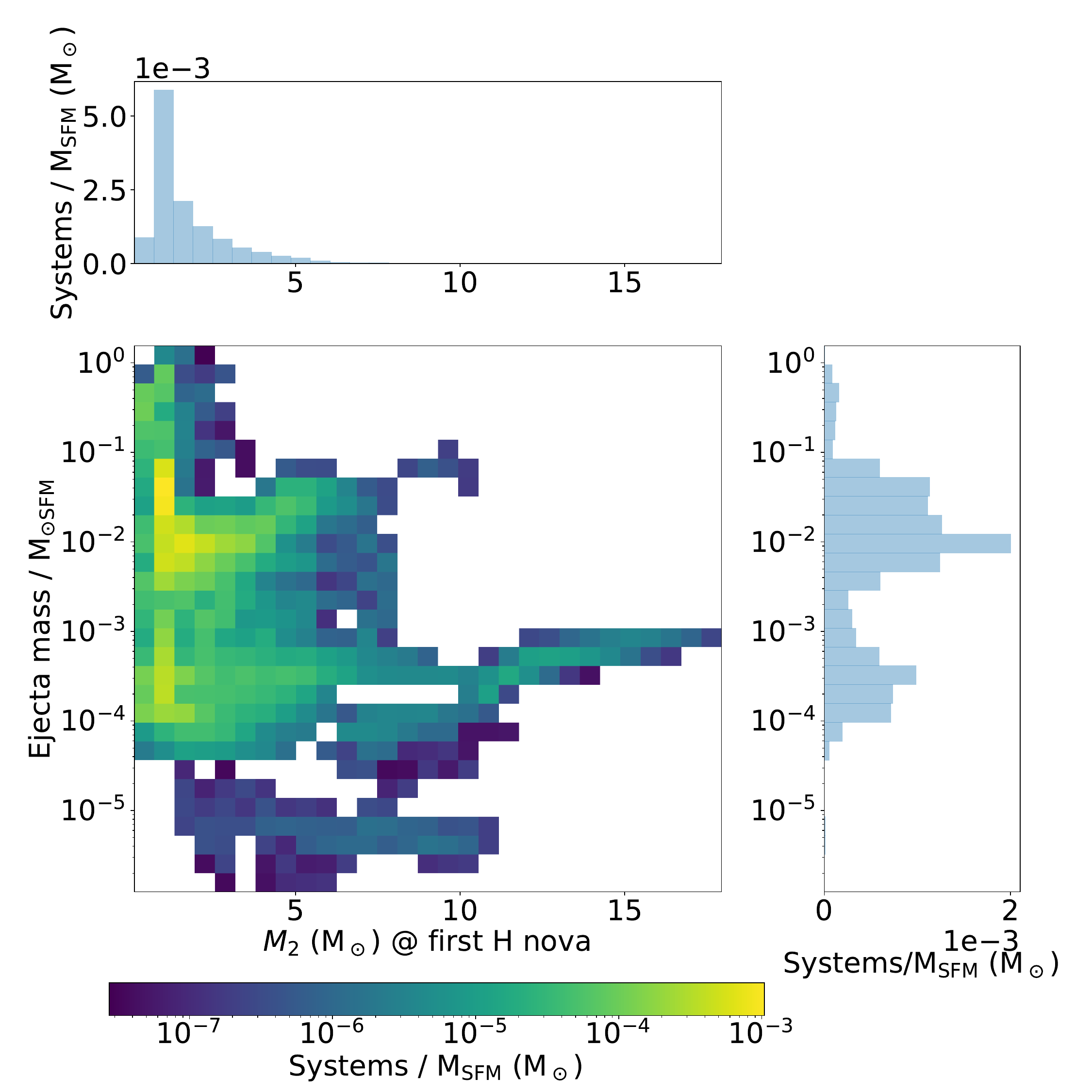}
\caption{}
\end{subfigure}

\caption{Total mass of nova ejecta produced in each nova system plotted against the initial system properties (left) and the system properties at the time of the first H nova (right). Distributions shown are from the $Z$~=~\tento{-3} population, and coloured by the number of systems per mass of star forming material.}
\label{fig:clawejectazm3}
\end{figure*}

\begin{figure*}
\vspace{-0.4cm}
\centering
\begin{subfigure}{0.45\textwidth}
\centering
\includegraphics[width=0.88\columnwidth]{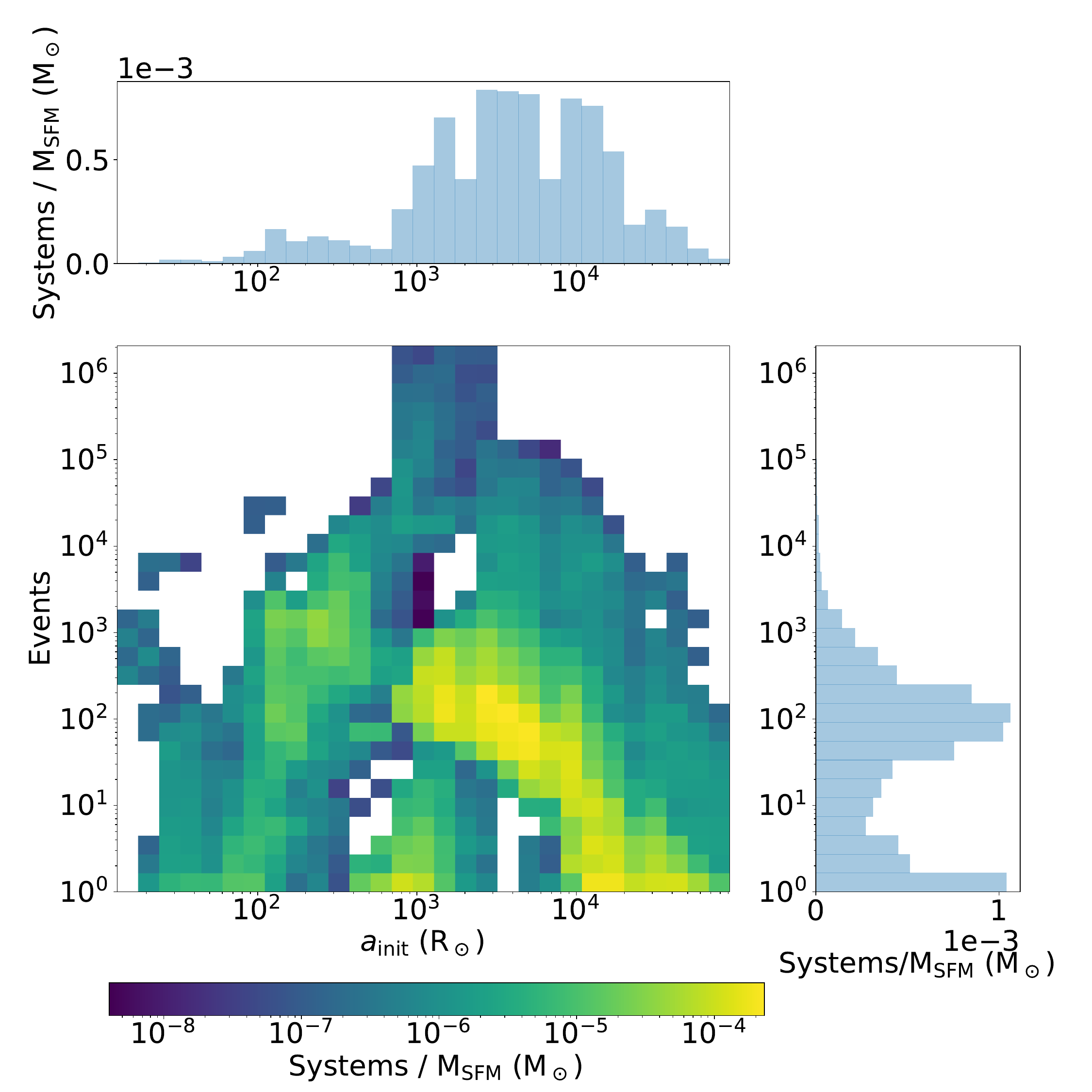}
\caption{}
\end{subfigure}%
\begin{subfigure}{0.45\textwidth}
\centering
\includegraphics[width=0.88\columnwidth]{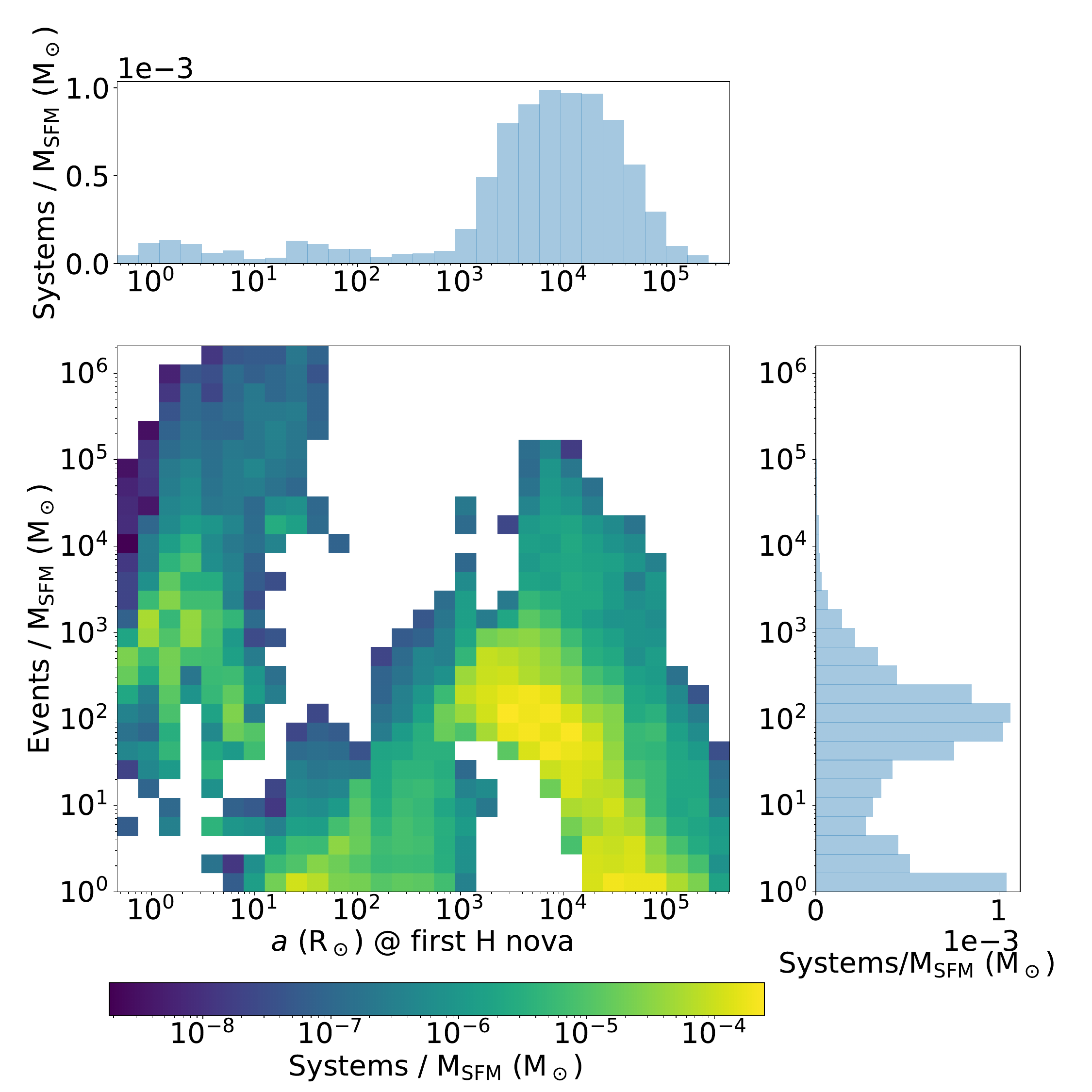}
\caption{}
\end{subfigure}

\begin{subfigure}{0.45\textwidth}
\centering
\includegraphics[width=0.88\columnwidth]{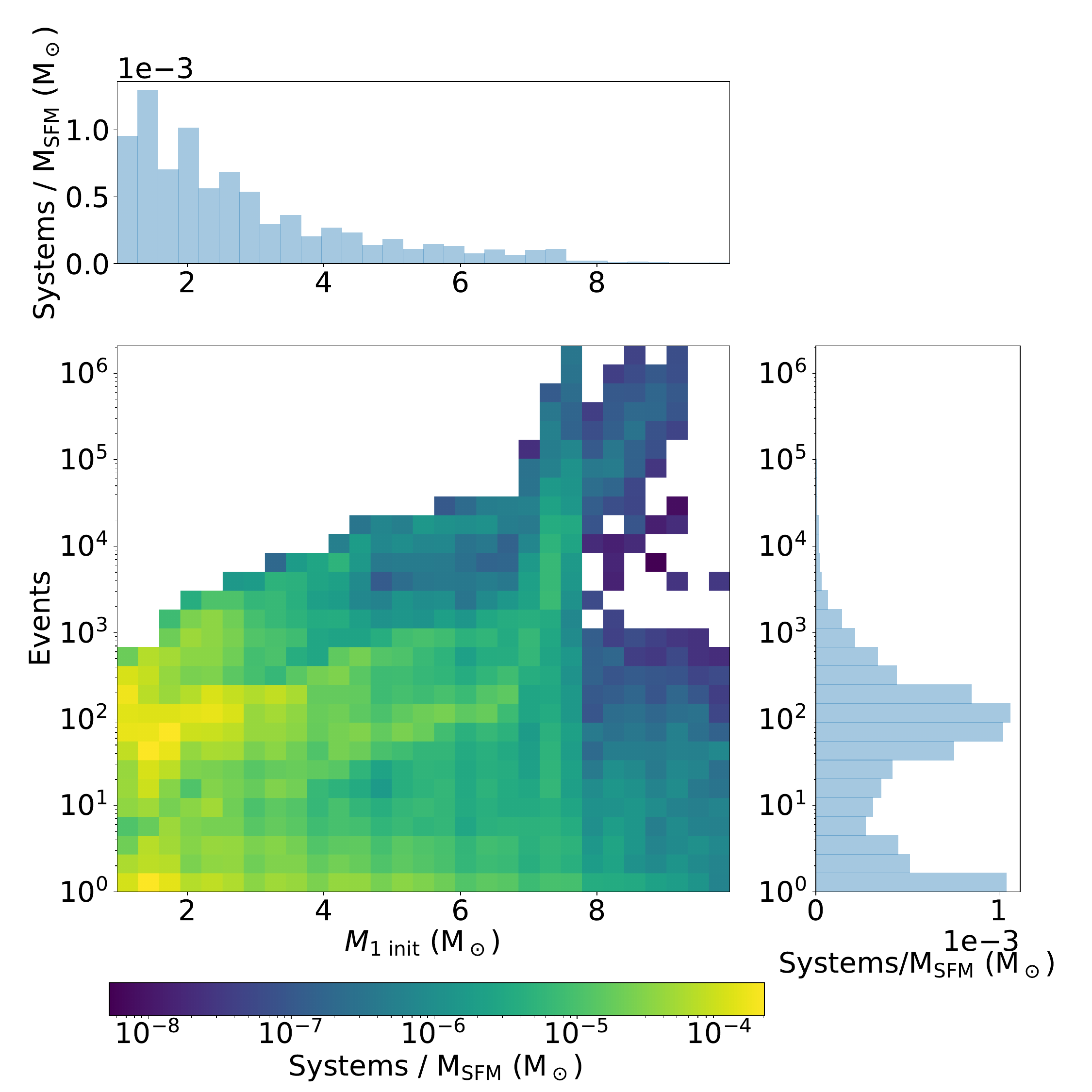}
\caption{}
\end{subfigure}%
\begin{subfigure}{0.45\textwidth}
\centering
\includegraphics[width=0.88\columnwidth]{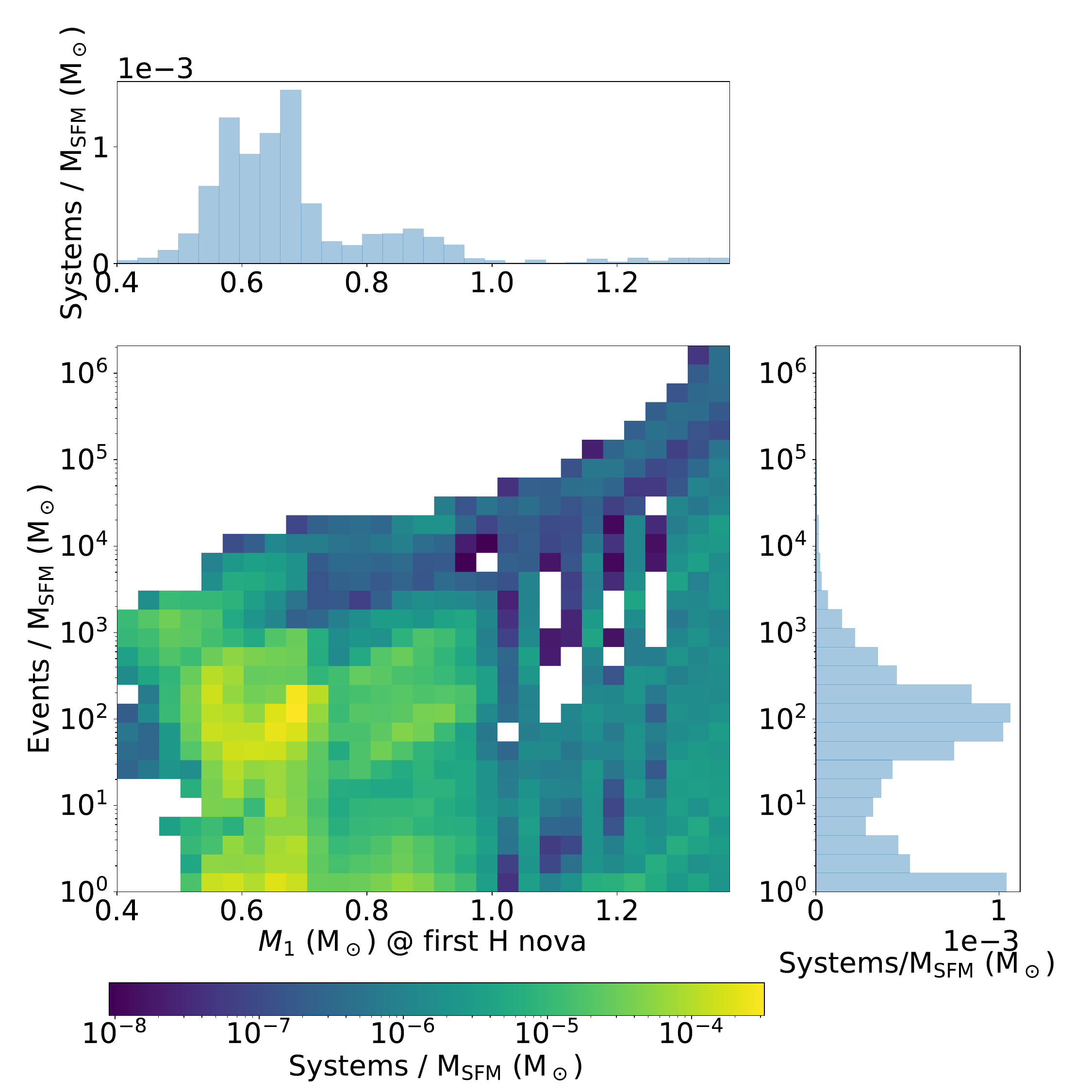}
\caption{}
\end{subfigure}

\begin{subfigure}{0.45\textwidth}
\centering
\includegraphics[width=0.88\columnwidth]{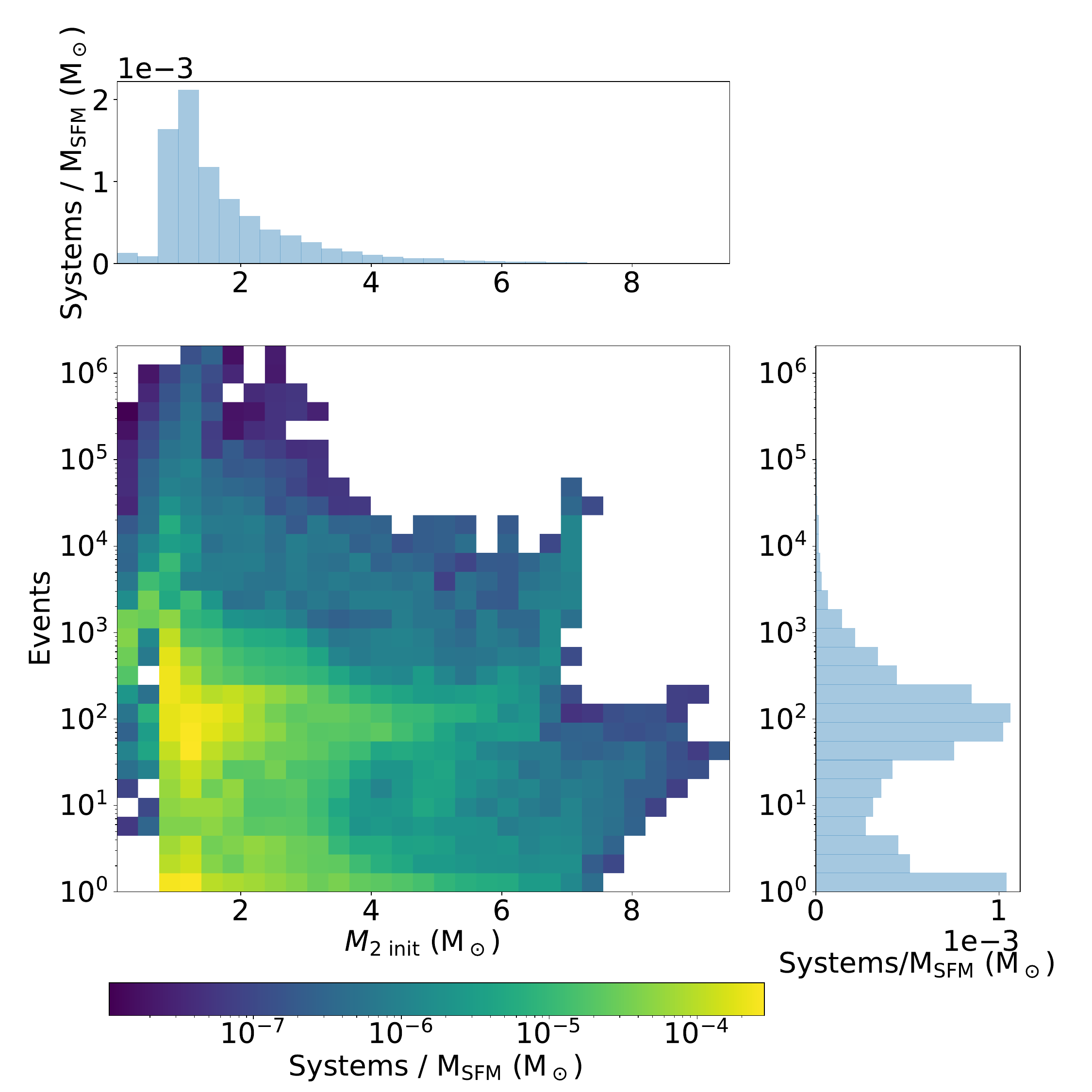}
\caption{}
\end{subfigure}%
\begin{subfigure}{0.45\textwidth}
\centering
\includegraphics[width=0.88\columnwidth]{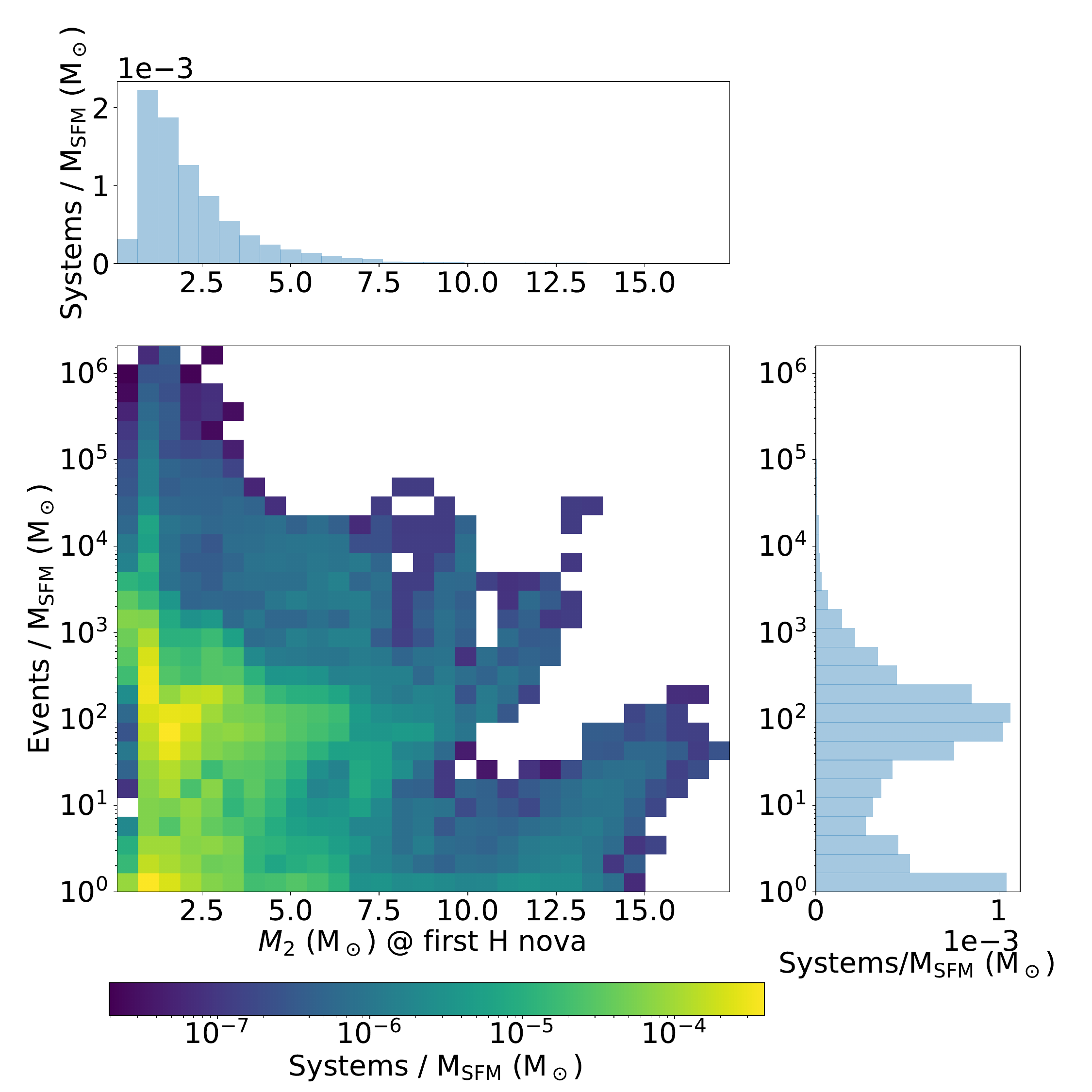}
\caption{}
\end{subfigure}

\caption{Total number of novae produced in each nova system plotted against the initial system properties (left) and the system properties at the time of the first H nova (right). Distributions shown are from the $Z$~=0.02 population, and coloured by the number of systems per mass of star forming material.}
\label{fig:claweventz0p02}
\end{figure*}

\begin{figure*}
\vspace{-0.4cm}
\centering
\begin{subfigure}{0.45\textwidth}
\centering
\includegraphics[width=0.88\columnwidth]{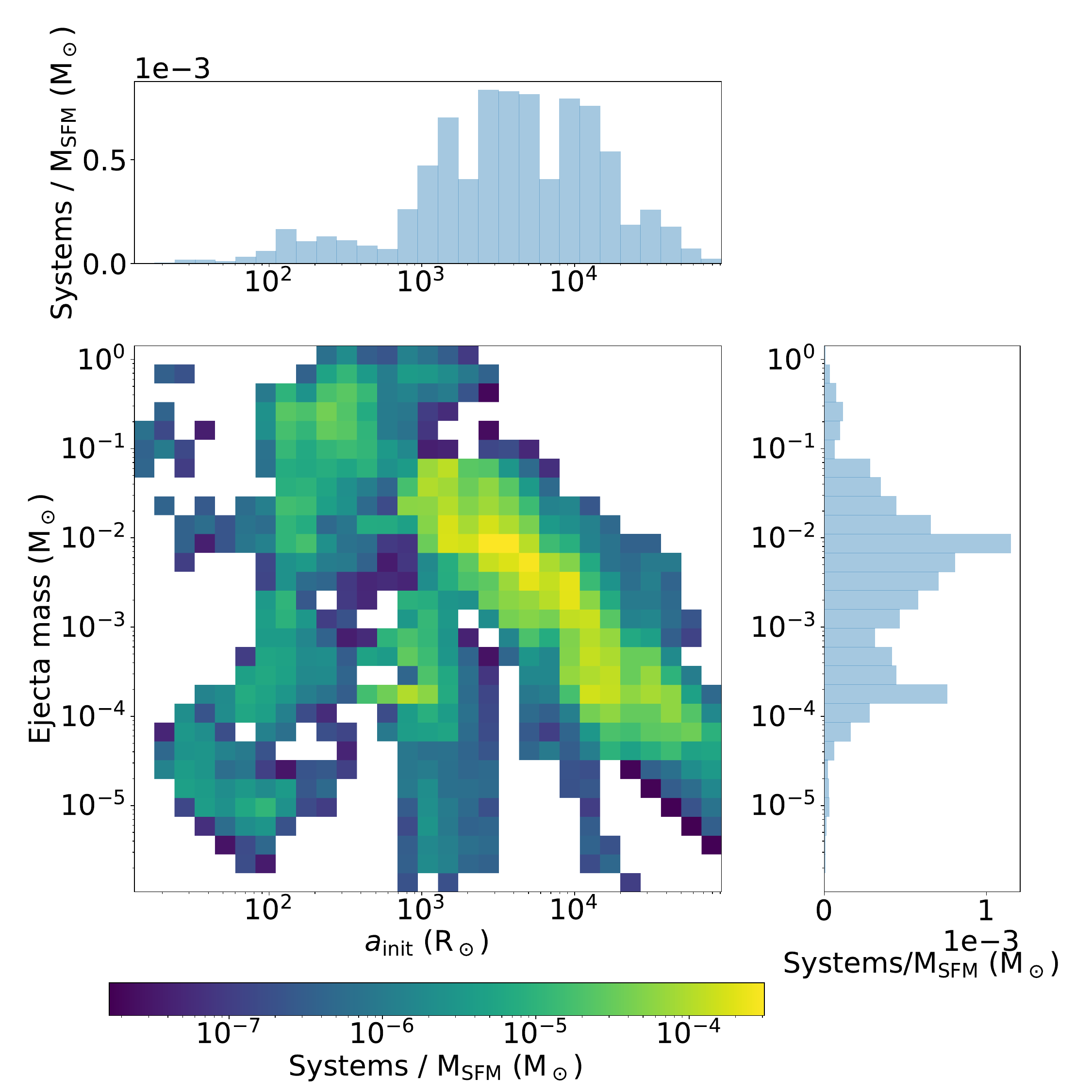}
\caption{}
\end{subfigure}%
\begin{subfigure}{0.45\textwidth}
\centering
\includegraphics[width=0.88\columnwidth]{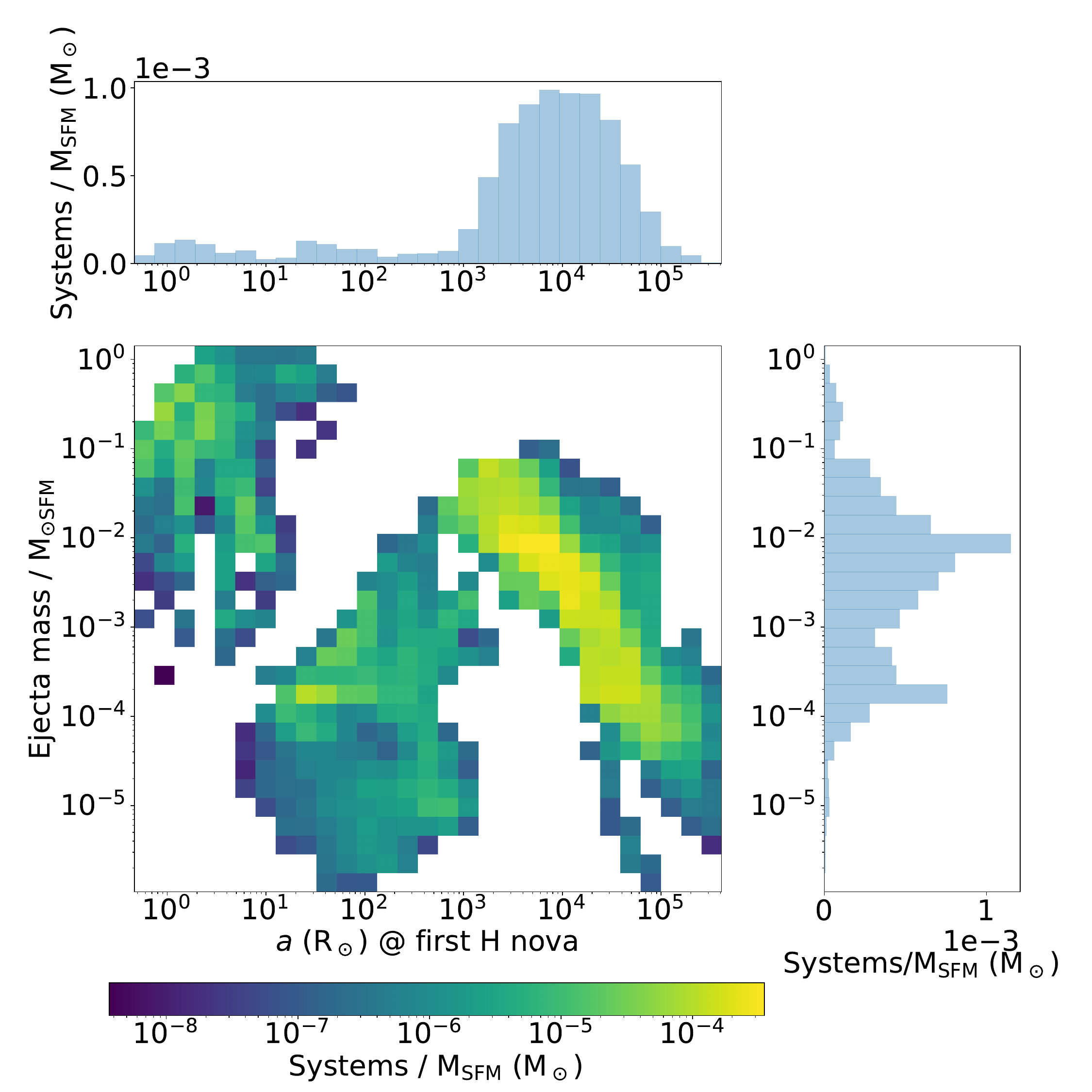}
\caption{}
\end{subfigure}

\begin{subfigure}{0.45\textwidth}
\centering
\includegraphics[width=0.88\columnwidth]{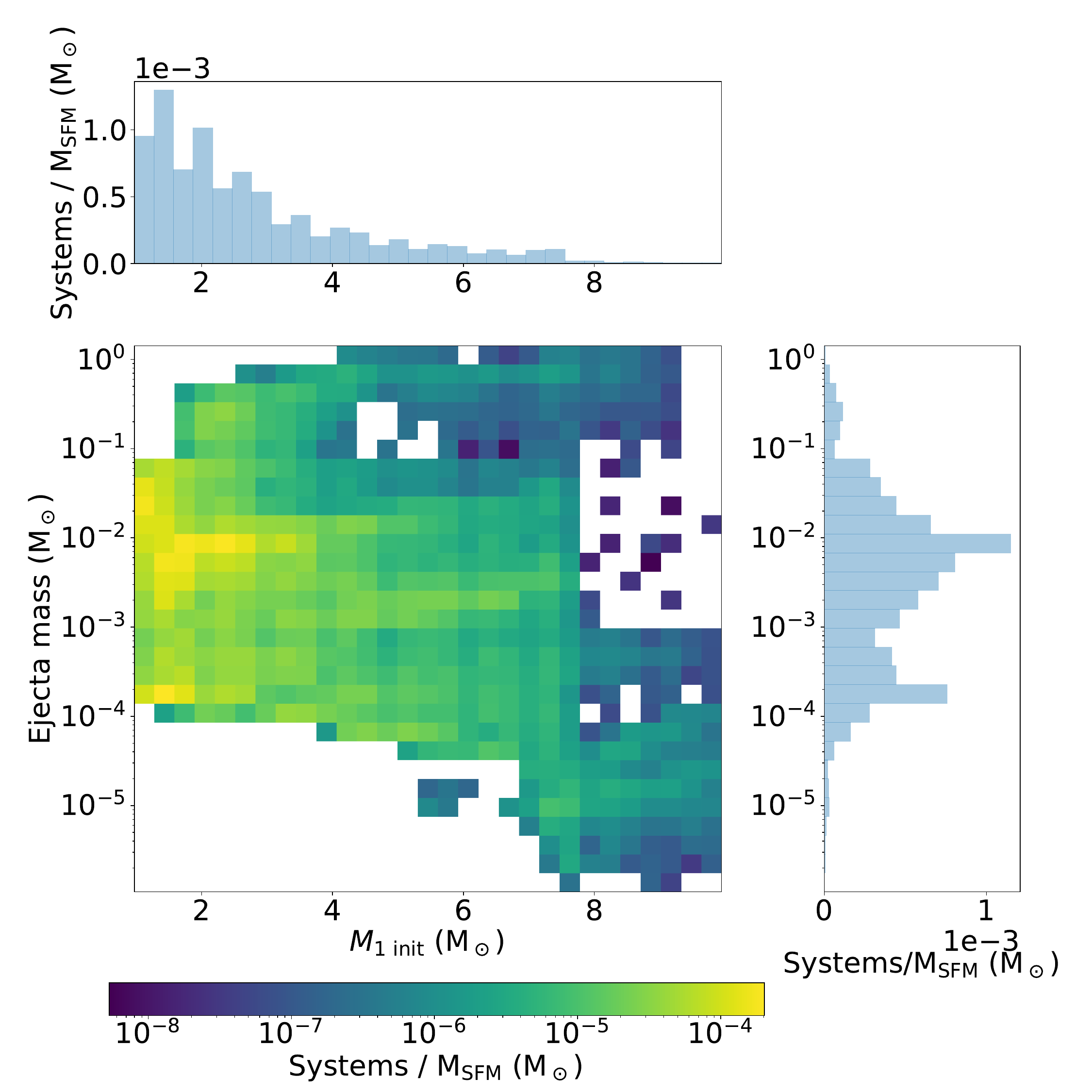}
\caption{}
\end{subfigure}%
\begin{subfigure}{0.45\textwidth}
\centering
\includegraphics[width=0.88\columnwidth]{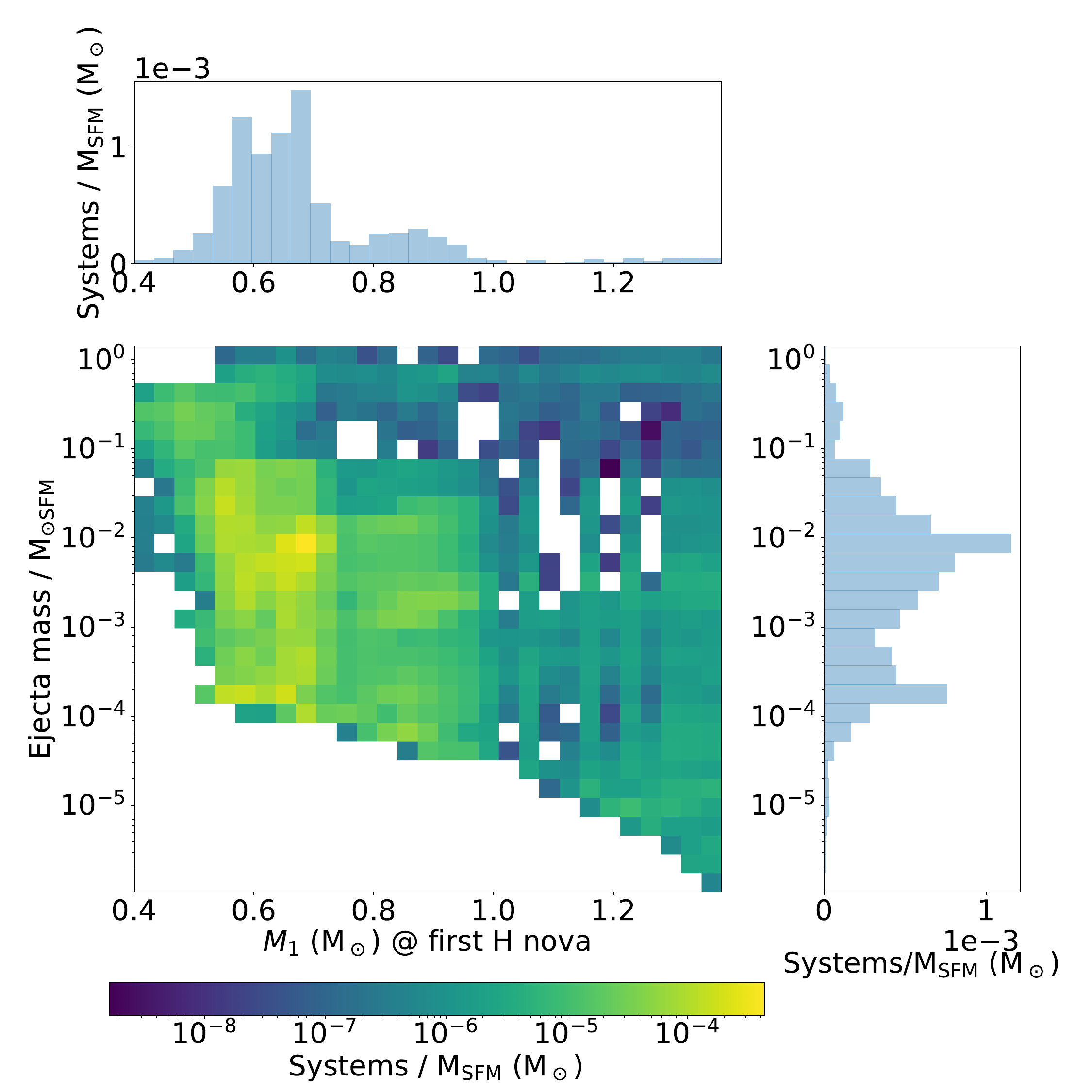}
\caption{}
\end{subfigure}

\begin{subfigure}{0.45\textwidth}
\centering
\includegraphics[width=0.88\columnwidth]{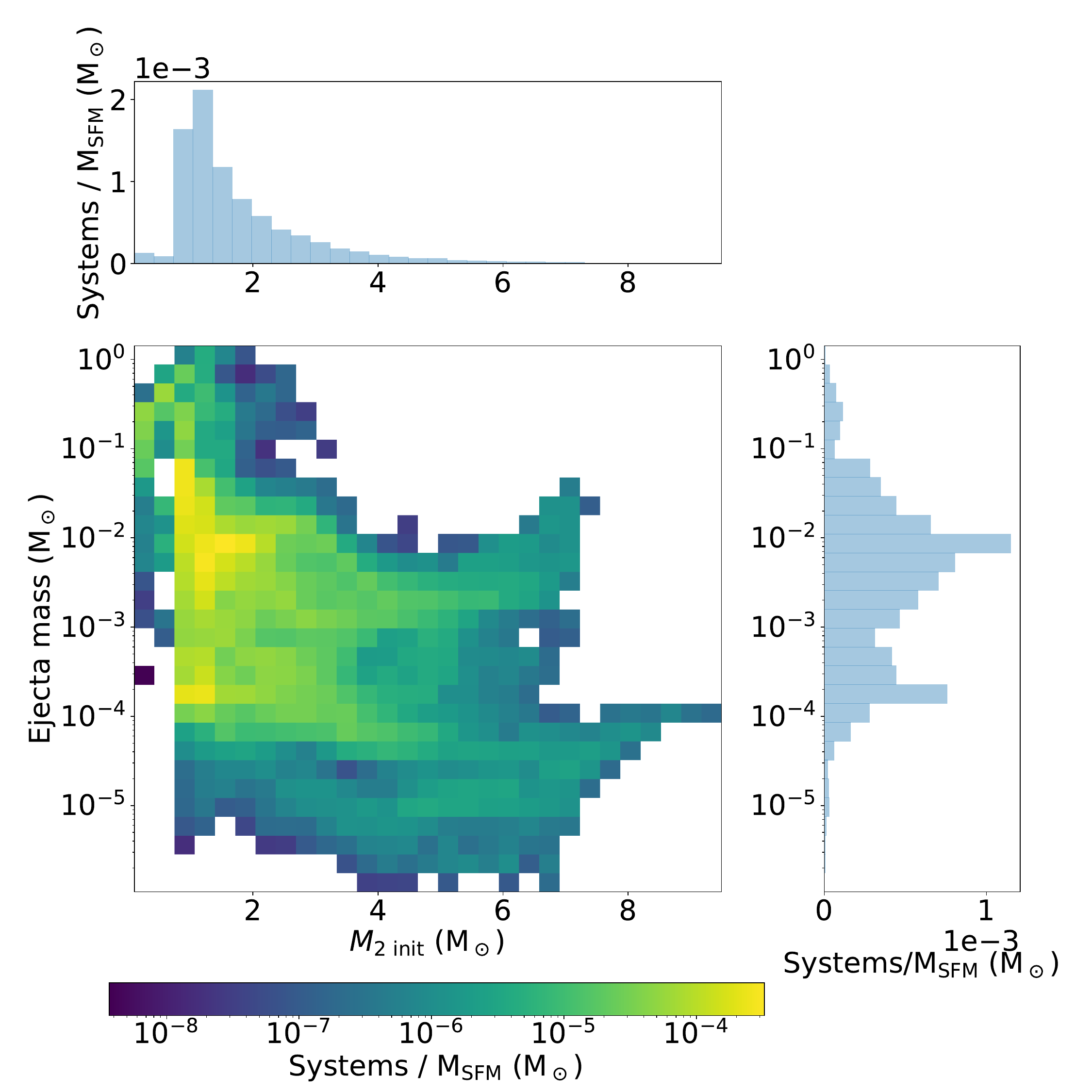}
\caption{}
\end{subfigure}%
\begin{subfigure}{0.45\textwidth}
\centering
\includegraphics[width=0.88\columnwidth]{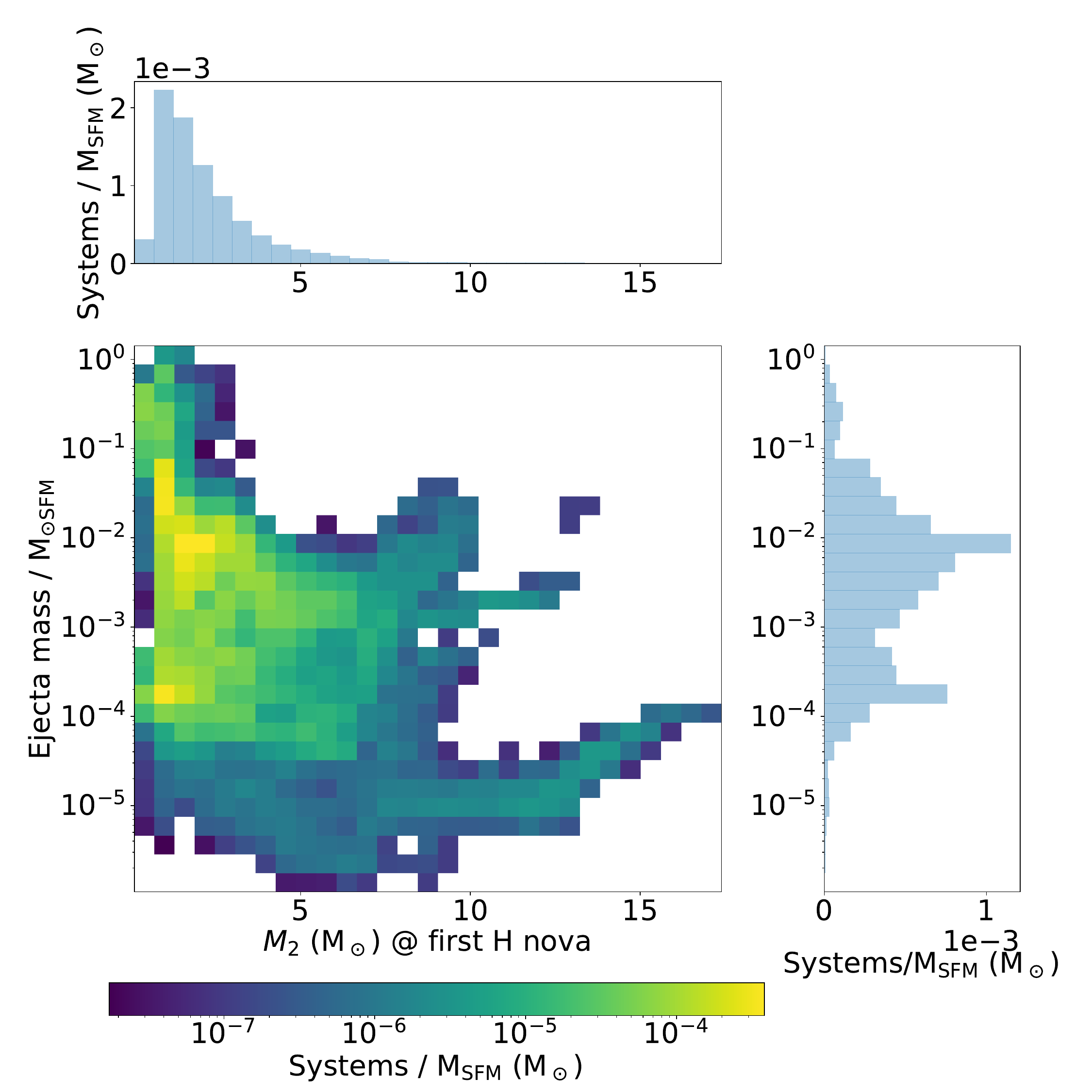}
\caption{}
\end{subfigure}

\caption{Total mass of nova ejecta produced in each nova system plotted against the initial system properties (left) and the system properties at the time of the first H nova (right). Distributions shown are from the $Z$~=~0.02 population, and coloured by the number of systems per mass of star forming material.}
\label{fig:clawejectaz0p02}
\end{figure*}

Figs. \ref{fig:claweventzm3}-\ref{fig:clawejectaz0p02} present another way of looking at the parameter space surrounding nova ejecta contributions. These figures show the number of novae produced (Figs. \ref{fig:claweventzm3} and \ref{fig:claweventz0p02}) and mass ejected (Figs. \ref{fig:clawejectazm3} and \ref{fig:clawejectaz0p02}) over each nova system's lifetime against that system's properties (primary mass, secondary mass, and orbital separation) at birth (left panels) and at the time of the first nova eruption (right panels). Figs. \ref{fig:claweventzm3}-\ref{fig:clawejectaz0p02} provide a lot of information about the relevance of different nova systems and their progenitors in these figures. We shall begin with an examination of the $Z$~=~\tento{-3} simulation (Figs. \ref{fig:claweventzm3}-\ref{fig:clawejectazm3}), with attention to which features can be linked to specific evolutionary channels, before discussing the changes to these distributions that occur when increasing the metallicity.

\begin{figure*}
\sidecaption
\centering
\includegraphics[width=12cm]{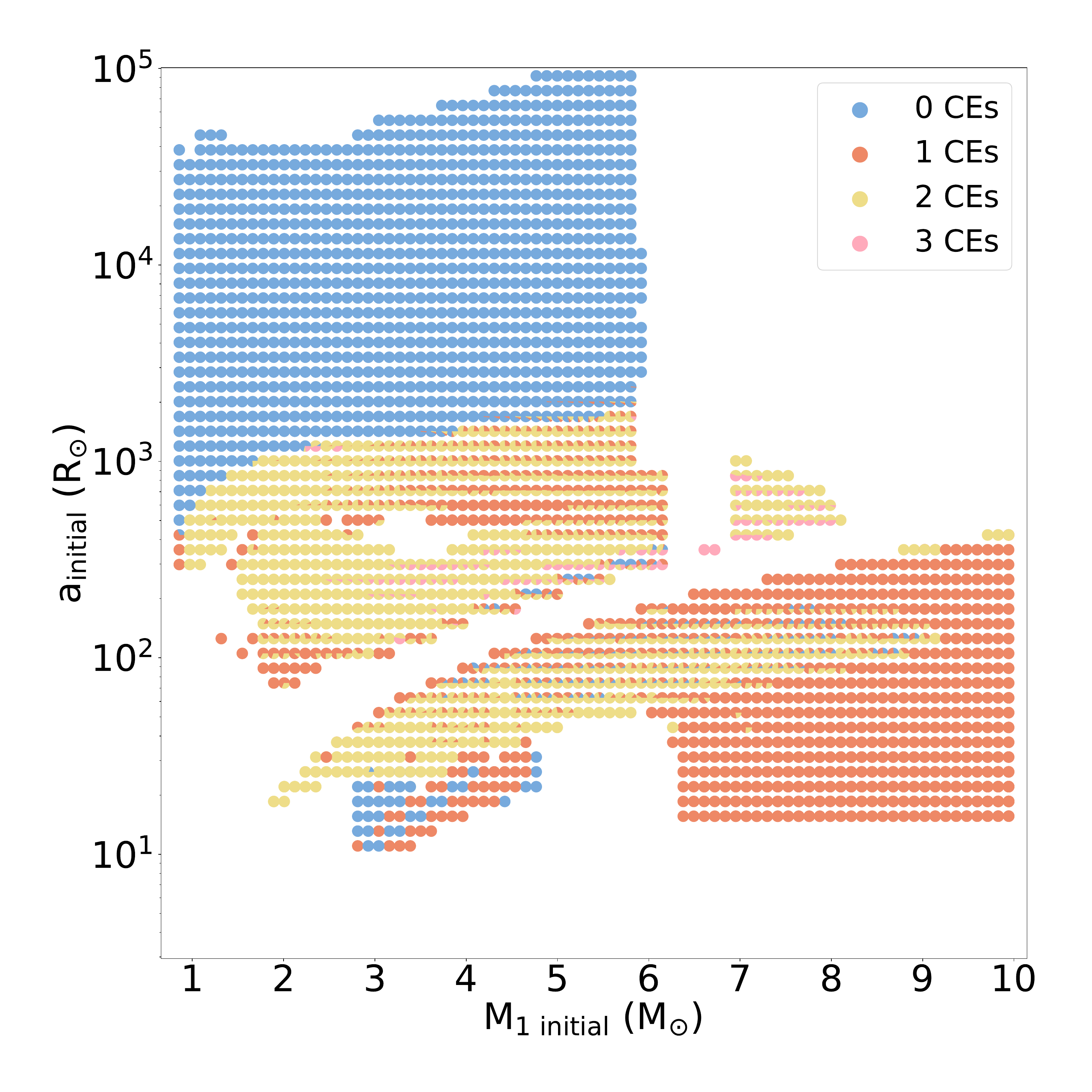} 
\caption{`Scatter-pie' plot for $Z$~=~\tento{-3} illustrating the distribution of number of CE events which occur throughout the lifetime of each nova system. Each point is a pie chart looking `into the page' to summarise the variation due to $M_{2\rm \ initial} \ (\rm M_{\odot})$.}
\label{fig:piescatzm3}
\end{figure*}

The behaviour at initial orbital separations (>\tento{3} R\solar) is relatively simple due to these systems being limited to relatively simple binary evolution channels. Fig. \ref{fig:piescatzm3} shows the common-envelope (CE) `scatter-pie' plot for $Z$~=~\tento{-3}, showing that systems above approximately \tento{3} R\solar\ do not undergo CE interactions. The majority of these nova systems, and all of those at the longest separations, are powered by wind or wind-RLOF \citep{abate2013wind} accreting WDs that have not experienced dramatic orbital shrinkage due to CE evolution or stable pre-nova-phase mass transfer via RLOF. This is reflected in the similarity in this regime between panels a and b in both the nova event and nova ejecta distributions (Figs. \ref{fig:claweventzm3} and \ref{fig:clawejectazm3}). As the orbital separation increases, the number of novae and the amount of ejecta both decline as the mass transfer becomes less conservative, resulting in a less efficient build-up to the critical ignition mass.

The orbital separation-space becomes far more complex at shorter separations, where CE physics plays a significant role in the evolution of many nova systems, resulting in diverse evolutionary channels. In terms of the number of nova events, the most extreme systems (producing over \timestento{5}{4} novae each) are readily identifiable in panels a and b of Fig. \ref{fig:claweventzm3}. These systems exhibit significant orbital shrinkage by the time of the first nova, from just under \tento{3} R\solar\ to between 1 and 10 R\solar. These extreme systems are rare, involving a high-mass WD (panel d) accreting from a relatively low-mass donor star (<3 M\solar, panels e-f) at a low accretion rate, resulting in low accretion efficiencies that allow the WD to undergo many nova eruptions without reaching \Mchand.

When considering the mass of ejecta produced (Fig. \ref{fig:clawejectazm3}), rather than the number of novae, the most extreme nova systems look quite different, although once again CE evolution is important. The main block of the extremely high ($\gtrsim$ 0.1 M\solar) ejecta-mass systems exists between initial separations of \tento{2} and \tento{3} R\solar, with a smaller, distinct block existing below \tento{2} R\solar. By the time of the first H nova, these two blocks have merged to form a single block with orbital separations less than \tento{2} R\solar, with the majority less than \tento{1} R\solar. The initial primary masses and WD masses for these systems are remarkably diverse; almost the entire initial mass and \Mwd\ parameter space is represented, reflecting the ability of even massive WDs to process significant amounts of material under the right conditions (specifically, when accretion efficiencies are very close to or below zero). The secondary masses are more constrained, restricted to <4 M\solar\ stars both in initial mass and when considering the mass at the time of the first H nova.

Panels e-f of Fig. \ref{fig:claweventzm3} and \ref{fig:clawejectazm3} highlight the existence of a population of Algol nova systems, defined here as systems with donor stars of mass greater than 10 M\solar\ at the time of the first nova. We resolve approximately 1300 examples of these systems in our $Z$~=~\tento{-3} simulation, with donor stars as massive as 18 M\solar. The bulk of these systems occur at orbital separations between \tento{2} and \tento{3} R\solar\ at the time of the first nova, with two small (<10 systems resolved) satellite populations around \timestento{2}{3} and between 20-30 R\solar. Significant pre-nova widening is evident in most of these systems, with almost all beginning with orbital separations between 10 and $\lesssim$ \tento{2} R\solar. The initial primary masses for these systems range between 6 and 10 M\solar, and the WD in these systems is always greater than 0.9 M\solar\ at the time of the first nova. The number of novae these systems produce can be as high as \tento{3} and as low as 1, but the range in ejecta masses is far smaller, with these systems producing between \tento{-4} and \tento{-3} M\solar\ each.

%%%%CLAWevents Zm3

%%%%%%%CLAwejecta%%%%%%%zm3

\begin{figure*}
\sidecaption
\centering
\includegraphics[width=12cm]{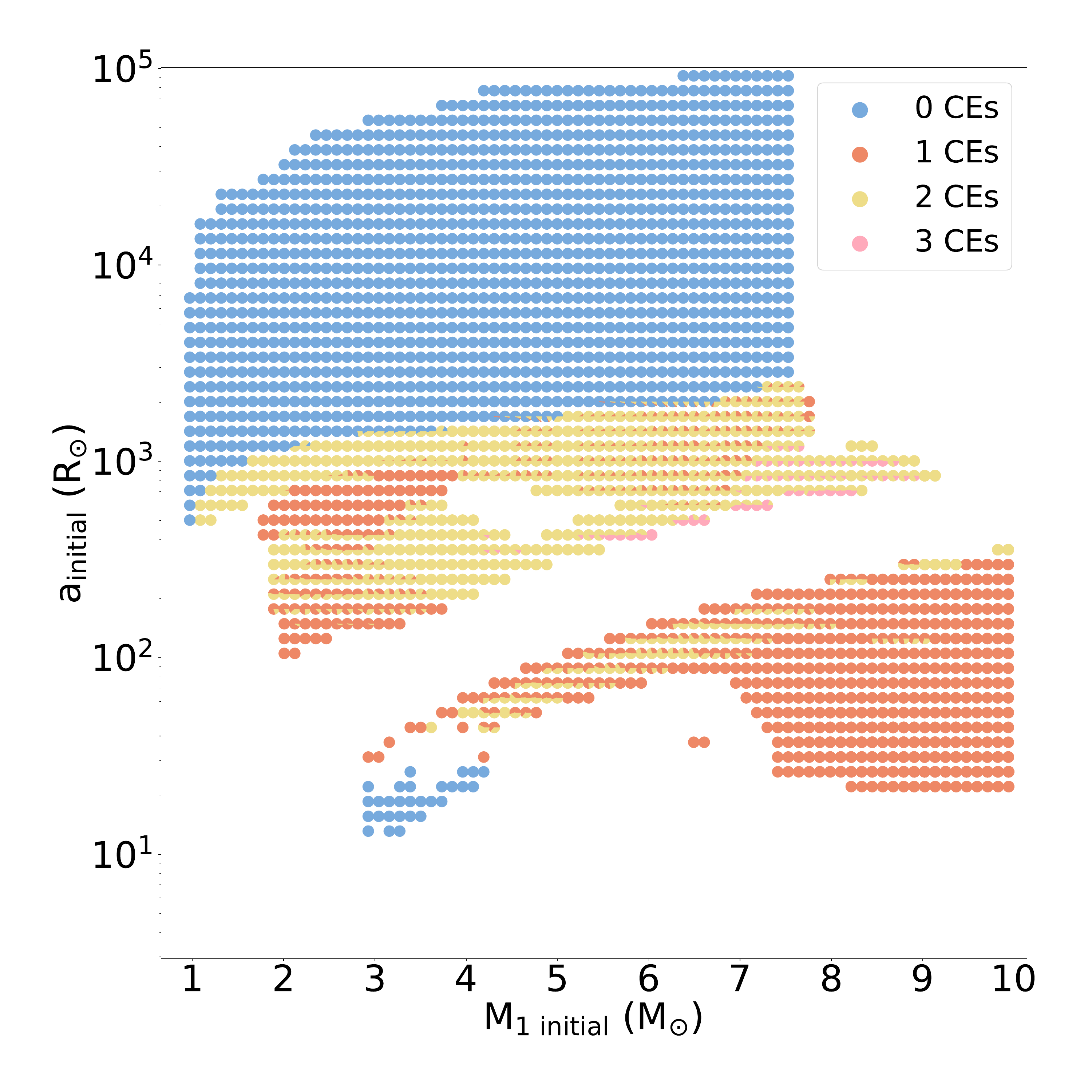} 
\caption{`Scatter-pie' plot for $Z$~=~0.02., equivalent to Fig. \ref{fig:piescatzm3}.}
\label{fig:piescatz0p02}
\end{figure*}

Figs. \ref{fig:claweventz0p02} and \ref{fig:clawejectaz0p02} present the $Z$~=~0.02 distributions for the event counts and ejecta masses per system respectively, with the `scatter-pie' shown in Fig. \ref{fig:piescatz0p02} summarising the CE space. The morphological differences between Figs. \ref{fig:piescatzm3} and \ref{fig:piescatz0p02} are discussed in terms of the initial primary mass and orbital separation with metallicity in \cite{kemp2022}, and a complete analysis of the variations in CE properties with metallicity in the context of novae is beyond the scope of this work. However, it was found in \cite{kemp2022} that variation in CE behaviour is unlikely to be the primary driver of metallicity-dependent variation in nova rates. Inspection of Figs. \ref{fig:piescatzm3} and \ref{fig:piescatz0p02} supports this finding, as it appears that there are few significant variations in the distributions of CEs, at least in terms of the number that occur prior to the onset of the nova phase. Most importantly, the initial orbital separation above which no CEs occur is almost identical.

In agreement with this similarity around separations~$\gtrsim$~\tento{3} R\solar, we find that the morphology in Figs. \ref{fig:claweventz0p02} and \ref{fig:clawejectaz0p02} in the long initial orbital separation regime is similar in location to that at $Z$~=~\tento{-3} (Figs. \ref{fig:claweventzm3} and \ref{fig:clawejectazm3}). The primary difference is that there is more scatter in this regime at solar metallicity, with more novae able to be produced in systems with initial orbital separations between \tento{-4} and \tento{-5} R\solar\ in particular. 
% This increased scatter may be due to the larger stellar radii associated with higher metallicity stars promoting more diverse binary evolution.

The extreme nova and nova ejecta producing systems for the two metallicity cases behave similarly. The same basic structures can be found in the same parts of the parameter space, although at $Z$~=~0.02 the space is better `filled out' (Figs. \ref{fig:claweventz0p02} and \ref{fig:clawejectaz0p02}). This is probably due to larger radii permitting a wider range of systems to enter the nova phase, promoting more diverse evolutionary channels.

The previously described Algol nova systems are still present at solar metallicity (approximately 800 examples resolved at $Z$~=~0.02), although these systems now typically produce an order of magnitude fewer novae and nova ejecta. This is in spite of the fact that the parameter-space these systems occupy is very similar to that at $Z$~=~\tento{-3}. 

This conundrum is resolved by examination of the evolution of these systems during the nova phase itself. The nova phase of the Algol nova systems typically occurs immediately before or during the start of core He burning in the donor and ends with a CE event as the donor star grows to increasingly larger radii. The productivity of an Algol nova system therefore primarily depends on how much mass is transferred prior to this CE event, meaning that it is a balance of how quickly mass can be lost from the donor and how quickly that star expands. For example, the longer orbital-separation systems (with separations close to \tento{3} R\solar) tend to have a longer nova phase, resulting in more novae per system and more ejecta. At higher metallicity, the donor stars expand more rapidly during core helium burning due to their higher opacity while also losing less mass due to their lower core masses and luminosities. The result is less mass transferred prior to the CE terminating the nova phase and, therefore, fewer novae and less ejecta, despite the initial parameter-space being similar.

%%%CLAW events z0p02

%%%%%%%CLAwejecta%%%%%%%z0p02

%%%CE FIGURES
%%%ZM3
%%%Z0p02

\section{Results: Elemental and isotopic abundances}
\label{sec:chap5res_abund}

In this section, we will address the contributions of novae to the nucleosynthesis of elements other than Li\footnote{Li is dealt with in \cite{kemp2022li}.} in the MW, as well as which nova sites -- distinguished in terms of the WD mass -- contribute significantly to which isotopes. Novae occurring on different WD masses produce different nucleosynthesis and have different delay-time distributions and overall importance, all important factors that are folded into our simulations.

% The varying contributions of different WD masses to different isotopes accounts for the different delay-time behaviour of nova ejecta contributions on different WD masses due to stellar and binary stellar evolution, as well as differences in reaction rates due to different thermonuclear conditions on the WD, inherited from the models producing the underlying yield profiles \citep{jose1998,starrfield2009,starrfield2020,jose2020}.

\begin{figure*}
\centering
\includegraphics[width=1.9\columnwidth]{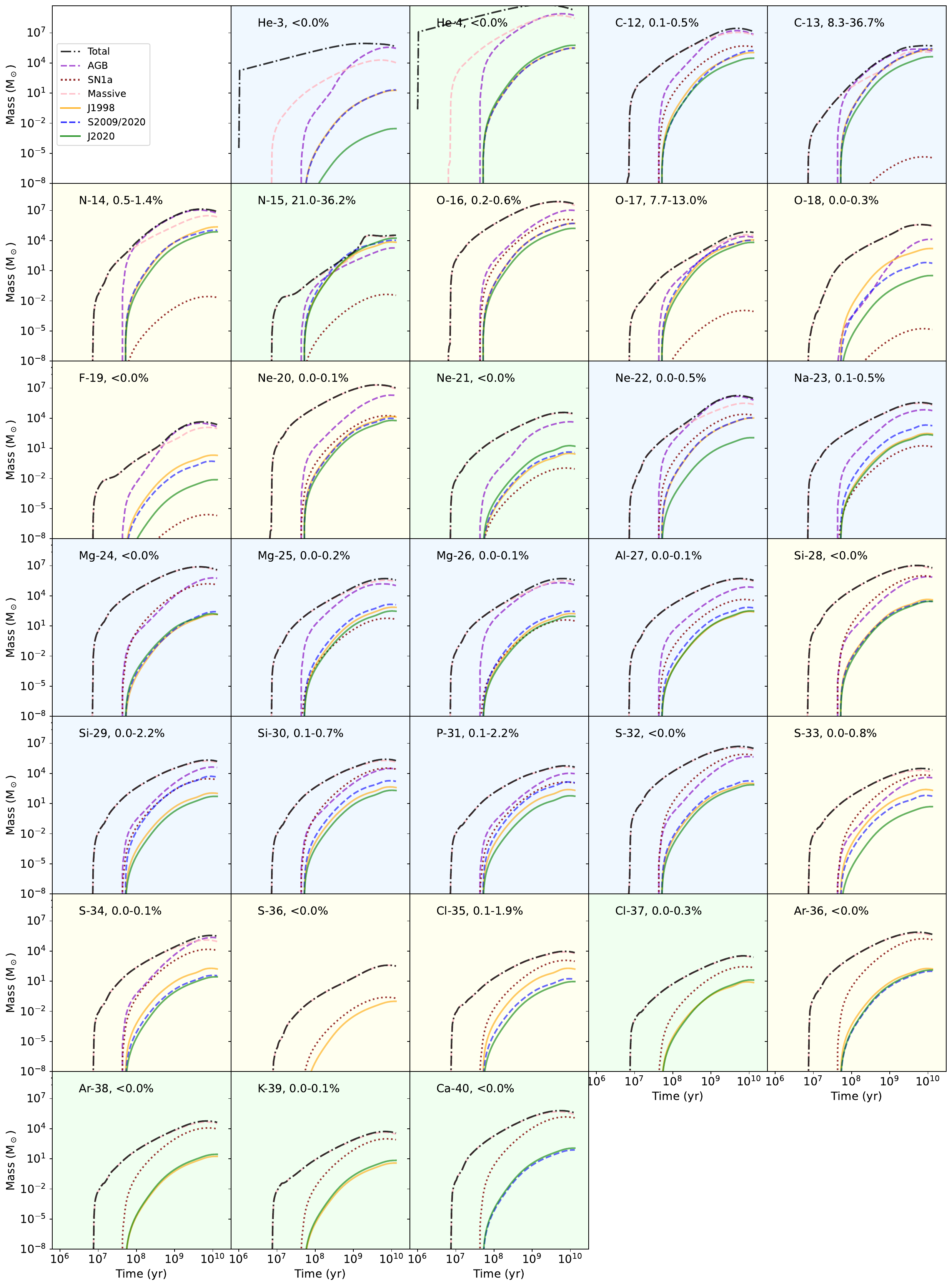}
\caption[Contributions to each isotope from stellar sources.]{Contributions to each isotope from stellar sources. The background color of each panel highlights which nova yield profile produces the most of the relevant isotope by [Fe/H]~=~0 (yellow for J1998, blue for S2009/2020, and green for J2020). Besides \iso{13}C, \iso{15}N, and \iso{17}O, nova contributions for most isotopes are small (<2\%) compared to other stellar sources.}
\label{fig:iso_masssource}
\end{figure*}

Fig. \ref{fig:iso_masssource} summarises the contributions from novae, AGBs, type Ia supernovae, and massive stellar sources for a range of isotopes, highlighting which nova model produced the most of each isotope at [Fe/H]=0. Also shown in each panel is the fraction of each isotope (at [Fe/H]~=~0) produced by novae. For example, according to our models 8.3-36.7\% of Galactic \iso{13}C is produced by novae, where the range reflects the variation between different nova yield profiles. Note that not all isotopes have reported contributions from every yield profile: \iso{36}S is only reported in the S2009/2020 profile, \iso{37}Cl, \iso{38}Ar, and \iso{39}K are not reported in the S2009/2020 profile, and \iso{40}Ca is not reported in the J1998 yield profile.

\begin{figure}
\centering
\includegraphics[width=1\columnwidth]{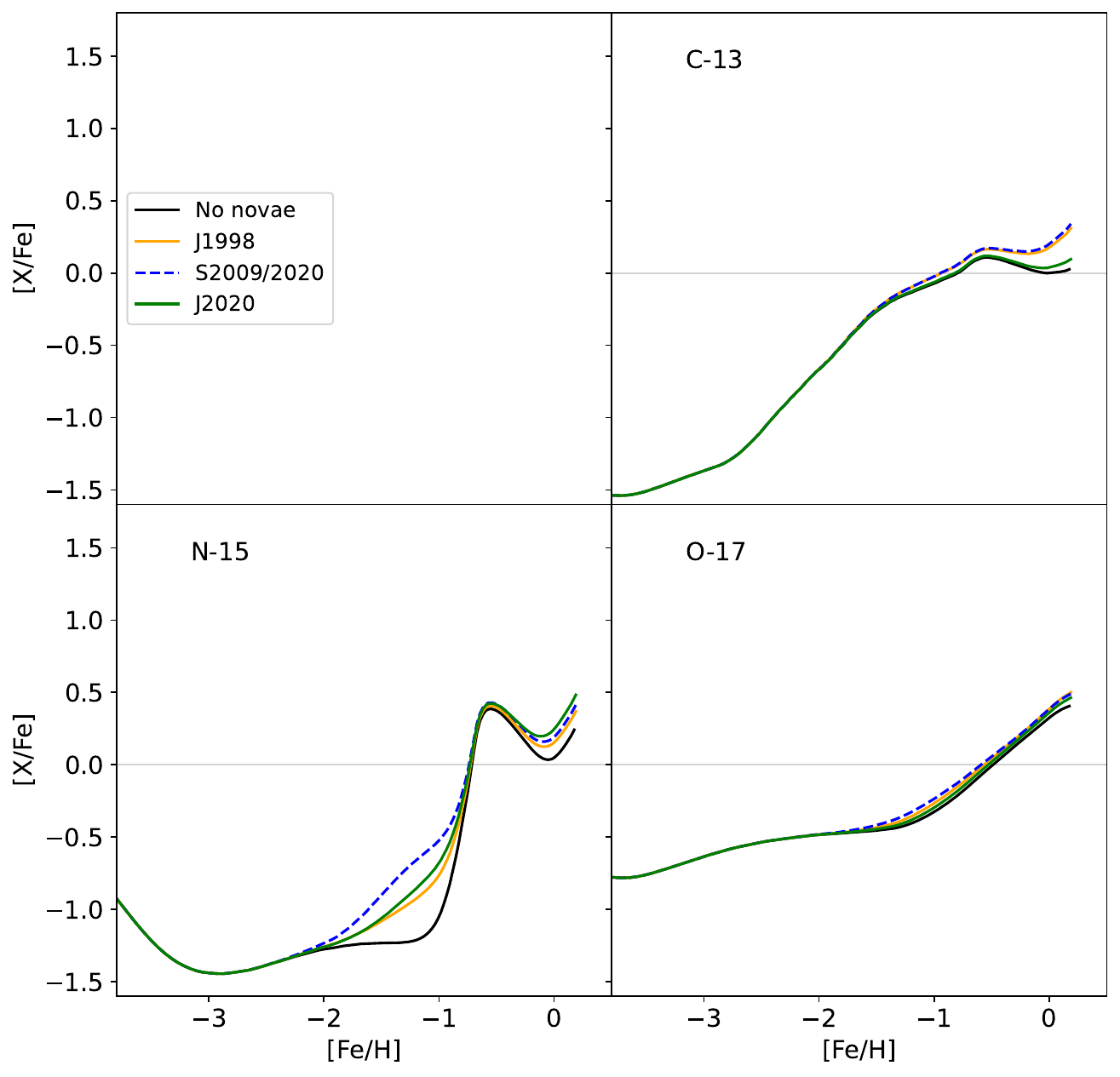}
\caption[Abundances relative to iron for the isotopes where the inclusion of novae in the model makes a difference to the overall Milky Way abundance.]{[X/Fe] abundances for the isotopes where the inclusion of novae in the model makes a difference to the overall Milky Way abundance. The inclusion of novae has negligible influence on other isotopes (excluding Li, see \citealt{kemp2022li})}
\label{fig:iso_sqr_selected}
\end{figure}

\begin{figure}
\centering
\includegraphics[width=1\columnwidth]{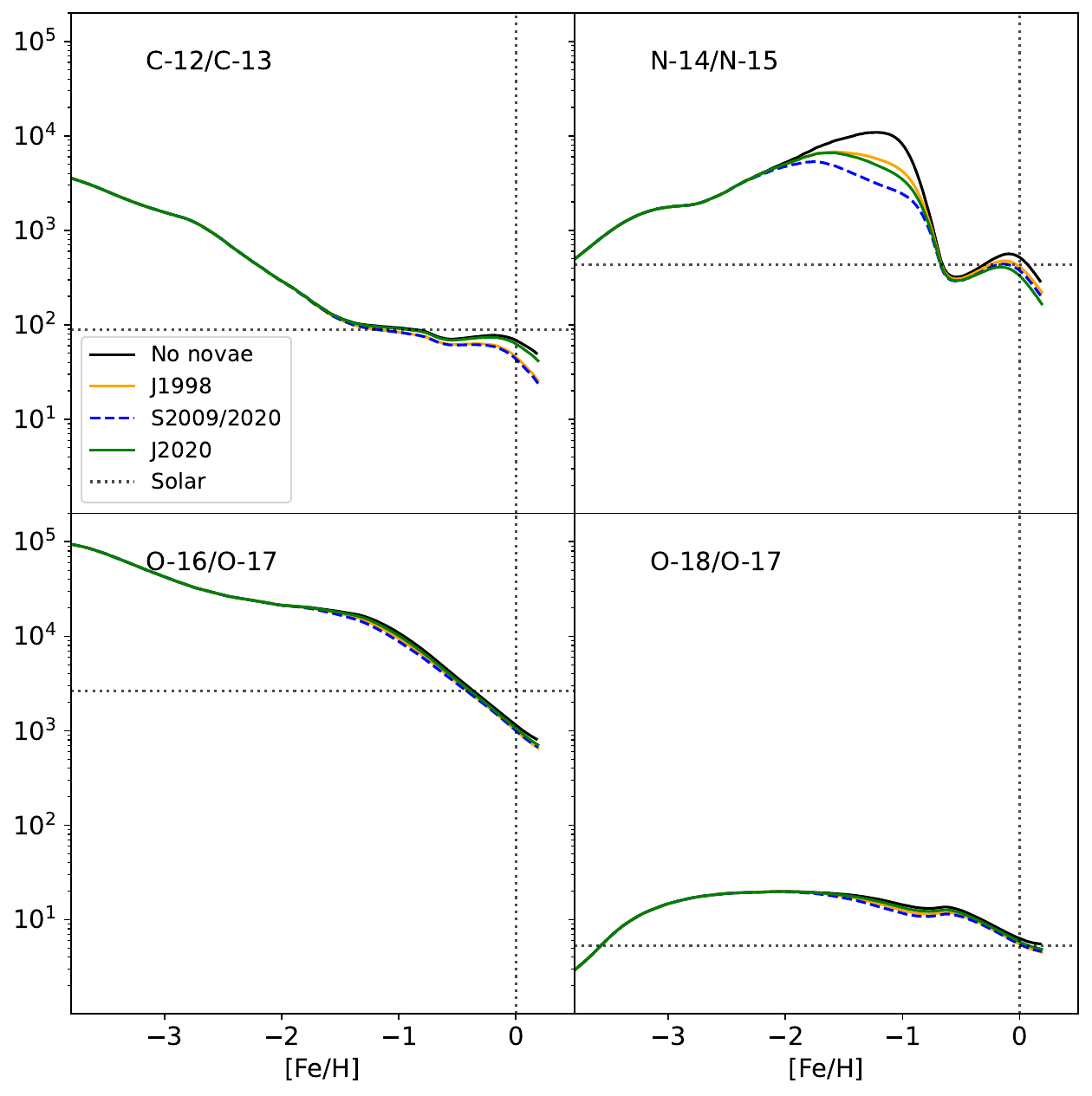}
\caption{Isotopic ratios where the inclusion of novae in the model makes a non-negligible difference to the overall Milky Way abundance.}
\label{fig:iso_rat_selected}
\end{figure}

Nova contributions are small (<2\%) for most isotopes relative to contributions from other stellar sources. Besides \iso{7}Li \citep{kemp2022li}, notable exceptions include \iso{13}C, \iso{15}N, and \iso{17}O, although even for these isotopes nova contributions are (at most) around 35\%. Fig. \ref{fig:iso_sqr_selected} shows the evolution of [\iso{13}C/Fe], [\iso{15}N/Fe], and [\iso{17}O/Fe] as a function of [Fe/H], while Fig. \ref{fig:iso_rat_selected} shows the evolution of \iso{12}C/\iso{13}C, \iso{14}N/\iso{15}N, \iso{16}O/\iso{17}O, and \iso{18}O/\iso{17}O ratios.

Figures \ref{fig:iso_sqr_selected} and \ref{fig:iso_rat_selected} demonstrate that nova contributions to \iso{13}C and \iso{15}N produce non-negligible variations in both the logarithmic abundance ratios relative to iron and their ratios relative to their dominant elemental isotopes. At [Fe/H]~=~0, the predicted Galactic [\iso{13}C/Fe] and [\iso{15}N/Fe] ratios show minor discrepancies (at most 0.2 dex).
Nova contributions to \iso{17}O cause only barely discernable variation in these figures.

Of these isotopes, \iso{15}N shows the most interesting behaviour, with [\iso{15}N/Fe] showing significant deviations between [Fe/H]~=~-2 and -1, up to 0.7 dex. 
% From Fig. \ref{fig:mwcalib1}, our model galaxy reaches [Fe/H] = -1 after just 1 Gyr of evolution. Thus, in this metallicity regime it is almost exclusively novae contributions from relatively massive WDs that are responsible for this behaviour, supporting.
From a close examination of the \iso{15}N panel in Fig. \ref{fig:iso_masssource}, it can be seen that beyond a few hundred Myr nova contributions to \iso{15}N overtake AGB contributions and are comparable to massive stellar contributions by 1 Gyr. Beyond [Fe/H]~=~-0.8 (a galaxy age of almost 2 Gyr), production of \iso{15}N from massive stars resurges, causing the large spike in Fig. \ref{fig:iso_sqr_selected}. The effects can be seen equally clearly in the ratios of \iso{14}N/\iso{15}N. It is worth mentioning that the evolution of our isotopic ratios is similar in shape to those presented in Figure 31 of \cite{kobayashi2020origin}, although their \iso{14}N/\iso{15}N ratio peaks somewhat earlier. Unfortunately, there are no observational data on \iso{14}N/\iso{15}N abundances relevant to this metallicity regime.

\begin{figure*}
\centering
\includegraphics[width=1.9\columnwidth]{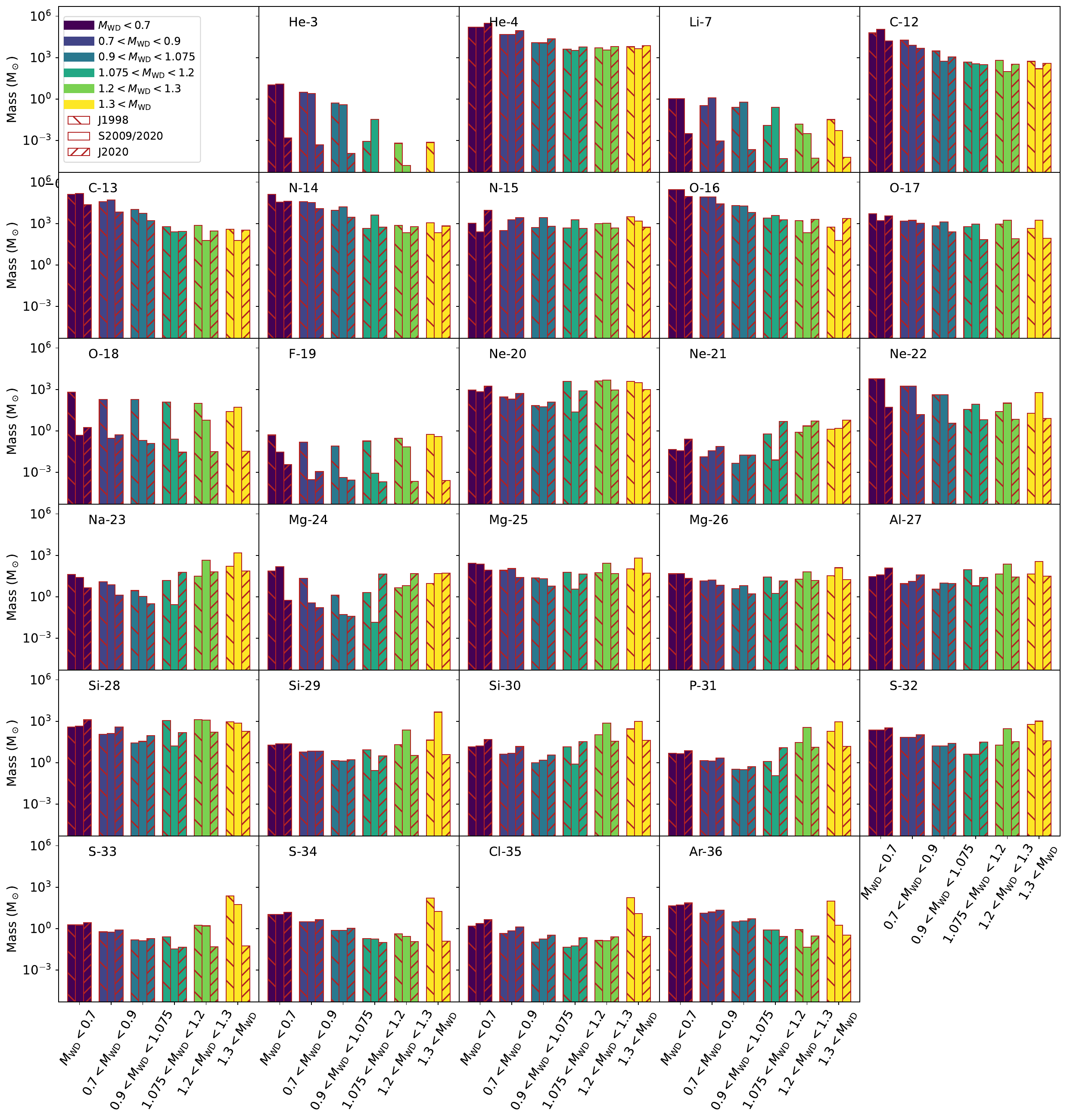}
\caption{Bar plot summarising the contributions to each isotope by [Fe/H]~=~0 for each nova yield profile, distinguishing between nova progenitor WD masses. Figs. \ref{fig:iso_mwdhistj1998}, \ref{fig:iso_mwdhists20092020}, and \ref{fig:iso_mwdhistj2020} present the full evolution history for each isotope, also distinguishing progenitor masses.}
\label{fig:iso_mwdbarsum}
\end{figure*}

Previously, we asserted that for different isotopes, different parts of the nova \Mwd\ parameter space would be important, primarily depending on the interplay between \Mwd-dependent reaction rates and the \Mwd-dependent ejecta masses, which govern the ejecta delay-time distributions for each of the mass brackets. Figs. \ref{fig:iso_mwdhistj1998}, \ref{fig:iso_mwdhists20092020}, and \ref{fig:iso_mwdhistj2020} show the  isotopic contributions from novae from different parts of the \Mwd\ parameter space -- using the J1998, S2009/2020, and J2020 yield profiles respectively -- as a function of Galaxy age. Fig. \ref{fig:iso_mwdbarsum} presents a slice from each of these figures at [Fe/H]~=~0 in side-by-side bar charts, allowing easier comparison between the three nova yield profiles.

There is a wealth of information in these figures, showing which parts of the nova parameter-space are most important for each isotope at different times during the Galaxy's evolution. Before proceeding, the reader should be reminded that the following discussion is strictly concerned with comparing different parts of the nova \Mwd\ parameter space. Galactic production of most of the isotopes discussed here is dominated by non-nova stellar sources at all times during the evolution of the Milky Way (Fig. \ref{fig:iso_masssource}).

We will begin by examining some of the main trends and underlying physics surrounding the J1998 case (Fig. \ref{fig:iso_mwdhistj1998}), and will proceed with our discussion from there. The first thing to note is that there is a systematic difference in the delay-times for different WD masses. Contributions from more massive progenitors kick in earlier, as these WDs form first. For this reason, in some cases the relative importance of different \Mwd-regimes shifts as a function of Galaxy age. For example, if we consider \iso{15}N, the lowest mass WDs (<0.7 M\solar) contribute significantly by 10 Gyr, but lag behind the contributions of more massive WDs at times earlier than 1 Gyr. 

Nova productivity of the low mass-number isotopes from \iso{3}He to \iso{19}F is eventually dominated by these <0.7 M\solar\ WDs, with their high ejecta masses but slower delay-time distributions. At earlier times, these isotopes are generally dominated by nova contributions from WDs with masses between 0.7 and 1.075 M\solar, with more massive WDs also being significant to the early production of \iso{15}N and  \iso{19}F. From \iso{20}Ne onwards, the eventual dominance of low-mass WDs is no longer assured. \iso{20}Ne and \iso{21}Ne contributions are dominated by the >1.075 O/Ne WD regime, before <1.075 M\solar\ WDs return to prominence in the production of \iso{22}Ne, formed as a decay product of \iso{22}Na. \iso{23}Na production is dominated by the most massive WDs (>1.3 M\solar), but the second most productive regime by 10 Gyr is the <0.7 M\solar\ mass WDs. \iso{24}Mg, \iso{25}Mg, and \iso{26}Mg are all eventually dominated by <0.7 M\solar\ WD contributions, while from \iso{27}Al to \iso{36}Ar (excepting \iso{36}S), WDs more massive than 1 M\solar\ dominate. \iso{36}S shows entirely different behaviour, with lower mass WDs being the most important contributors and massive WDs being the least.

Most isotopes do not exhibit a nice monotonic progression, with either the low-mass or high-mass WDs being dominant, but contributions from WD masses between these extremes being far less prominent. Each of these panels reflects the balance of several highly non-linear influences, and the somewhat chaotic behaviour in most of the isotopes reflects that. In general, the massive WDs benefit from faster delay-times and higher reaction rates (manifesting as higher mass fractions in our yield profiles), but are disadvantaged by the lower mass of nova ejecta that they produce (e.g., Fig. \ref{fig:hist_mwdzm3}). Massive white dwarfs contribute at significantly higher levels than other nova sites for \iso{30}Si, \iso{31}P, \iso{33}S, \iso{34}S, \iso{35}Cl, and \iso{37}Cl.

Fig. \ref{fig:iso_mwdhists20092020} is analogous to Fig. \ref{fig:iso_mwdhistj1998}, presenting our Milky Way model relying on the S2009/2020 yield profile. For now, we shall forgo a detailed description of the differences in the quantities of each isotope produced, and instead focus on changes to the qualitative behaviour of each isotope. Using this yield profile, <0.7 M\solar\ WDs remain significantly weaker producers of \iso{15}N than the other mass brackets throughout the simulation, with the 0.9-1.075 \Mwd\ mass bracket dominant. Other than massive WDs now dominating the production of \iso{18}O and \iso{19}F, the other low mass-number isotopes behave similarly to the J1998 case. \iso{20}Ne-\iso{34}S also show similar behaviour, although there is far more spread between different \Mwd\ contributions for \iso{23}Na, and \iso{34}S shows less prominent >1.3 M\solar\ contributions in the S2009/2020 model. Yields of isotopes from \iso{36}S-\iso{39}K were not reported by \cite{starrfield2009,starrfield2020}, while \iso{40}Ca was reported by \cite{starrfield2009,starrfield2020} but not by \cite{jose1998}. \iso{40}Ca becomes dominated by low-mass contributions beyond 1 Gyr, with high-mass WD contributions remaining low throughout.

Finally, the third of our nova yield profiles is presented in Fig. \ref{fig:iso_mwdhistj2020}. The J2020 profile is unique in that the yields do not rely on any premixing assumptions, instead combining 3D and 1D modelling to treat the evolution of the eruption \citep{jose2020}. However, this profile also includes distinct nova yields for only two different WDs: a 1.0 M\solar\ C/O WD and a 1.25 M\solar\ O/Ne WD. The evolution of low mass-number isotopes (\iso{3}He-\iso{19}F) is, qualitatively, mostly similar to the J1998 case, with low-mass WDs dominating. The main difference is that novae from massive WDs do not contribute to \iso{3}He at all. \iso{20}Ne-\iso{23}Na behave similarly to the J1998 and S2009/2020 yield profiles, although <0.7M\solar\ contributions to \iso{20}Ne are higher in J2020. Nova contributions from massive WDs dominate \iso{24}Mg production, differing from the J1998 and S2009/2020 profiles. \iso{25}Mg and \iso{26}Mg remain dominated by low-mass WD contributions. From \iso{27}Al-\iso{36}Ar the behaviour diverges significantly from the J1998 and S2009/2020 models; where previously massive WDs were dominant in most of this regime, the J2020 profile finds that low-mass WDs are dominant. This may be due to the poor \Mwd\ resolution of this profile; the high reaction rates obtainable on the most massive WDs are essential to their ability to overcome their low ejecta masses compared to low-mass WDs. For this profile, we are applying the yields computed by \cite{jose2020} for a 1.25 M\solar\ WD to all our massive WDs, while in the J1998 and S2009/2020 profiles we had the benefit of nova yields computed at 1.15, 1.25, and 1.35 M\solar. However, despite this limitation, the behaviour of \iso{38}Ar, \iso{39}K, and \iso{40}Ca appears similar to the J1998 and S2009/2020 cases.

Finally, we compare the quantitative variations at [Fe/H]~=~0, considering the different yield profiles and the different aspects of the \Mwd\ parameter space. Aspects of our previous discussion on the \Mwd\ parameter space can be seen more easily in Fig. \ref{fig:iso_mwdbarsum}, which allows for easier inspection of the relative productivities of the different yield profiles. Recall that the actual amount of raw mass ejected by the binary models from each of these simulations is the same; what changes between the models are the mass fractions in that nova ejecta.

Many of the previously discussed trends can be seen more easily in Fig. \ref{fig:iso_mwdbarsum}, such as the preference among the low mass-number elements -- excepting \iso{15}N and \iso{19}F -- for lower mass WDs, while new features become apparent. Three orders of magnitude more \iso{18}O is produced in the J1998 model from WDs less than 1.2 M\solar\ than either of the other models, with only S2009/2020's >1.3 M\solar\ WD contributions comparable to J1998's production for this isotope. Similar behaviour (although not as extreme) is exhibited by \iso{19}F, with the J1998 model again predicting far more material than the other models, excepting the most massive WD regime. Other isotopes where there is particular variation between the different models are \iso{3}He (especially at high \Mwd), \iso{7}Li (specifically the low values in the J2020 model, which were discussed briefly in \citealt{kemp2022li}), and the >1.3 M\solar\ regime of \iso{29}Si, \iso{33}S, \iso{34}S, \iso{35}Cl, and \iso{36}Ar.

Also illustrated in Fig. \ref{fig:iso_mwdbarsum} is that variation in productivity can often be non-monotonic with increasing WD mass regime. The low mass-number isotopes (\iso{3}He-\iso{19}F) are the most consistent, with all -- once again excepting \iso{15}N and \iso{19}F -- models showing an unambiguous decline in productivity with increasing mass regime. We find no examples of isotopes where productivity increases monotonically and unambiguously with increasing WD mass, but there are many examples where increasing the WD mass first reduces and then increases the productivity. The resulting `U' curve is not unexpected; from Figs. \ref{fig:hist_mwdzm3} and \ref{fig:hist_mwdz0p02}, we can see that, in a simple stellar population, a huge part of the ejecta mass of our population originates from the <0.7 M\solar\ regime. The sharp fall-off in nova ejecta with increasing \Mwd\ mass results in this initial decline before the ejecta mass-\Mwd\ distribution flattens -- to a very low value -- beyond 1 M\solar, allowing the increased nuclear efficiency of novae on the massive WDs to compensate for their low ejecta masses. This `U' behaviour is almost ubiquitous among the high mass-number isotopes, usually with the high-mass WDs having the highest contributions overall.

\section{Discussion and conclusions}
\label{sec:chap5_conc}
In this work, we explore the nucleosynthesis of stable isotopes by novae in the context of galactic chemical evolution, with attention to which parts of the \Mwd-parameter space contribute to which elements. 

We discussed the influences that determine which parts of the nova parameter space produce the most nova ejecta, focusing on how this differs from which nova systems produce the most nova events. Lower mass WDs have higher ejecta masses per eruption due to their higher critical ignition masses, a fact that disadvantages them when considering the number of novae that such a system produces but does not when considering the ejecta mass. While some nova systems with massive WDs can produce amounts of nova ejecta comparable to the most productive low-mass nova systems, they are far fewer in number, as they are heavily selected against by the initial mass function.

When comparing with AGB stars, massive stars, and type Ia supernovae, we find that novae do not dominate the elemental abundances other than Li (see \citealt{kemp2022li}). However, we find that novae do contribute sufficient \iso{13}C and \iso{15}N (up to 35\% by [Fe/H]~=~0) to noticeably affect the abundances of these isotopes at certain times during the evolution of the Galaxy. The evolution of [\iso{15}N/Fe] and \iso{14}N/\iso{15}N from [Fe/H]~=~-2 to -1 in particular shows significant variation (up to 0.7 dex) dependent on the inclusion of novae and the nova yield profile adopted. Unfortunately, there are no observations constraining \iso{15}N abundances at this metallicity, nor are there likely to be in the future. The variation due to novae in the predicted abundance ratios of \iso{13}C and \iso{15}N at solar metallicity, where we do have observations courtesy of isotopic analyses of meteorites \citep[e.g.,][]{lodders2009}, is far smaller, at most around 0.2 dex.

Recent work by \cite{bekki2024} has found that novae on O/Ne WDs contribute significantly to the Galactic [P/Fe] abundance ratio. In this work, as previously discussed, we find that nova contributions to P -- and most other elements -- are negligible compared with other stellar sources. \cite{bekki2024}'s parametric model for novae borrows many of its parameters from \cite{kemp2022}, notably the high contribution of O/Ne WDs to the total number of novae. They combine this with an estimate for the total number of novae, an assignment for the ejecta mass per eruption, and a gradient for a metallicity-dependent nova rate to form their model for nova ejecta, relying on yields from \cite{jose1998}'s massive WD simulations for the mass fraction of P in each eruption. 

There are other details to the model, but these are the most relevant to explaining their high levels of P from novae compared to this work. \cite{bekki2024} consider a range of nova ejecta, but the lowest considered is \tento{-5} M\solar\ per nova event. As can be seen in Figs. \ref{fig:hist_dmproczm3} and \ref{fig:hist_dmprocz0p02}, this is actually very high for O/Ne WDs; most have at least an order of magnitude less ejecta per eruption, and this is particularly true for the most massive WDs (where P production will be the highest, as assumed by \citealt{bekki2024}). As discussed previously in this work, the high importance of ONe WDs to nova counts is inextricably linked to their low critical ignition masses, which translates to lower ejecta masses. it is almost certainly the decoupling of these two factors (by taking the high fraction of O/Ne WDs implied by \cite{kemp2022}'s nova rate calculations but also assuming a relatively high ejecta mass for each eruption) that leads to the high contributions of novae to Galactic [P/Fe] in \cite{bekki2024}.

These kinds of parametric models are typical of how novae have been included within GCE models in the past. Indeed, many of the assumptions necessary to the nova model \cite{bekki2024} are very similar to \cite{romano2003}, where a parametric model was used to test the effect of novae on C/N/O abundance ratios, finding that novae have potential to account for secondary production of certain isotopic ratios. Similarly to \cite{bekki2024}, \cite{romano2003} also assumed \tento{4} novae per system and \timestento{2}{-5} solar masses of ejecta per nova, implying roughly 0.2 solar masses of material per nova system on average. Comparing these assumptions with Table \ref{tab:zsumtabupdated}, this is roughly 10 times more ejecta per nova system on average than we find in our models (between 0.02 and 0.03 M\solar\ per system, depending on the metallicity). This is likely the reason we find relatively little impact from novae on these C/N/O abundance ratios compared to older works.

More recent works have linked novae to explaining details of the [F/Fe] distribution, finding good agreement when nova contributions are increased by factors of a few and yields from massive O/Ne WDs are employed \citep{spitoni2018,grisoni2020}, but negligible impact without this artificial enhancement. We are in agreement with these works insofar as that our models, which do not include artificial enhancements, also find negligible contributions from novae.

\cite{kemp2022li} highlighted that theoretical nova models severely under-produce Li relative to observations of nova ejecta. A natural question to ask is whether they also under-produce other elements. No direct observations of isotopic abundances have been made in nova ejecta; therefore, we must rely upon elemental abundances to provide a sanity check. \cite{helton2010} provide detailed spectroscopic analysis of the dusty nova V1065 Cen, reporting a O/Ne white dwarf composition and logarithmic He~($-0.89\pm0.11$), N~($-2\pm0.11$), O($-1.54\pm0.17$), Ne~($-1.57\pm0.09$), Mg~($-2.75\pm0.11$), S~($-3.14\pm0.37$), Ar~($-3.90\pm0.13$), and Fe~($-0.89\pm0.13$) abundances. In comparison, converting the reported yields from the `ONe5' model of a 1.15 M\solar O/Ne WD from \cite{jose1998} to abundances results in: He~(-0.7); N~(-1.72, including \iso{14,15}N contributions); O~(-1.6, including \iso{16,17,18}O contributions); Ne~(-1.5); Mg~(-3.4, including \iso{24,25,26}Mg contributions); and S~(-4.0). S and -- to a lesser extent -- Mg are both under-predicted, but recomputing these abundances using \cite{jose1998}'s 1.25 M\solar\ model (`ONe6') results in abundances of Mg~(-3.0) and S~(-2.2), reducing the discrepancy in Mg and bracketing the S abundance. This leads us to conclude that, at least where observations are available, the theoretical yields are not necessarily in tension for most elements. It is beyond the scope of this discussion to repeat this assessment for a comprehensive set of novae with elemental yields. However, we note that \cite{jose1998} compare their theoretical mass fractions with the observational yields of five different novae (V693 CrA 1981, V1370 Aql 1982, QU Vul 1984, PW Vul 1984, and V1688 Cyg 1978), finding that -- allowing for the possibility of different mixing fractions between novae -- their theoretical yields match relatively well.

In this work, we only considered the evolution of stable isotopes. We therefore can make no comment on the importance of novae to the production of radioactive isotopes in the Milky Way. However, of the radio-isotopes expected to be produced by novae, only \iso{26}Al has a significantly long half-life ($t_{1/2} \approx$ \timestento{7}{5} years). We note that recent GCE modelling by \cite{vasini2022} find that including novae is necessary to recover the mass of \iso{26}Al derived from COMPTEL and INTEGRAL $\gamma$-ray observations \citep{schonfelder1984,diehl1995,prantzos1996,winkler1996,diehl2010}.

We examined the relative importance of different nova sites for all considered isotopes, where different sites are distinguished according to the WD mass. We found that, by the time the Galaxy has reached [Fe/H]~=~0, low-mass WDs dominate the production of most low mass-number isotopes ($\leq$\iso{19}F), while massive WDs dominate most of the more massive isotopes. Agreement between the three nova yield profiles is relatively good for most isotopes. Isotopes where the agreement is relatively poor -- at least in certain \Mwd\ regimes -- include \iso{3}He (especially at high \Mwd), \iso{7}Li, \iso{18}O, \iso{19}F, and the >1.3 M\solar\ regime of \iso{29}Si, \iso{33}S, \iso{34}S, \iso{35}Cl, and \iso{36}Ar.

Ultimately, we find that novae play a non-negligible role in the origin of the elements. They appear to dominate Galactic Li production when considering observed Li yields \citep{kemp2022li}. Further, we find that novae may produce up to 35\% of Galactic \iso{13}C and \iso{15}N by [Fe/H]~=~0, and that novae may -- briefly -- be the dominant sources of \iso{15}N between [Fe/H] $= -2$ and $-1$, producing at a level at least comparable with massive stars. We have not formally explored the affect that uncertain binary evolution parameters could have on the evolution of \iso{13}C and \iso{15}N, and we regret that such an analysis must be left to future work. However, we comment that the impact of novae on these isotopes is far less than it was for Li, and so it is certainly possible that there are combinations of uncertain binary parameters that could reduce nova contributions to \iso{13}C and \iso{15}N significantly.

\begin{acknowledgements}

A.~R.~C. is supported in part by the Australian Research Council through a Discovery Early Career Researcher Award (DE190100656).
B.~C. acknowledges support from the National Science Foundation (NSF, USA) under grant No. PHY-1430152 (JINA Center for the Evolution of the Elements).
R.~G.~I. thanks the STFC for funding, in particular Rutherford fellowship ST/L003910/1 and consolidated grant ST/R000603/1.
Parts of this research were supported by the Australian Research Council Centre of Excellence for All Sky Astrophysics in 3 Dimensions (ASTRO 3D), through project number CE170100013.
\end{acknowledgements}

\bibliographystyle{aa}
\bibliography{bibfile.bib}

\appendix
\section{Additional figures}

\begin{figure*}
\centering
\includegraphics[width=1.9\columnwidth]{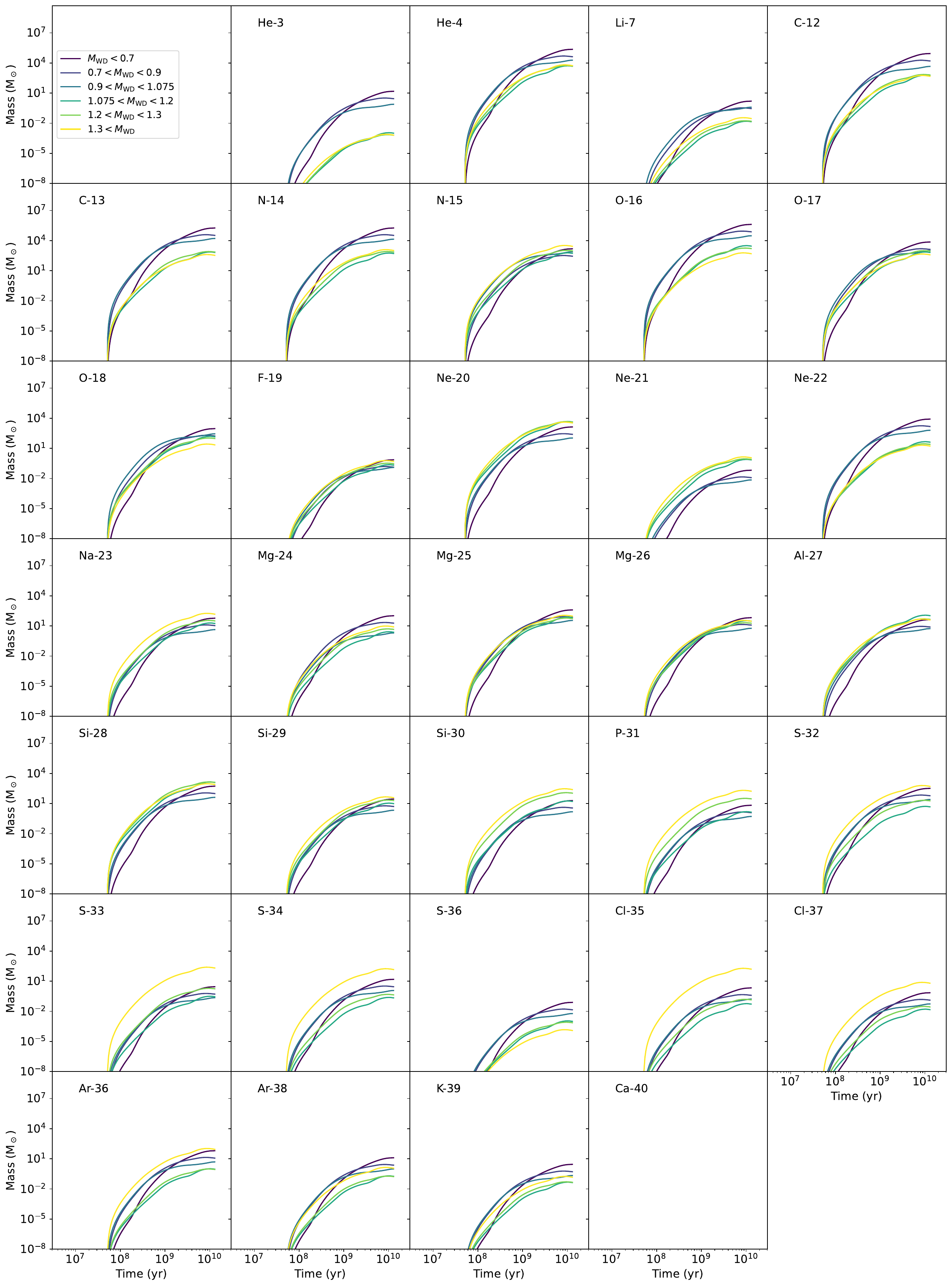}
\caption{Contributions to each isotope according to the J1998 yield profile, distinguishing between nova progenitor WD masses.}
\label{fig:iso_mwdhistj1998}
\end{figure*}

\begin{figure*}
\centering
\includegraphics[width=1.9\columnwidth]{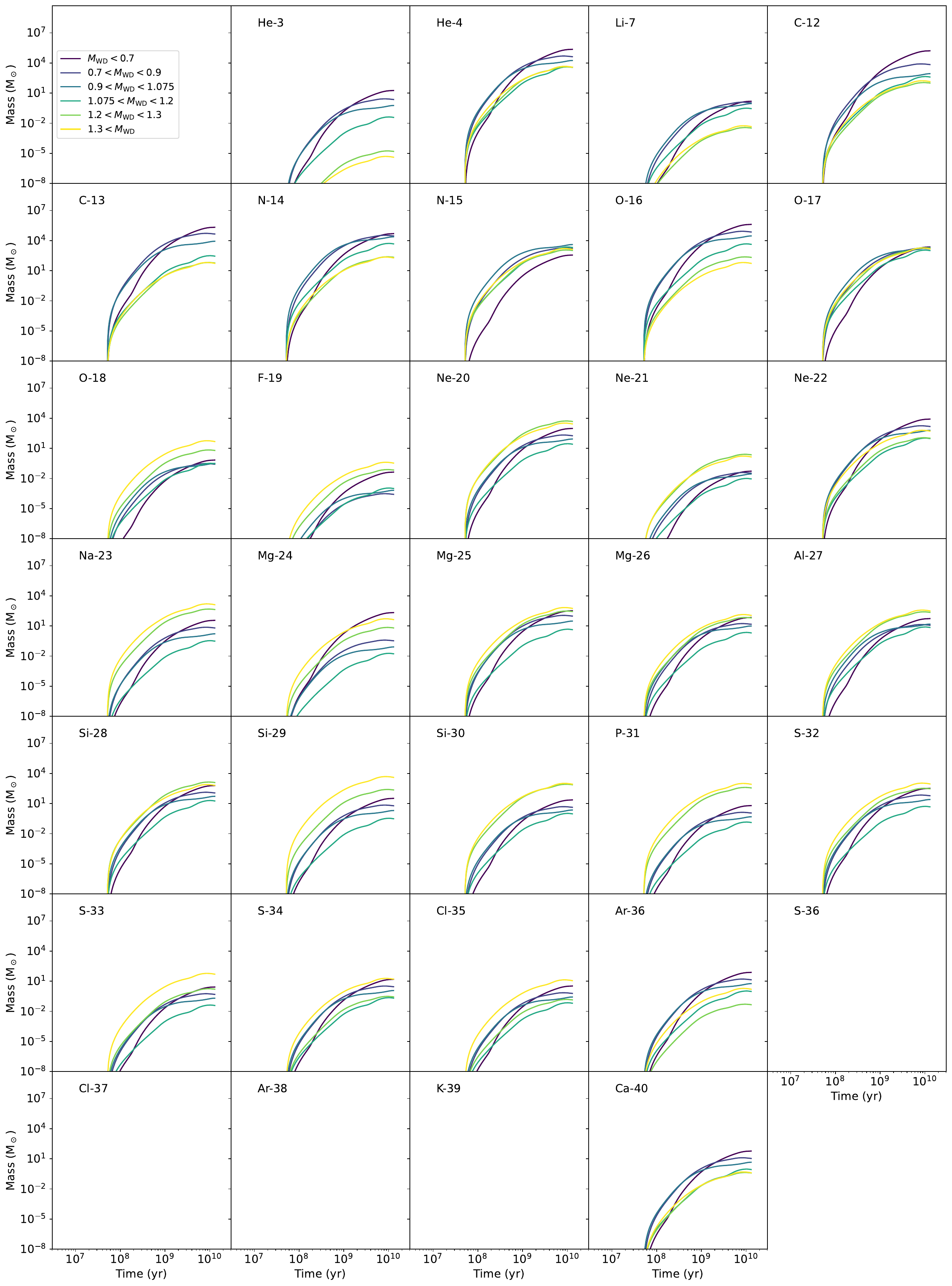}
\caption{Contributions to each isotope according to the S2009/2020 yield profile, distinguishing between nova progenitor WD masses.}
\label{fig:iso_mwdhists20092020}
\end{figure*}

\begin{figure*}
\centering
\includegraphics[width=1.9\columnwidth]{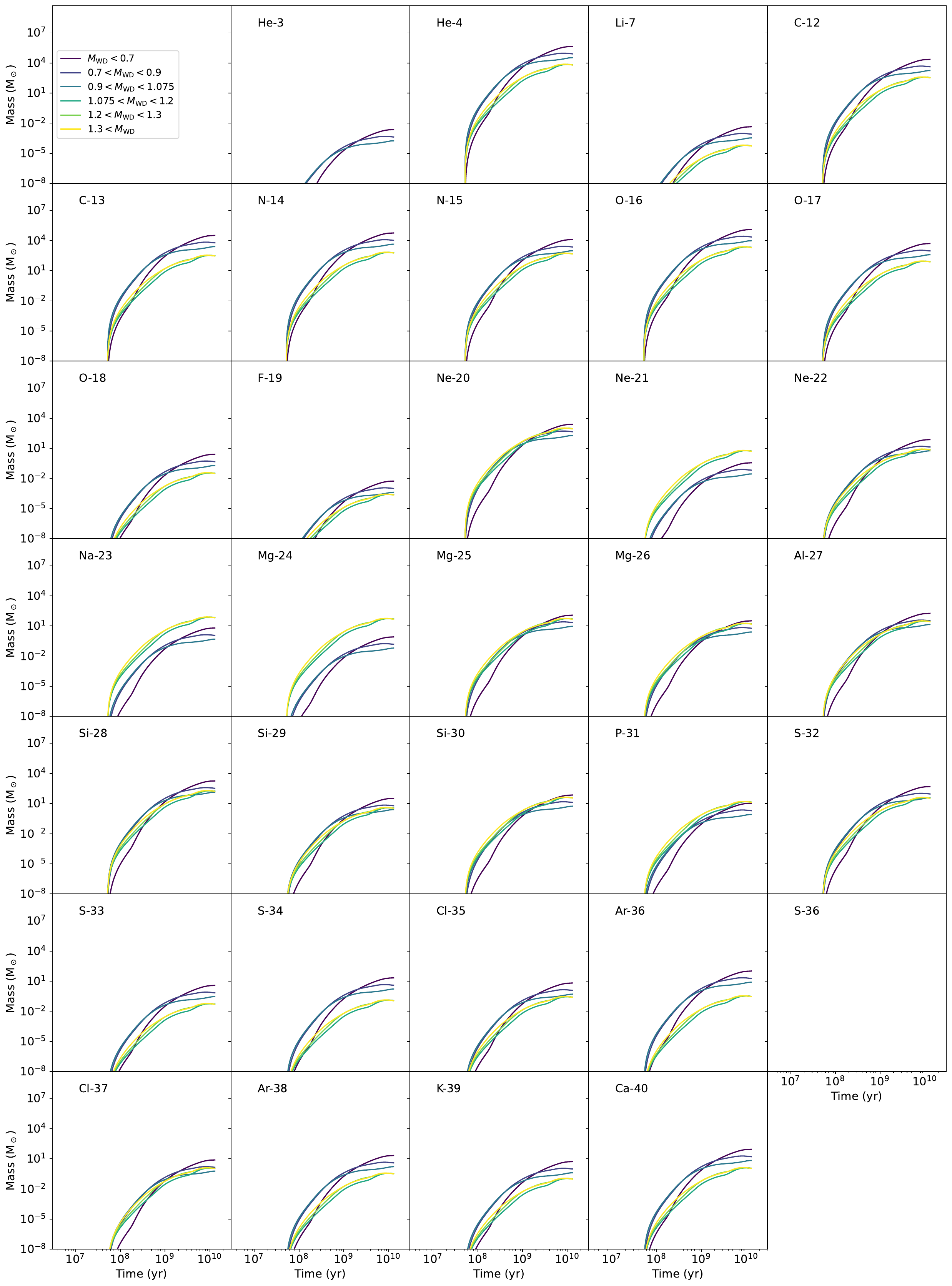}
\caption{Contributions to each isotope according to the J2020 yield profile, distinguishing between nova progenitor WD masses.}
\label{fig:iso_mwdhistj2020}
\end{figure*}

\end{document}